\PassOptionsToPackage{table}{xcolor}
\documentclass[acmsmall,screen,nonacm]{acmart}

\setcopyright{none}
\renewcommand\footnotetextcopyrightpermission[1]{}

\usepackage{amsmath}
\usepackage{booktabs}
\usepackage{array}
\usepackage{graphicx}
\usepackage{multirow}
\usepackage{makecell}
\usepackage{enumitem}
\usepackage{wrapfig}
\usepackage{listings}
\usepackage[most]{tcolorbox}
\usepackage{tikz}
\usetikzlibrary{positioning,arrows.meta,fit,backgrounds,calc}
\usepackage{microtype}
\usepackage{xurl}
\usepackage{needspace}

\graphicspath{{figures/arxiv/}}

\setlist{nosep,leftmargin=*}
\definecolor{findingblue}{RGB}{33,79,122}
\definecolor{findingback}{RGB}{243,247,251}
\definecolor{rowgray}{gray}{0.94}
\definecolor{negcell}{RGB}{248,215,210}   
\definecolor{poscell}{RGB}{214,228,244}   
\definecolor{vulnback}{RGB}{252,236,233}  
\definecolor{fixback}{RGB}{232,245,236}   
\definecolor{ruleback}{RGB}{238,238,245}  
\definecolor{tprblue}{RGB}{44,110,165}    
\definecolor{fprred}{RGB}{192,57,43}      
\newcommand{\nc}[1]{\cellcolor{negcell}#1}
\newcommand{\pc}[1]{\cellcolor{poscell}#1}
\newcommand{\zebra}{\rowcolors{2}{rowgray}{white}}

\newtcolorbox{findingbox}{enhanced,breakable,boxrule=0pt,frame hidden,
  borderline west={1.1mm}{0pt}{findingblue},colback=findingback,
  left=1.6mm,right=1.4mm,top=0.6mm,bottom=0.6mm,before skip=5pt,after skip=5pt}
\newcommand{\finding}[2]{\begin{findingbox}\small\textbf{Finding #1.} #2\end{findingbox}}

\lstdefinestyle{snip}{basicstyle=\fontsize{5.8}{6.9}\selectfont\ttfamily,columns=fullflexible,
  keepspaces=true,showstringspaces=false,breaklines=false,frame=none,aboveskip=0pt,belowskip=0pt,
  xleftmargin=0pt,escapeinside={(*}{*)}}
\newcommand{\head}[1]{\vspace{2pt}\noindent\textbf{#1}\hspace{0.4em}}
\newcommand{\dF}{\Delta\mathrm{F1}}
\newcommand{\RQ}[1]{\textbf{RQ#1}}

\begin{document}

\title{Fixed Detectors, Moving Code: How Well Do Static Security Analyzers Hold Up over a Decade of C/C++ Evolution?}

\author{Junhua Tang}
\affiliation{%
  \institution{University at Buffalo, SUNY}
  \city{Buffalo}
  \state{NY}
  \country{USA}}
\email{junhuata@buffalo.edu}

\author{Zhengsong Zhang}
\affiliation{%
  \institution{University at Buffalo, SUNY}
  \city{Buffalo}
  \state{NY}
  \country{USA}}
\email{zhengson@buffalo.edu}

\author{Haipeng Cai}
\affiliation{%
  \institution{University at Buffalo, SUNY}
  \city{Buffalo}
  \state{NY}
  \country{USA}}
\email{haipengc@buffalo.edu}

\begin{abstract}
Static application security testing (SAST) tools are adopted on the expectation that a detector set up today will keep finding vulnerabilities in code written later, including previously unknown vulnerabilities. This expectation is rarely tested. In this paper, we test it on 20,000 vulnerable and fixed C/C++ cases from 1,116 open-source projects with source years 2016--2025, under three CodeQL releases, three Infer releases, and one Semgrep configuration. We conduct three studies in which we progressively control for changes in code composition to isolate SAST tool performance differences associated with code evolution over time: yearly populations, project- and label-matched year pairs, and within-project trajectories. We then trace the largest within-project drops to repair diffs, rule and query predicates, alert locations, and execution logs, and we account for failed, degraded, and costly runs. 
Under our changed-file case-label scoring, the overall F1 of these tools ranges from 0.066 to 0.478, yet every overall count-based AUROC lies between 0.495 and 0.526. For CodeQL, higher F1 in newer releases is accompanied by nearly parallel increases in recall and false-positive rate, rather than better vulnerable/fixed separation.
Across the full project-matched year-pair analysis, the mean later-minus-earlier F1 for CodeQL R2024 is negative at every source-year gap, whereas Infer v1.1 and v1.3 are positive at every gap. 
In the selected drops, only 6 of 39 CodeQL findings and 2 of 620 Infer findings touch a changed repair line, 
and 78.96\% of Semgrep cases carry parser or timeout diagnostics although every run reports success. 
Overall, our results show that apparent temporal changes in SAST effectiveness can reflect broader alerting, shifting analysis coverage, or incidental findings as well as genuine losses in defect coverage. Temporal sustainability should therefore be evaluated at the level of a fixed detector condition, with project controls, repair-level evidence, and explicit accounting of failed and degraded analyses.
\end{abstract}



\maketitle

\section{Introduction}\label{sec:introduction}

Static application security testing (SAST) analyzes source or compiled code without running it, so it can flag a weakness before the code ships. The value of that early warning lies in the future. A detector adopted this year is expected to keep flagging the same classes of defects in code that will be written next year, and it matters most when a vulnerability has no identifier, exploit, or patch yet. In one large study, zero-day attacks went undisclosed for 312 days on average~\cite{bilge2012before}, and vulnerable code commonly survives in open-source projects for years~\cite{alexopoulos2022vulnerabilities}. A SAST tool cannot know a future zero-day by name, but its queries, abstract domains, and patterns may still cover the defect, provided that the tool can parse, build, and analyze the code in which the defect appears.

Both sides of that interaction change. Programs evolve continuously~\cite{lehman1980programs}: languages gain idioms, build systems are replaced, and dependencies move. Historical snapshots can become unbuildable; in one study only 38\% of the change history of 100 Java projects could still be compiled~\cite{tufano2017there}. Analyzers change as well, through new engines, query packs, rule sets, and recovery behavior. Noirot Ferrand et al.\ ran 114 CodeQL versions on 3,993 CVEs and found that 21 detected CVEs were no longer detected after a version change~\cite{ferrand2026longitudinal}. Most SAST evaluations, however, are cross-sectional. They measure how many known vulnerabilities a fixed tool set finds at one point in time and report that much is missed~\cite{li2023comparison,lipp2022empirical,habib2018many,bennett2024semgrep}. They do not say whether a fixed detector keeps working as the code it analyzes grows older.

Answering that question is harder than it looks. Suppose a detector scores F1=0.30 on code from 2016 and 0.10 on code from 2024. The drop may reflect aging, or it may reflect that the 2024 sample contains different projects, a different label balance, larger files that the parser handles only partially, or more runs that timed out and were silently dropped. The same pitfalls are well documented for learning-based security systems, where random splits and mixed time periods inflate results~\cite{pendlebury2019tesseract,arp2022dos,risse2024uncovering}. SAST adds two more. An alert can disappear because the relevant file was never scanned, and a positive score can come from an alert unrelated to the labelled vulnerability.

This paper studies the \emph{temporal sustainability} of SAST: under a fixed, documented detector condition, do effectiveness, executability, and cost remain adequate as the analyzed code moves forward in time? We evaluate 20,000 C/C++ vulnerable and fixed cases from 1,116 projects and source years 2016--2025 under seven detector conditions: CodeQL releases from 2021, 2024, and 2026; Infer v1.1.0, v1.2.0, and v1.3.0; and one Semgrep configuration with 170 rules. Three complementary designs separate a change in the code population from a change over time: yearly populations (Study~A), exactly project- and label-matched year pairs (Study~B), and balanced within-project trajectories (Study~C). We then examine the largest within-project drops at the level of repair diffs, alert locations, rule or query predicates, traces, and logs, and we analyze execution failures and cost on deduplicated targets. The main findings are as follows.
\begin{itemize}
\item \emph{A higher score is not better detection} (\S\ref{sec:study-a}). F1 ranges from 0.066 (Infer v1.1) to 0.478 (Semgrep), yet every overall count-based AUROC lies between 0.495 and 0.526. Newer CodeQL releases more than double F1 by raising recall and FPR together, Infer's 2024 release outperforms its 2026 release, and one generic Semgrep rule produces 63\% of all qualifying findings.



\item 
\textit{Temporal degradation is detector-condition specific} 
(\S\ref{sec:study-b}). Fixed historical detector conditions exhibit later-year degradation where post-release observations are available. Across the broader project- and label-matched analysis, CodeQL R2024 has negative mean \(\Delta F1\) at all nine source-year gaps (up to $-0.16$), whereas Infer v1.1 and v1.3 are positive at every gap and Semgrep changes little at most gaps. 
Differences between year pairs are larger than most gap means, so no tool-wide ``F1 lost per year'' exists. 



\item \textit{Aggregate effectiveness and within-project temporal trajectories need not agree} (\S\ref{sec:study-c}). Later CodeQL releases have higher F1 but more projects with negative descriptive F1 slopes (37, 74, and 85 of 202), and most flat Infer trajectories are flat only because F1 is zero in every year (144 to 176 of 202 projects).

\item 
\textit{Several selected extreme F1 drops are driven by off-repair or scope-misaligned findings} 
(\S\ref{sec:failure-analysis}). In the selected drops, only 6 of 39 CodeQL findings and 2 of 620 Infer findings touch a changed repair line, no Semgrep positive reaches the labelled repair, and 11 of 20 Semgrep vulnerable/fixed groups are scored on disjoint files. The genuine misses we found arise from rule predicates that do not encode the defect and from code that never reaches the analysis.
\item \emph{A successful run can hide a degraded analysis} (\S\ref{sec:operations}). Execution status changes in both directions across CodeQL releases (281 vs.\ 177 targets), Infer turns thousands of empty or failed captures into scored successes, 78.96\% of Semgrep cases carry parser or timeout diagnostics under a successful status, and two range-analysis queries alone can exhaust a 12-hour budget.
\end{itemize}

\head{Contributions.} This paper makes four contributions.
\begin{itemize}
\item A definition of SAST sustainability that covers effectiveness, executability, and cost, and that ties the property to prospective vulnerability discovery (\S\ref{subsec:sustainability}).
\item A three-estimand design that separates yearly ecosystems, matched populations, and within-project trajectories, with explicit units of inference (\S\ref{subsec:studies}).
\item A large longitudinal evaluation of seven detector conditions from three analyzer families over 20,000 cases, 45 year pairs, 100 resamples, and 202 projects (\S\ref{sec:study-a}--\S\ref{sec:study-c}).
\item A triangulation of scores with repair diffs, alert locations, rule and query predicates, traces, controlled local experiments, and execution ledgers, yielding concrete guidance for tool builders, practitioners, and benchmark designers (\S\ref{sec:failure-analysis}--\S\ref{sec:discussion}).
\end{itemize}

\head{Paper organization.} 
Section~\ref{sec:background} explains what can age in a SAST tool and gives a motivating example. Section~\ref{sec:methodology} describes the corpus, the seven detector conditions, the scoring rule, and the three studies in enough detail to reproduce them. Sections~\ref{sec:study-a}--\ref{sec:operations} answer the five research questions, one tool at a time, and report every per-year, per-gap, and per-cohort table we have. Each results section ends with a short boxed finding. Section~\ref{sec:discussion} draws the results together, Section~\ref{subsec:threats} discusses threats to validity, and Section~\ref{sec:related-work} relates the study to prior work.

\section{Background and Motivation}\label{sec:background}

\subsection{SAST Families and What Can Age}\label{subsec:sast}

SAST covers lightweight pattern matching, data-flow and taint analysis, abstract interpretation, and database-backed semantic queries~\cite{bessey2010few,sadowski2018lessons,distefano2019scaling}. Each family depends on interfaces that evolve over time. Table~\ref{tab:families} summarizes the three analyzers we study and the components through which time can affect them.

\begin{table}[!tb]
\caption{The three analyzers, their analysis style, and the components through which code evolution or tool evolution can change their results.}
\label{tab:families}
\centering\footnotesize
\setlength{\tabcolsep}{3pt}
\zebra
\begin{tabular}{@{}p{0.07\linewidth}>{\raggedright\arraybackslash}p{0.35\linewidth}>{\raggedright\arraybackslash}p{0.19\linewidth}>{\raggedright\arraybackslash}p{0.33\linewidth}@{}}
\toprule
\textbf{Tool} & \textbf{Analysis style} & \textbf{Input it needs} & \textbf{Time-sensitive components}\\
\midrule
CodeQL & Relational database of the program; declarative QL queries over syntax, control and data flow~\cite{demoor2007ql,avgustinov2016ql} & Extraction from a traced build or buildless extraction & Extractor and compiler discovery; query packs and their models; query evaluation cost\\
Infer & Compositional analysis with procedure summaries, rooted in separation logic and bi-abduction~\cite{calcagno2011compositional,calcagno2015moving,ohearn2018continuous} & Captured compilation (build interception) & Build capture and dependency availability; checker set and models; analysis budget\\
Semgrep & Structural pattern matching with constant propagation and intrafile taint rules~\cite{semgrep2026glossary} & Source files, no build & Parser coverage of constructs; rule definitions and severities; per-rule timeouts\\
\bottomrule
\end{tabular}
\end{table}

The table makes one point that the rest of the paper relies on. A SAST result is the product of an engine, a set of rules or queries, an extraction or parsing step, and an execution policy. When any of these components meets code it was not built for, the result can change even if nobody touched the defect-detection logic. We therefore treat the whole bundle, which we call a \emph{detector condition}, as the object under study.

\subsection{Sustainability and Zero-Day Relevance}\label{subsec:sustainability}

Let $D_r$ be a detector condition (engine release, rules or queries, mode, and execution policy) and $C_t$ the population of code from source year $t$. We call $D_r$ \emph{effectiveness-sustainable} when $\mathrm{Perf}(D_r,C_t)$ does not degrade with $t$ after known composition changes are controlled; \emph{reliability-sustainable} when $D_r$ keeps parsing, building, extracting, and finishing on $C_t$; and \emph{cost-sustainable} when its time and memory remain within practical budgets. Concept drift, a change in the relation between inputs and labels over time~\cite{gama2014survey}, is one cause of degradation. SAST can also degrade without any learned decision boundary, simply because the code no longer reaches the analysis.

Our full longitudinal analyses characterize how a fixed detector condition responds to code from different source years. We use aging more specifically when later code follows the frozen detector condition in time; comparisons involving code that predates the condition characterize temporal sensitivity but not prospective aging.

The link of our study to zero-days is prospective rather than evidentiary. A zero-day has no case label when it is discovered, so no retrospective corpus measures zero-day discovery directly. Our corpus tests a necessary condition: whether detector logic and execution still apply to defect-bearing code as that code moves away from the context in which the detector condition was built. A sustainable detector may still miss a zero-day. A detector that has lost applicability has one more, measurable reason to miss it.

\subsection{Motivating Example}\label{subsec:motivating}

HAProxy illustrates why an aggregate score can mislead. In our corpus, the 2019 and 2023 HAProxy cells each contain one vulnerable and one fixed case per resampling repetition. Under CodeQL R2021, F1 falls from 1.0 in 2019 to 0.0 in 2023, which looks like a clear loss. The 2019 alerts, however, are in \texttt{src/proto\_htx.c}, while the paired fix changes \texttt{src/mux\_h2.c}; once only alerts in the repaired file are counted, the 2019 F1 is already zero. Under R2024 the 2023 cell has F1=$2/3$, which looks like a gain. Here an overflow-style warning on \texttt{if (unlikely(cl + c < cl))} appears on both the vulnerable and the fixed snapshot (the line only moves from 57 to 64), because the repair leaves that check alone and instead rejects an empty Content-Length header. The cell therefore has one true positive and one false positive, and its FPR is 1.0. Under R2026 the same cell loses the true positive but keeps the false positive, so its F1 is zero again. None of these changes says anything about the CVE. Section~\ref{sec:failure-analysis} shows that such cases are common among the largest changes.

\section{Study Design}\label{sec:methodology}

\subsection{Research Questions}\label{subsec:rqs}

We ask five research questions. RQ1--RQ3 and RQ5 are quantitative; RQ4 is explanatory.
\begin{itemize}
\item \RQ{1} (ecosystem effectiveness). How do effectiveness and coverage vary across source years and detector conditions?
\item \RQ{2} (project-controlled change). After project and label composition are matched, how does later-year performance differ from earlier-year performance as the year gap grows?
\item \RQ{3} (within-project change). As the same project evolves, how often and how strongly does detector performance change?
\item \RQ{4} (failure mechanisms). Which scope, rule or query, parsing, extraction, and execution mechanisms explain the largest within-project drops?
\item \RQ{5} (reliability and cost). How do execution failures, degraded analyses, and measured costs differ across conditions, and how often does the same target change status?
\end{itemize}

\subsection{Corpus and Temporal Identity}\label{subsec:dataset}

The frozen corpus contains 20,000 C/C++ cases: 1,000 vulnerable and 1,000 fixed cases for each source year from 2016 to 2025, drawn from 1,116 canonical projects. The source year is the year of the analyzed snapshot.

A case is a single commit snapshot of a C/C++ repository, with the files changed by the associated security fix and the fix diff. A fixed case is the fix commit named by a vulnerability record; a vulnerable case is the commit the record marks as vulnerable or, if none is given, the parent of the fix commit. Cases come from NVD (15,776), PrimeVul (3,854), and OSS-Fuzz/OSV (370) records. Each year holds 1,000 vulnerable and 1,000 fixed cases drawn at random from these candidates. A vulnerability record (vulnerability ID such as a CVE number, repository, and commit) listed by several sources was kept only once among the vulnerable or fixed cases; different CVEs fixed by the same commit are kept as separate cases.

Projects hosted under several URLs (e.g., GitHub and git.kernel.org copies of Linux) are merged into one canonical project, while related but distinct codebases such as ImageMagick and ImageMagick6 are kept separate.

\subsection{Detector Conditions}\label{subsec:subjects}

We chose CodeQL, Infer, and Semgrep because they are widely used, open to inspection, and built on three different analysis architectures (Table~\ref{tab:families}), and we picked release points that span the source years of the corpus. Table~\ref{tab:conditions} lists the seven conditions. Because engine, rules, extraction, and recovery change together, our comparisons across releases are descriptive. They do not isolate a pure engine-version effect, and we never read them as an estimate of what a version upgrade does.

\begin{table}[!tb]
\caption{The seven detector conditions. 
A condition comprises an analyzer release and its associated rules, extraction, and execution/recovery policy; target-specific recovery differences are tracked as provenance. 
}
\label{tab:conditions}
\centering\footnotesize
\setlength{\tabcolsep}{4pt}
\zebra
\begin{tabular}{@{}lllll@{}}
\toprule
\textbf{Condition} & \textbf{Engine} & \textbf{Release date / bundle} & \textbf{Rules or queries} & \textbf{Execution}\\
\midrule
CodeQL R2021 & CLI 2.7.3 & bundle 2021-12-08 & cpp queries 0.0.4 & traced (manual) build\\
CodeQL R2024 & CLI 2.20.0 & bundle v2.20.0 & cpp queries 1.3.0 & buildless (build-mode none)\\
CodeQL R2026 & CLI 2.25.2 & --- & cpp queries 1.7.0 & buildless, release-level recovery\\
Infer v1.1 & 1.1.0 & 2021-03-26 & bundled defaults & capture + analyze\\
Infer v1.2 & 1.2.0 & 2024-06-21 & bundled defaults & capture + analyze\\
Infer v1.3 & 1.3.0 & 2026-05-12 & bundled defaults & capture + analyze\\
Semgrep & 1.161.0, OSS mode & --- & 170 local rules & scan of changed files\\
\bottomrule
\end{tabular}
\end{table}

\head{CodeQL.} The 2021 condition uses traced compilation through a saved build wrapper. The 2024 condition uses buildless extraction, in which CodeQL's autobuild step discovers a compiler and extracts the source without running the project's build. R2026 is a single release-level condition. Its native security suite (\texttt{security-extended} in cpp queries 1.7.0) has 98 queries. During recovery, some R2026 targets were rerun with a different extraction scope, a reduced query set, or cached results; we keep these differences as provenance of one condition rather than treating them as extra releases. The query lists that each historical release actually resolved were saved, and they differ: for example, the R2021 list lacks \texttt{cpp/missing-check-scanf} and \texttt{cpp/use-after-free}, which the R2024 list contains, and neither list selects \texttt{MissingNullTest} or \texttt{InconsistentNullnessTesting}. These differences matter in \S\ref{sec:failure-analysis}. 


\head{Infer.} Each Infer condition uses the native binary, models, and default checkers shipped with its release. The command sequence is the same for all three. \texttt{infer capture --keep-going} runs over the full checked-out source tree into a version-specific output directory. \texttt{infer analyze --keep-going} then runs with eight jobs and the buffer-overrun analysis enabled explicitly; when the case's changed-file index is nonempty, it is passed with \texttt{--reactive}, and an empty index is recorded as an explicit fallback to full analysis. Finally, \texttt{infer report} runs against the same output. The baseline stage budgets are 3,600 seconds for workspace setup, 1,800 for project setup, 7,200 each for capture and analysis, 120 for reporting, 600 for cleanup, and 21,600 end to end. Recovery configurations are versioned supplements with larger budgets and explicit repairs. Recovery rounds 1--9 allow 21,600 seconds for capture and 43,200 for analysis (72,000 end to end), and rounds 10--13 allow 43,200 and 86,400 (144,000 end to end). The later rounds supply 95 v1.1 and 215 v1.3 selected results. Several completed candidate runs can exist for one case. The selection rule, inherited from the original v1.2 workflow, keeps the candidate with the highest changed-file score within each identity-aware reuse group. Results are reused across cases only when commit, Git tree, input fingerprint, changed files, and labels all match; a shared commit alone is not enough. 


\head{Semgrep.} Semgrep runs in OSS mode with 170 C/C++ rules: 141 search rules and 29 taint rules. Their metadata marks 16 as community rules and 154 as Pro-authored, so rule provenance and engine mode are separate properties. By severity, 109 rules are INFO, 38 WARNING, 21 ERROR, and 2 CRITICAL; only the 61 rules at WARNING or above count toward the score. The configuration is not purely syntactic: it includes taint rules and semantic rules such as \texttt{null-library-function} and \texttt{local-variable-new-delete}. Each case runs \texttt{semgrep scan} over the directory of exported changed files, with JSON output and recorded cache reuse. All 20,000 raw outputs record version 1.161.0 and OSS mode. The runner supports per-rule time limits, a whole-scan deadline, and parallel jobs, but the historical records do not keep the effective values per case, so we do not treat the current script defaults as measured settings.

\subsection{Scope, Prediction, and Metrics}\label{subsec:measurement}

All pipelines score findings inside a resolved changed-file scope. We score within the changed files because the vulnerable/fixed label refers to the code touched by a commit. Counting alerts anywhere in a repository would let any unrelated warning decide the label, and changed-file scanning resembles the diff-focused way SAST warnings are consumed in code review~\cite{charoenwet2024secure}.
\begin{itemize}
\item For CodeQL, the score is the number of security-relevant findings whose primary location lies in the case's changed-file set.
\item For Infer, the normalized report has one entry per issue, and an issue counts when its top-level file matches a path in the case's changed-file list (the union of old and new paths, matched exactly or by a fixed suffix rule). The score counts findings, not files, traces, or procedures; the bug trace is kept as evidence but does not enter the score.
\item For Semgrep, the score is the number of WARNING-or-higher findings in the changed-file set selected for the case. The scope is resolved from target metadata and the before/after commit relation, never from the role alone: 7,801 vulnerable cases use the \emph{changed\_before} file set, and 2,062 vulnerable cases and all 9,881 primary fixed cases use the \emph{changed\_after} set. These names select a file set; they do not by themselves say whether the scanned code is pre-fix or post-fix.
\end{itemize}


A score above zero predicts the case as \textit{vulnerable}. We then compute standard binary-classification metrics—precision, recall (TPR), F1, and FPR—against the vulnerable/fixed case labels. Thus, a true positive means that a vulnerable case has at least one qualifying in-scope finding; it does not imply CWE agreement, repair overlap, or detection of the labeled vulnerability mechanism. 

We additionally report AUROC over the raw finding counts, using average ranks for ties~\cite{fawcett2006introduction}, to measure whether vulnerable cases tend to receive more findings than fixed cases; AUROC is undefined when one class is absent. When a denominator is zero, precision, recall, F1, and FPR are recorded as zero, which matters in small cells (i.e., when an evaluation stratum contains few cases). We read F1 together with FPR and AUROC because F1 can rise when a tool reports more findings on both labels. A successful run with no findings (including an empty report) is an evaluated negative, whereas a failed or missing run has no prediction and enters only reliability denominators. Overall Study-A metrics are pooled over all cases rather than averaged over years.

\subsection{Three Complementary Studies}\label{subsec:studies}

\begin{figure}[!tb]
\centering
\begin{tikzpicture}[font=\scriptsize,
  panel/.style={draw=black!25, rounded corners=2pt, minimum width=4.45cm, minimum height=2.02cm, anchor=north west},
  cell/.style={draw=black!45, fill=black!4, rounded corners=1pt, inner sep=1.5pt, align=center},
  obox/.style={draw=findingblue!60, fill=poscell, rounded corners=1.5pt, inner sep=2pt, align=center, text width=4.05cm},
  hdr/.style={font=\scriptsize\bfseries, anchor=north west},
  arr/.style={-{Stealth[length=1.5mm]}, black!55, line width=0.5pt}]
\node[panel] (PA) at (0,0) {};
\node[hdr] at (0.08,-0.06) {Study A: observed ecosystem (RQ1)};
\foreach \i/\y in {0/2016,1/2017,2/2018,3/$\cdots$,4/2025}{\node[cell, minimum width=0.72cm, minimum height=0.5cm] (a\i) at (0.52+\i*0.85,-0.68) {\y};}
\node[font=\tiny, black!60] at (2.22,-1.03) {all eligible cases of each source year $t$};
\node[obox] (oa) at (2.22,-1.55) {$\mathrm{Perf}(D_r,C_t)$ per year\\(19,344--19,744 cases per condition)};
\node[panel] (PB) at (4.73,0) {};
\node[hdr] at (4.81,-0.06) {Study B: matched year pairs (RQ2)};
\node[cell, text width=1.75cm] (bx) at (5.85,-0.7) {year $x$\\[-1pt]$p$: $q_p$\,V + $q_p$\,F};
\node[cell, text width=1.75cm] (by) at (8.08,-0.7) {year $y$\\[-1pt]$p$: $q_p$\,V + $q_p$\,F};
\draw[arr,{Stealth[length=1.5mm]}-{Stealth[length=1.5mm]}] (bx.east) -- (by.west);
\node[font=\tiny, black!60] at (6.97,-1.1) {same projects, same $q_p$ per label};
\node[obox] (ob) at (6.95,-1.55) {$\dF=\mathrm{F1}(C_y)-\mathrm{F1}(C_x)$\\45 pairs $\times$ 100 repetitions};
\node[panel] (PC) at (9.46,0) {};
\node[hdr] at (9.54,-0.06) {Study C: within-project paths (RQ3)};
\draw[black!45] (10.0,-0.98) -- (13.4,-0.98);
\draw[black!45] (10.0,-0.98) -- (10.0,-0.36);
\foreach \x/\v in {10.35/0.62,11.25/0.42,12.15/0.55,13.05/0.18}{\node[circle, fill=findingblue, inner sep=1.1pt] at (\x,-0.98+\v*0.8) {};}
\draw[dashed, red!70!black, line width=0.6pt] (10.2,-0.45) -- (13.25,-0.8);
\node[font=\tiny, red!70!black, anchor=west] at (12.55,-0.48) {OLS};
\node[font=\tiny, black!60, anchor=north] at (11.7,-0.99) {one project, years with both labels};
\node[obox] (oc) at (11.68,-1.64) {slope per project: $\geq$3 years (202)\\or $\geq$5 years (61, nested)};
\node[draw=red!45, fill=vulnback, rounded corners=1.5pt, align=left, text width=6.65cm, inner sep=2.5pt, anchor=north west] (q) at (0,-2.12)
  {\textbf{RQ4} Largest Study-C drops $\rightarrow$ repair diffs, alert and trace locations, rule and query predicates, location ablation, local controls};
\node[draw=green!45!black, fill=fixback, rounded corners=1.5pt, align=left, text width=6.65cm, inner sep=2.5pt, anchor=north east] (o) at (13.91,-2.12)
  {\textbf{RQ5} Deduplicated execution targets $\rightarrow$ final status, status transitions, diagnostics, phase time, CPU, and memory};
\end{tikzpicture}
\caption{Study design. Studies A--C answer RQ1--RQ3 with different estimands; the qualitative analysis (RQ4) starts from Study-C extremes, and the operational analysis (RQ5) uses execution ledgers. V/F: vulnerable/fixed; $q_p$: matched capacity of project $p$.}
\label{fig:design}
\end{figure}
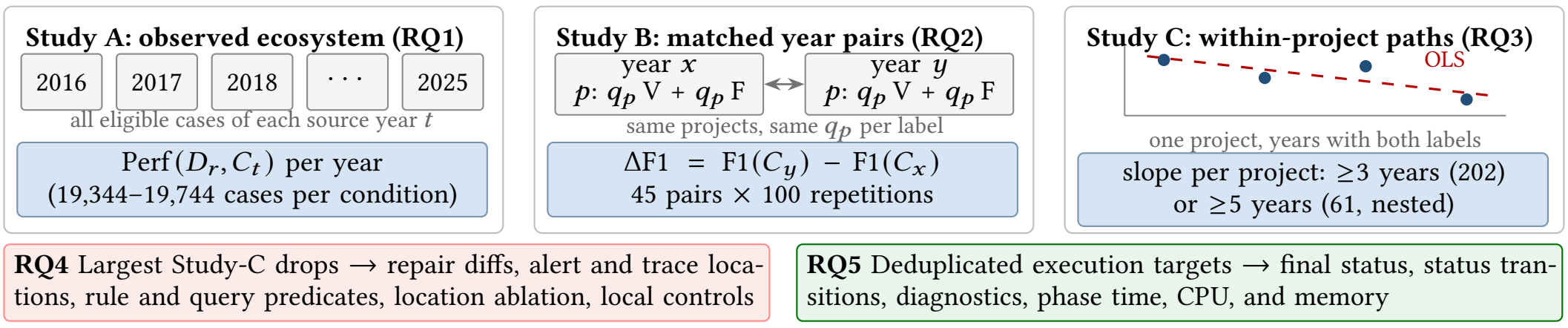

Figure~\ref{fig:design} summarizes the design. The three studies answer different questions and use different weights, so their numbers are not interchangeable. None of them reruns a detector: Studies B and C resample the stored Study-A results.

\head{Study A: observed ecosystem.} For each condition, Study~A evaluates the complete eligible population of each source year. It describes what a user would see on the corpus as it is, including changes in projects and defect types. It does not identify a temporal effect, because the project distribution $P(\mathrm{project}\mid C_t)$ changes with $t$: a difference between two years may come from different projects rather than from older or newer code.

\head{Study B: matched year pairs.} Study~B asks whether a detector still performs differently on code from two years once project composition is held fixed. For each of the 45 pairs $(x,y)$ with $2016\le x<y\le 2025$, a project is eligible if it has both labels in both years; it does not need to appear in any year in between, and it need not span the decade. Its matched capacity $q_p$ is the minimum count over its four project--year--label cells, $q_p=\min(n_{p,x,V},n_{p,x,F},n_{p,y,V},n_{p,y,F})$. Each of 100 deterministic repetitions draws $q_p$ vulnerable and $q_p$ fixed cases without replacement from each year, so both sides contain exactly the same projects with the same number of cases per label, $P(\mathrm{project},\mathrm{label}\mid C_x)=P(\mathrm{project},\mathrm{label}\mid C_y)$~\cite{stuart2010matching}. Within a repetition, the selected cases are pooled over projects in each year, so projects contribute in proportion to their matched capacity. We compute $\dF$ per repetition, average the repetitions within a pair, and weight pairs equally within each gap $\Delta t=y-x$. Across the 45 pairs there are 1,928 eligible project--pair combinations. Gap~1 has nine pairs, and each longer gap has one pair fewer, down to the single 2016$\rightarrow$2025 pair at gap~9. Matching does not stop one project from dominating a pair: the largest project supplies between 15.0\% and 92.8\% of the matched capacity, depending on the pair. Study~B deliberately does not restore 1,000 cases per label and year; it trades scale for comparability. Thus, Study B estimates a case-weighted temporal difference under identical project-and-label composition; it does not give every matched project equal influence.

\head{Study C: within-project trajectories.} Study~C asks whether a project's own F1 changes as the project evolves. A \emph{usable year} for a project is a source year in which it has at least one case of each label; usable years need not be consecutive. Study~C keeps projects with at least three usable years (primary cohort: 202 projects and 869 project-year cells) and, as a stricter sensitivity check, projects with at least five (nested cohort: 61 projects and 396 cells). Each project-year draws $\min(n_V,n_F)$ cases per label in each of 100 repetitions, which gives 15,598 selected cases per repetition in the primary cohort and 13,126 in the sensitivity cohort. We average repetitions per project-year and fit an ordinary least-squares F1 slope over the actual source years of each project, which weights projects equally; a slope needs at least two years with a defined metric. Slopes with absolute value at most $10^{-12}$ ($10^{-10}$ for Semgrep) count as zero, to absorb floating-point noise. Annual cohort metrics are pooled over projects within each repetition and then averaged over the 100 repetitions, so they weight cases; the cohort summaries in this paper are equal-weight means of these ten annual values. We report both views. 
These slopes are descriptive summaries rather than per-project significance tests; some project-year cells are small, which is why we report the nested $\ge$5-year cohort and the underlying support alongside slope directions.

\head{Common rules for Studies B and C.} Memberships were fixed from the corpus catalog before any detector result was compared, and the same memberships are used for all seven conditions. Balancing happens before the gates of \S\ref{subsec:populations}, including the corpus-wide label-conflict gate, so the frozen memberships stay identical for all conditions. A case that a gate removes is dropped without replacement, so evaluated counts can be slightly unbalanced and can vary across repetitions. The 2.5--97.5\% ranges we report describe resampling variation for a fixed pool of cases. They are not confidence intervals for a wider population of projects, because pairs and repetitions share projects and cases. The 100 repetitions reuse stored results, so they cannot measure run-to-run nondeterminism, and a case that appears in 100 repetitions is not 100 independent detections. Table~\ref{tab:support} shows how many unique evaluable cases each study uses for Infer; the counts overlap and must not be added.

\begin{table}[!tb]
\caption{Unique evaluable cases used by each study under the three Infer conditions. Studies B and C reuse already scored cases; the cohorts are nested, so rows must not be added. The full Study-B membership ledger holds 4,517,600 case references across repetitions and pairs.}
\label{tab:support}
\centering\footnotesize
\setlength{\tabcolsep}{6pt}
\zebra
\begin{tabular}{@{}lrrrr@{}}
\toprule
\textbf{Study} & \textbf{v1.1} & \textbf{v1.2} & \textbf{v1.3} & \textbf{Common intersection}\\
\midrule
Study A & 19,461 & 19,700 & 19,581 & 19,450\\
Study B & 17,368 & 17,557 & 17,464 & 17,357\\
Study C, primary ($\geq$3 years) & 15,669 & 15,828 & 15,759 & 15,658\\
Study C, sensitivity ($\geq$5 years) & 13,108 & 13,234 & 13,197 & 13,097\\
\bottomrule
\end{tabular}
\end{table}

\subsection{Evaluation Populations}\label{subsec:populations}

Table~\ref{tab:populations} shows how each condition goes from 20,000 planned cases to its Study-A population. Because the gates differ by tool, we also report results on the common intersection of each tool's releases: 19,169 cases for CodeQL and 19,450 for Infer. Effectiveness is counted on case rows and reliability on deduplicated targets (19,780--19,820 per CodeQL release for 20,000 cases).

\begin{table}[!tb]
\caption{How each detector condition (a tool release or configuration) goes from 20,000 planned cases to its Study-A population. All counts are cases.}
\label{tab:populations}
\centering\footnotesize
\setlength{\tabcolsep}{4pt}
\zebra
\begin{tabular}{@{}lrrr>{\raggedright\arraybackslash}p{0.47\linewidth}@{}}
\toprule
\textbf{Condition} & \textbf{Planned} & \textbf{Evaluable} & \textbf{Study A} & \textbf{What is excluded}\\
\midrule
CodeQL R2021 & 20,000 & 19,709 & 19,455 & 104 timeout, 182 database-failure, and 5 unmapped cases; 254 label conflicts\\
CodeQL R2024 & 20,000 & 19,563 & 19,344 & 332 timeout, 100 database-failure, and 5 unmapped cases; 219 label conflicts\\
CodeQL R2026 & 20,000 & 20,000 & 19,744 & 256 label conflicts; no execution failure after recovery\\
Infer v1.1 & 20,000 & 19,760 & 19,461 & 232 build failures, 8 timeouts; identity gate (47); 252 label conflicts\\
Infer v1.2 & 20,000 & 20,000 & 19,700 & identity gate (48); 252 label conflicts\\
Infer v1.3 & 20,000 & 19,881 & 19,581 & 108 build failures, 11 timeouts; identity gate (48); 252 label conflicts\\
Semgrep & 20,000 & 20,000 & 19,744 & 256 label conflicts; 109 zero-target cases kept\\
\bottomrule
\end{tabular}
\end{table}

\begin{figure}[!tb]
\centering
\includegraphics[width=\linewidth]{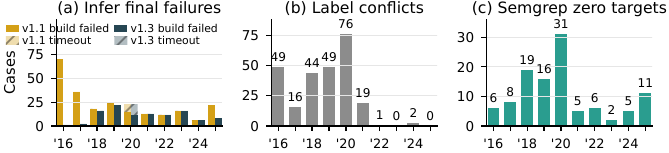}
\caption{Cases removed or flagged per source year (x-axis). (a) Final Infer build failures and timeouts per 2,000 planned cases; v1.2 has none. (b) Corpus records excluded for all tools because the same repository, vulnerability, and commit occur under both labels. (c) Primary Semgrep cases in which no file was scanned; they stay in the population with score zero.}
\label{fig:exclusions}
\end{figure}

\head{Label-conflict gate.} Some CVEs are fixed by several commits, so an intermediate commit is both the fixed version of one patch and the vulnerable parent of the next. Using corpus metadata only, we find 256 records (137 vulnerable, 119 fixed) whose repository, vulnerability, and commit appear under both labels; 253 of them come from 2016--2021, with the most in 2020 (76; Figure~\ref{fig:exclusions}(b)). These 256 ambiguous case identities are excluded from label-based evaluation for all tools. They were executed like every other case and stay in the execution and reliability counts. The gate removes 254, 219, and 256 cases from the CodeQL R2021, R2024, and R2026 populations and 252 from each Infer population; the remaining conflicting cases had already been removed by an execution failure or the identity gate.

\head{CodeQL.} CodeQL registers 19,820 canonical targets for the 20,000 cases, because several cases can share one execution target. The two historical releases each connect to 19,780 of them. Of the 40 registration differences, 35 are aliases of other targets and 5 cannot be resolved; the latter are the five unmapped cases in Table~\ref{tab:populations}. Measured over cases, the coverage of the three releases is 98.55\%, 97.82\%, and 100\%; most missing cases come from timeouts and database failures, and only five from unresolved registrations. After the label-conflict gate, the class sizes are 9,719 vulnerable and 9,736 fixed cases for R2021, 9,650 and 9,694 for R2024, and 9,863 and 9,881 for R2026.

\head{Infer.} Only build failures and timeouts count as execution failures for Infer. The identity gate is a separate, version-independent rule inherited from the original workflow: it drops 40 catalog identities whose directory leaf does not follow the expected naming convention and 8 that are not the chosen representative of a truncated composite key. The gate removes 48 cases in v1.2 and v1.3 but 47 in v1.1, because one of the 48 identities is already a v1.1 build failure. After the label-conflict gate, the class sizes are 9,703 vulnerable and 9,758 fixed cases (v1.1), 9,823 and 9,877 (v1.2), and 9,761 and 9,820 (v1.3). Figure~\ref{fig:exclusions}(a) shows that failures concentrate in a few years: v1.1 loses 70 of the 2,000 cases of 2016 to build failures, and all timeouts of both releases fall in 2020. In 2021--2024 the two releases have identical yearly failure counts.

\head{Semgrep.} Every Semgrep case has a resolved scope and a result record. Apart from the shared label-conflict gate, Semgrep has no primary exclusion, leaving 9,863 vulnerable and 9,881 fixed cases. Diagnostics and timeouts never remove a case. The 109 primary cases (113 in the catalog) in which Semgrep scanned zero files are kept with score zero; Figure~\ref{fig:exclusions}(c) shows their distribution, again peaking in 2020. Table~\ref{tab:sg-catalog} gives the confusion counts before and after the quarantine for all rules and for the security-category rules alone.

\begin{table}[!tb]
\caption{Semgrep confusion counts for the whole catalog and for the primary population, with all 61 qualifying rules and with security-category rules only. The four differences between catalog and primary counts (37, 31, 100, 88) sum to the 256 label-conflict records.}
\label{tab:sg-catalog}
\centering\footnotesize
\setlength{\tabcolsep}{6pt}
\zebra
\begin{tabular}{@{}lrrrrr@{}}
\toprule
\textbf{Population} & \textbf{Cases} & \textbf{TP} & \textbf{FP} & \textbf{FN} & \textbf{TN}\\
\midrule
All rules, catalog & 20,000 & 4,626 & 4,764 & 5,374 & 5,236\\
All rules, primary & 19,744 & 4,589 & 4,733 & 5,274 & 5,148\\
Security-category rules, catalog & 20,000 & 4,550 & 4,671 & 5,450 & 5,329\\
Security-category rules, primary & 19,744 & 4,515 & 4,642 & 5,348 & 5,239\\
\bottomrule
\end{tabular}
\end{table}

\subsection{Reliability and Cost Measurement}\label{subsec:reliability-method}

Reliability uses deduplicated execution targets, so that a target shared by several cases is not counted as several failures. For CodeQL, the unified ledger has 60,000 case-by-release rows and 59,380 release-by-target records; these are neither 60,000 vulnerabilities nor 59,380 distinct snapshots. We classify final outcomes as success, timeout, database or extraction failure, build failure, or missing, and we compare exact target identities (target, commit, role, project, year) across CodeQL releases. An intermediate failed attempt counts as a final failure only if it is explicitly linked to the selected attempt. For cost, we use recorded phase wall time, CPU time, and peak resident set size (RSS). Missing and cached values are never treated as zero, and CPU time is never inferred from wall time. These measurements do not control machine load or cache state, so we use them to describe operating ranges, not to rank tools by speed.

\subsection{Qualitative Failure Analysis}\label{subsec:qualitative-method}

For RQ4 we take, for each tool, the primary Study-C cohort and compute every project's annual F1 averaged over the repetitions. We then rank all early-to-late year differences within each project and keep every pair that ties for the largest decline. The selection therefore follows endpoint differences, and slopes play no role in it. We compare the same endpoints across releases. For CodeQL and Infer, the primary ties come from the 2021 release (four CodeQL ties under R2021 and one Infer tie under v1.1), and the other releases add their own ties as supplementary examples; they are not an exhaustive list of those releases' ties. Semgrep has one condition.

For each selected case we inspect the target identity, the scored snapshot side, the changed-file scope, the vulnerable-to-fixed diff, alert and trace locations, diagnostics, and the definitions of relevant rules or queries. A positive counts as \emph{mechanism-supported} only if the labelled repair, the scanned source, the rule semantics, and the finding or trace location all agree; otherwise it is \emph{label-level}, and if the evidence cannot settle the relation it is \emph{unresolved}. For zero-finding cases we separate three explanations that can overlap: the repair lies outside the scanned files (a scope limitation), the scanned repair file carries a diagnostic (an analysis limitation), or the scanned repair file has no diagnostic and still no finding, which can motivate a rule--target hypothesis. When classifying false negatives, scope mismatch takes precedence. We also apply deterministic location ablations that keep only already-scored alerts in a verified repair file, or on a changed repair line, without changing labels, findings, or repetitions. Table~\ref{tab:queues} lists the evidence in each queue.

\begin{table}[!tb]
\caption{Qualitative evidence queues (RQ4). Parent-to-fix relations were checked as first-parent comparisons; for CodeQL, the 13 relations cover 26 of the 27 cases.}
\label{tab:queues}
\centering\footnotesize
\setlength{\tabcolsep}{8pt}
\zebra
\begin{tabular}{@{}lrrr@{}}
\toprule
& \textbf{CodeQL} & \textbf{Infer} & \textbf{Semgrep}\\
\midrule
Selected project-year pairs & 8 & 9 & 9\\
Unique cases & 27 & 34 & 42\\
Case $\times$ release records & 81 & 102 & 42\\
Verified parent-to-fix relations & 13 & 17 & 19 of 20\\
Scored findings inspected & 39 & 620 & 73\\
Before/after file pairs reviewed & --- & --- & 67\\
\bottomrule
\end{tabular}
\end{table}

For CodeQL we also ran bounded local experiments: a compiler-visibility control, a source-slice query test, and a saved-build-wrapper control (\S\ref{subsec:no-finding} and \S\ref{subsec:reliability}). None of them reruns a historical target or changes a reported score. Reliability mechanisms are graded by evidence: a direct diagnostic, a locally validated mechanism, or a uniquely linked historical cause. We state the grade with each claim, and ``unresolved'' always means that the specific historical attribution is not isolated, not that no mechanism is known.

\subsection{Validity Safeguards}\label{subsec:validity-method}

Membership was frozen before conditions were compared, the same Study-B/C samples are reused for every condition, and all 100 repetitions are retained. The two Study-C cohorts are nested and never summed. Cases, targets, findings, project-years, and repetitions are distinct units, and each claim names its unit. An audit of the CodeQL metric, denominator, weighting, deduplication, coverage, and provenance computations passed all 1,728 checks. The Infer pipeline reproduced all ten reference Study-B/C tables of the original v1.2 workflow (four Study-B and six Study-C outputs) and validated all 60,000 case-by-release identities. The CodeQL and Infer values after the label-conflict gate come from a separate recomputation that reproduces every earlier Study-A/B/C value exactly when the gate is switched off. For Semgrep, we recomputed every derived metric from its confusion counts, and the rule-version identifiers embedded in all 73 queue findings match the rule definitions we inspected. All 27 selected R2026 CodeQL result files are available (six of them recovered from a result archive) and match their target path, commit, and normalized findings. Language-model agents assisted with artifact navigation, source and log inspection, table construction, and running the bounded local control experiments, and review agents checked those artifacts. The labels, the frozen results, and every reported metric come from the study artifacts rather than from the agents, and every qualitative claim is tied to inspectable source, diff, rule, query, or log evidence.

\section{RQ1: Ecosystem-Scale Effectiveness (Study A)}\label{sec:study-a}

\begin{figure}[!tb]
\centering
\includegraphics[width=\linewidth]{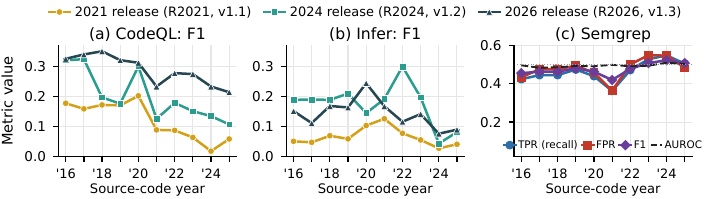}
\caption{Study-A annual profiles by source-code year (x-axis). (a, b) F1; colors and markers denote the release year and mean the same in every release plot of this paper. (c) Semgrep F1, TPR, FPR, and count-based AUROC; the dotted line marks 0.5. Lines connect yearly populations with different projects, not trajectories of the same cases.}
\label{fig:study-a}
\end{figure}

Table~\ref{tab:study-a} reports the Study-A results for all seven conditions, and Figure~\ref{fig:study-a} shows the annual profiles. To read the table, take the CodeQL R2026 row: on 19,744 evaluated cases, 49.2\% of positive predictions fall on vulnerable cases, 20.7\% of vulnerable cases are flagged, and 21.3\% of fixed cases are flagged as well, so the raw-count AUROC of 0.4969 is no better than chance. The last column repeats F1 on a sensitivity population that removes coverage differences.

\begin{table}[!tb]
\caption{Study-A overall results (changed-file scoring). AUROC ranks raw finding counts. ``Sens.\ F1'' is F1 on the common intersection of the tool's releases (CodeQL 19,169 and Infer 19,450 cases) or, for Semgrep, after removing 109 zero-target cases.}
\label{tab:study-a}
\centering\footnotesize
\setlength{\tabcolsep}{3.4pt}
\zebra
\begin{tabular}{@{}lrrrrrrrrrrr@{}}
\toprule
\textbf{Condition} & \textbf{Eval.} & \textbf{TP} & \textbf{FP} & \textbf{FN} & \textbf{TN} & \textbf{Prec.} & \textbf{Recall} & \textbf{F1} & \textbf{AUROC} & \textbf{FPR} & \textbf{Sens.\ F1}\\
\midrule
CodeQL R2021 & 19,455 & 686 & 729 & 9,033 & 9,007 & .4848 & .0706 & .1232 & .4979 & .0749 & .1249\\
CodeQL R2024 & 19,344 & 1,265 & 1,264 & 8,385 & 8,430 & .5002 & .1311 & .2077 & .5005 & .1304 & .2091\\
CodeQL R2026 & 19,744 & 2,042 & 2,105 & 7,821 & 7,776 & .4924 & .2070 & .2915 & .4969 & .2130 & .2913\\
Infer v1.1 & 19,461 & 345 & 363 & 9,358 & 9,395 & .4873 & .0356 & .0663 & .4992 & .0372 & .0663\\
Infer v1.2 & 19,700 & 1,003 & 505 & 8,820 & 9,372 & \textbf{.6651} & .1021 & .1770 & \textbf{.5257} & .0511 & .1778\\
Infer v1.3 & 19,581 & 830 & 872 & 8,931 & 8,948 & .4877 & .0850 & .1448 & .4981 & .0888 & .1446\\
Semgrep & 19,744 & 4,589 & 4,733 & 5,274 & 5,148 & .4923 & \textbf{.4653} & \textbf{.4784} & .4951 & .4790 & .4801\\
\bottomrule
\end{tabular}
\end{table}

\subsection{CodeQL: Higher F1 without Better Separation}\label{subsec:study-a-codeql}

Across the CodeQL releases, F1 more than doubles, from 0.1232 to 0.2077 to 0.2915. At first sight this looks like steady tool improvement. The confusion components say otherwise. Recall rises from 0.0706 to 0.2070, but FPR rises in parallel from 0.0749 to 0.2130, precision stays close to 0.5, and AUROC never moves more than 0.0031 away from 0.5. In absolute terms, R2026 flags 4,147 cases against 1,415 for R2021, adding 1,356 true positives and 1,376 false positives. The newer releases flag more cases on both labels; they do not rank vulnerable cases above fixed ones any better.

The ordering survives on the 19,169-case common intersection (F1 0.1249, 0.2091, 0.2913), so it does not come from which targets each release happened to complete. That intersection is itself a selected population, though: it keeps only targets on which every release succeeded and may therefore favor easier targets. It checks the ranking but cannot replace the reliability denominators.

The annual profiles in Figure~\ref{fig:study-a}(a) are not smooth functions of source age. R2021 stays between about 0.16 and 0.20 until 2020, then falls sharply and reaches about 0.02 in 2024. R2024 swings between years and ends near 0.11 in 2025. R2026 is the highest curve in every year but ends below its own 2016 value. Because each year has a different project mix, these shapes motivate the controlled studies of the next two sections rather than answer the aging question.

\finding{A1}{Newer CodeQL releases reach higher F1 mainly by flagging more cases on both labels: recall and FPR rise together and AUROC stays at chance. A higher score does not mean better discrimination.}

\subsection{Infer: The Middle Release Performs Best}\label{subsec:study-a-infer}

Infer does not improve with release date. v1.1 finds 345 true positives against 9,358 false negatives (recall 0.0356). v1.2 nearly triples the true positives to 1,003 while keeping false positives at 505, which gives the best Infer precision (0.6651), F1 (0.1770), and AUROC (0.5257). v1.3 then loses ground on both sides: 830 true positives and 872 false positives. Its FPR of 0.0888 is 74\% higher than that of v1.2, while its recall is lower. Compared with v1.1, v1.3 raises recall and F1 without improving AUROC. The dominant source of low F1 for all releases is the false-negative count: between 8,820 and 9,358 vulnerable cases receive no changed-file finding at all. A zero score may mean that the snapshot has no issue, that the issue lies off the scored paths, or that capture was limited; it is not a diagnosis of which pattern a checker cannot recognize. Likewise, a finding on a fixed case is a false positive with respect to the label even when it points at a different real defect.

On the 19,450-case common intersection, the three releases keep the same order. Precision, recall, F1, AUROC, and FPR are 0.4873, 0.0356, 0.0663, 0.4992, and 0.0372 for v1.1; 0.6660, 0.1026, 0.1778, 0.5260, and 0.0512 for v1.2; and 0.4870, 0.0849, 0.1446, 0.4980, and 0.0889 for v1.3. Because the intersection keeps only cases that every release completed, it can favor cases that are easy to build; it is a check on support, not a correction for selection.

\begin{table}[!tb]
\caption{Annual Study-A metrics for Infer. Shading marks each column's maximum (blue, including ties) and minimum (red). Primary cases per year: 1,879--1,992 (v1.1), 1,914--2,000 (v1.2), and 1,891--1,992 (v1.3).}
\label{tab:infer-annual}
\centering\scriptsize
\setlength{\tabcolsep}{2.3pt}
\begin{tabular}{@{}l|rrrrr|rrrrr|rrrrr@{}}
\toprule
& \multicolumn{5}{c|}{\textbf{v1.1}} & \multicolumn{5}{c|}{\textbf{v1.2}} & \multicolumn{5}{c}{\textbf{v1.3}}\\
\textbf{Year} & \textbf{Prec.} & \textbf{Rec.} & \textbf{F1} & \textbf{FPR} & \textbf{AUC} & \textbf{Prec.} & \textbf{Rec.} & \textbf{F1} & \textbf{FPR} & \textbf{AUC} & \textbf{Prec.} & \textbf{Rec.} & \textbf{F1} & \textbf{FPR} & \textbf{AUC}\\
\midrule
2016 & .5000 & .0267 & .0507 & .0265 & .5001 & .6524 & .1104 & .1889 & .0582 & .5268 & .4833 & .0898 & .1514 & .0950 & .4973\\
2017 & .5217 & .0247 & .0472 & .0226 & \pc{.5010} & .6271 & .1121 & .1902 & .0665 & .5229 & \nc{.4468} & .0637 & .1115 & .0787 & \nc{.4925}\\
2018 & .4865 & .0373 & .0693 & .0392 & .4990 & \nc{.5550} & .1138 & .1889 & .0911 & .5113 & .4975 & .1014 & .1685 & .1021 & .4998\\
2019 & .4545 & .0314 & .0587 & .0373 & \nc{.4971} & .5605 & .1291 & .2099 & \pc{.1002} & .5147 & .4657 & .0993 & .1637 & .1127 & .4934\\
2020 & .5047 & .0574 & .1032 & .0557 & .5008 & .6290 & .0818 & .1448 & .0479 & .5169 & \pc{.5000} & \pc{.1619} & \pc{.2446} & \pc{.1597} & \pc{.5007}\\
2021 & .4830 & \pc{.0724} & \pc{.1259} & \pc{.0772} & .4977 & .6832 & .1114 & .1916 & .0514 & .5305 & .4950 & .1009 & .1677 & .1025 & .4992\\
2022 & .4767 & .0421 & .0773 & .0453 & .4984 & .7541 & \pc{.1874} & \pc{.3002} & .0601 & \pc{.5645} & \pc{.5000} & .0656 & .1160 & .0645 & \pc{.5007}\\
2023 & .4833 & .0292 & .0551 & .0312 & .4990 & .8235 & .1120 & .1972 & .0240 & .5447 & .4910 & .0827 & .1415 & .0857 & .4985\\
2024 & \nc{.4375} & \nc{.0141} & \nc{.0272} & \nc{.0181} & .4980 & \pc{.9565} & \nc{.0220} & \nc{.0431} & \nc{.0010} & \nc{.5105} & .4940 & \nc{.0412} & \nc{.0760} & \nc{.0422} & .4995\\
2025 & \pc{.5250} & .0212 & .0408 & .0192 & \pc{.5010} & .7679 & .0430 & .0814 & .0130 & .5150 & \pc{.5000} & .0491 & .0895 & .0492 & .4999\\
\bottomrule
\end{tabular}
\end{table}

\begin{figure}[!tb]
\centering
\includegraphics[width=\linewidth]{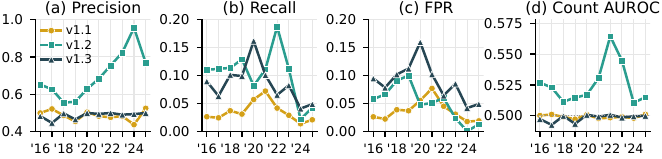}
\caption{Annual Study-A precision, recall, FPR, and count AUROC for the three Infer releases by source-code year (x-axis). The F1 curves are in Figure~\ref{fig:study-a}(b); the dotted line in (d) marks 0.5.}
\label{fig:infer-annual}
\end{figure}

Table~\ref{tab:infer-annual} and Figure~\ref{fig:infer-annual} give the annual values behind Figure~\ref{fig:study-a}(b). Each release peaks in a different year: v1.1 reaches F1 0.1259 in 2021, v1.2 0.3002 in 2022, and v1.3 0.2446 in 2020. All three are far lower for the 2024 code than at their peaks. The two figures also show that recall and FPR of v1.1 and v1.3 move together from year to year. v1.3 has its highest recall (0.1619) and its highest FPR (0.1597) in the same year, 2020, and its annual AUROC never exceeds 0.5007; v1.1's annual AUROC lies between 0.4971 and 0.5010. v1.2 is the only release that separates the labels at all: its annual AUROC stays between 0.5105 and 0.5645, the highest single value of any Infer cell (2022). The v1.2 row for 2024 shows how one metric can mislead in a sparse cell. Precision reaches 0.9565 only because almost no fixed case is flagged (FPR 0.0010), while recall is 0.0220. Coverage also differs by year (Figure~\ref{fig:exclusions}(a)): in 2016, v1.1 has 1,879 primary cases, against 1,949 for v1.2 and 1,948 for v1.3. These differences do not explain the annual pattern away, because project composition and issue behavior also change from year to year.

\finding{A2}{Infer's 2024 release, not its newest, has the best label-level discrimination, and the newest release broadens fixed-side alerting without keeping the vulnerable-side gain. Release date does not order sustainability; the unit of evaluation is the whole detector condition.}

\subsection{Semgrep: High F1 from a Proxy That Barely Separates Labels}\label{subsec:study-a-semgrep}

Semgrep has the highest F1 (0.4784), but its recall (0.4653) is slightly lower than its FPR (0.4790), and its AUROC is 0.4951: a vulnerable case is no more likely to be flagged than a fixed one. Put as row shares, Semgrep flags 46.5\% of vulnerable cases and 47.9\% of fixed cases. Table~\ref{tab:sg-annual} breaks the results down by year. The annual F1 is not monotonic: 0.4558 in 2016, 0.4201 in 2021, 0.5234 in 2024, and 0.5083 in 2025. Figure~\ref{fig:study-a}(c) shows why the curve rises after 2021. TPR and FPR move up and down together, while AUROC never leaves 0.484--0.511.

\begin{table}[!tb]
\caption{Annual Study-A results for Semgrep (primary population). ``Zero'' is the number of zero-target cases in the year; $\Delta_0$ is the F1 change when they are removed. Shading marks each metric column's maximum (blue) and minimum (red).}
\label{tab:sg-annual}
\centering\footnotesize
\setlength{\tabcolsep}{3.2pt}
\zebra
\begin{tabular}{@{}lrrrrrrrrrrrr@{}}
\toprule
\textbf{Year} & \textbf{Cases} & \textbf{TP} & \textbf{FP} & \textbf{FN} & \textbf{TN} & \textbf{Prec.} & \textbf{Recall} & \textbf{F1} & \textbf{FPR} & \textbf{AUROC} & \textbf{Zero} & \textbf{$\Delta_0$}\\
\midrule
2016 & 1,951 & 412 & 425 & 559 & 555 & .4922 & .4243 & .4558 & .4337 & .4967 & 6 & .0013\\
2017 & 1,984 & 442 & 473 & 550 & 519 & \nc{.4831} & .4456 & .4636 & .4768 & \nc{.4843} & 8 & .0012\\
2018 & 1,956 & 436 & 466 & 542 & 512 & .4834 & .4458 & .4638 & .4765 & .4869 & 19 & .0032\\
2019 & 1,951 & 462 & 484 & 511 & 494 & .4884 & .4748 & .4815 & .4949 & .4932 & 16 & .0030\\
2020 & 1,924 & 423 & 441 & 538 & 522 & .4896 & .4402 & .4636 & .4579 & .4921 & 31 & \pc{.0051}\\
2021 & 1,981 & 359 & 361 & 630 & 631 & .4986 & \nc{.3630} & \nc{.4201} & \nc{.3639} & .4993 & 5 & .0007\\
2022 & 1,999 & 472 & 501 & 528 & 498 & .4851 & .4720 & .4785 & .5015 & .4916 & 6 & .0010\\
2023 & 2,000 & 529 & 550 & 471 & 450 & .4903 & .5290 & .5089 & \pc{.5500} & .4948 & 2 & \nc{.0002}\\
2024 & 1,998 & 548 & 547 & 451 & 452 & .5005 & \pc{.5485} & \pc{.5234} & .5475 & \pc{.5106} & 5 & .0008\\
2025 & 2,000 & 506 & 485 & 494 & 515 & \pc{.5106} & .5060 & .5083 & .4850 & .5041 & 11 & .0010\\
\bottomrule
\end{tabular}
\end{table}

Two sensitivity checks leave this picture intact. First, removing the 109 zero-target cases leaves 19,635 cases. All of them were negatives (70 false negatives and 39 true negatives), so TP and FP do not change: recall rises to 0.4686, F1 to 0.4801, and AUROC to 0.4958. FPR rises slightly, to 0.4809, because its denominator shrinks. The yearly effect of the removal never exceeds 0.0051 (Table~\ref{tab:sg-annual}), so the result does not come from giving zero scores to empty scopes. Second, restricting the score to rules in the security category, on the same cases and scopes, yields TP=4,515 and FP=4,642 (Table~\ref{tab:sg-catalog}), that is, precision 0.4931, recall 0.4578, F1 0.4748, and FPR 0.4698.

The score is also concentrated. Table~\ref{tab:rules} lists the eight rules that emit the most findings and splits each rule's cases by label. Together they produce 94.2\% of the 78,696 qualifying findings. A single generic rule, \texttt{insecure-use-memset}, produces 49,606 of them (63.03\%) and fires in 7,769 cases, of which 3,837 are vulnerable and 3,932 are fixed. The rule matches any \texttt{memset} call and does not require sensitive data at the call site. Each of the top eight rules fires on similar numbers of vulnerable and fixed cases, even the two memory-safety rules (\texttt{function-use-after-free}: 125 vs.\ 125 cases; \texttt{double-free}: 73 vs.\ 74). A case can appear under several rules, because rules overlap.

\begin{table}[!tb]
\caption{The eight Semgrep rules with the most qualifying findings (primary population). ``Cases'' counts cases with at least one finding of the rule; TP/FP split them into vulnerable and fixed cases. Bars show both counts on one scale (\textcolor{tprblue}{\rule{2mm}{4.5pt}}~vulnerable, \textcolor{fprred}{\rule{2mm}{4.5pt}}~fixed).}
\label{tab:rules}
\centering\footnotesize
\setlength{\tabcolsep}{4pt}
\zebra
\begin{tabular}{@{}lrrrrrl@{}}
\toprule
\textbf{Rule (last ID component)} & \textbf{Findings} & \textbf{Share} & \textbf{Cases} & \textbf{TP} & \textbf{FP} & \textbf{Cases by label}\\
\midrule
\texttt{insecure-use-memset} & 49,606 & 63.03\% & 7,769 & 3,837 & 3,932 & \textcolor{tprblue}{\rule{1.580cm}{4.5pt}}\textcolor{fprred}{\rule{1.620cm}{4.5pt}}\\
\texttt{insecure-use-string-copy-fn} & 11,467 & 14.57\% & 2,454 & 1,208 & 1,246 & \textcolor{tprblue}{\rule{0.498cm}{4.5pt}}\textcolor{fprred}{\rule{0.513cm}{4.5pt}}\\
\texttt{insecure-use-strcat-fn} & 5,181 & 6.58\% & 863 & 440 & 423 & \textcolor{tprblue}{\rule{0.181cm}{4.5pt}}\textcolor{fprred}{\rule{0.174cm}{4.5pt}}\\
\texttt{incorrect-use-ato-fn} & 4,293 & 5.46\% & 889 & 442 & 447 & \textcolor{tprblue}{\rule{0.182cm}{4.5pt}}\textcolor{fprred}{\rule{0.184cm}{4.5pt}}\\
\texttt{unbounded-copy-to-stack-buffer} & 1,184 & 1.50\% & 380 & 201 & 179 & \textcolor{tprblue}{\rule{0.083cm}{4.5pt}}\textcolor{fprred}{\rule{0.074cm}{4.5pt}}\\
\texttt{function-use-after-free} & 1,002 & 1.27\% & 250 & 125 & 125 & \textcolor{tprblue}{\rule{0.051cm}{4.5pt}}\textcolor{fprred}{\rule{0.051cm}{4.5pt}}\\
\texttt{insecure-use-strtok-fn} & 738 & 0.94\% & 171 & 95 & 76 & \textcolor{tprblue}{\rule{0.039cm}{4.5pt}}\textcolor{fprred}{\rule{0.031cm}{4.5pt}}\\
\texttt{double-free} & 653 & 0.83\% & 147 & 73 & 74 & \textcolor{tprblue}{\rule{0.030cm}{4.5pt}}\textcolor{fprred}{\rule{0.030cm}{4.5pt}}\\
\bottomrule
\end{tabular}
\end{table}

\finding{A3}{Semgrep's F1 of 0.4784 is not a 47.84\% vulnerability-detection rate. The finding count barely separates labels (AUROC 0.4951), the annual curve is non-monotonic, and almost two thirds of qualifying findings come from one generic rule that fires equally on vulnerable and fixed code.}

\section{RQ2: Project-Matched Temporal Change (Study B)}\label{sec:study-b}

\begin{figure}[!tb]
\centering
\includegraphics[width=\linewidth]{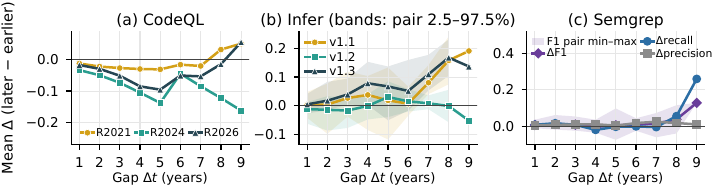}
\caption{Study-B mean change by gap (x-axis, years). (b) Bands show the 2.5--97.5\% quantiles of the year-pair means (undefined at gap~9, which has one pair). (c) The band shows the minimum and maximum year-pair $\dF$; Semgrep's $\Delta$recall drives the gap-9 jump. Pair means share projects and years, so the bands are descriptive.}
\label{fig:study-b}
\end{figure}

Study B controls one major alternative explanation for Study-A trends (i.e., project and label composition): 
For every year pair, the sampling procedure selects the same projects and equal vulnerable/fixed counts per project on both sides; subsequent evaluation exclusions may introduce small imbalances. 
Table~\ref{tab:study-b} reports the pair-equal mean $\dF$ for each gap, where negative values mean lower F1 in the later year, and Figure~\ref{fig:study-b} plots the same means together with the spread across year pairs. For example, the CodeQL R2024 entry at gap~5 ($-0.1376$) averages five year pairs, each of which was itself averaged over 100 repetitions. 
A progressively more negative gap profile would indicate a simple monotonic aging pattern; departures from monotonicity instead show that temporal degradation interacts with the particular years and cases being compared.

\begin{table}[!tb]
\caption{Study-B mean $\dF$ (later minus earlier) by source-year gap; red cells are negative, blue cells positive. Support columns: Semgrep's matched projects per pair and evaluated cases per year side.}
\label{tab:study-b}
\centering\footnotesize
\setlength{\tabcolsep}{3.6pt}
\begin{tabular}{@{}rr|rrr|rrr|r|rr@{}}
\toprule
& & \multicolumn{3}{c|}{\textbf{CodeQL}} & \multicolumn{3}{c|}{\textbf{Infer}} & & \multicolumn{2}{c}{\textbf{Support (Semgrep)}}\\
\textbf{Gap} & \textbf{Pairs} & \textbf{R2021} & \textbf{R2024} & \textbf{R2026} & \textbf{v1.1} & \textbf{v1.2} & \textbf{v1.3} & \textbf{Semgrep} & \textbf{Projects} & \textbf{Cases/side}\\
\midrule
1 & 9 & \nc{$-$.0123} & \nc{$-$.0332} & \nc{$-$.0170} & \pc{+.0025} & \nc{$-$.0115} & \pc{+.0049} & \pc{+.0073} & 32--113 & 464--1,492\\
2 & 8 & \nc{$-$.0218} & \nc{$-$.0495} & \nc{$-$.0293} & \pc{+.0043} & \nc{$-$.0142} & \pc{+.0178} & \pc{+.0124} & 24--89 & 393--1,374\\
3 & 7 & \nc{$-$.0267} & \nc{$-$.0734} & \nc{$-$.0511} & \pc{+.0270} & \nc{$-$.0165} & \pc{+.0394} & \pc{+.0091} & 23--53 & 290--754\\
4 & 6 & \nc{$-$.0297} & \nc{$-$.1049} & \nc{$-$.0849} & \pc{+.0379} & \nc{$-$.0005} & \pc{+.0785} & \nc{$-$.0032} & 27--55 & 99--580\\
5 & 5 & \nc{$-$.0309} & \nc{$-$.1376} & \nc{$-$.0954} & \pc{+.0188} & \pc{+.0307} & \pc{+.0674} & \pc{+.0057} & 27--34 & 92--352\\
6 & 4 & \nc{$-$.0159} & \nc{$-$.0445} & \nc{$-$.0505} & \pc{+.0075} & \pc{+.0153} & \pc{+.0518} & \pc{+.0094} & 21--28 & 110--284\\
7 & 3 & \nc{$-$.0201} & \nc{$-$.0844} & \nc{$-$.0532} & \pc{+.0805} & \pc{+.0087} & \pc{+.1099} & \pc{+.0148} & 19--27 & 72--174\\
8 & 2 & \pc{+.0321} & \nc{$-$.1221} & \nc{$-$.0144} & \pc{+.1583} & \nc{$-$.0009} & \pc{+.1680} & \pc{+.0382} & 14--20 & 49--92\\
9 & 1 & \pc{+.0510} & \nc{$-$.1618} & \pc{+.0555} & \pc{+.1912} & \nc{$-$.0517} & \pc{+.1370} & \pc{+.1284} & 12 & 60--62\\
\bottomrule
\end{tabular}
\end{table}

\subsection
{CodeQL: Release-Conditional Temporal Loss}
\label{subsec:study-b-codeql}

R2024 is negative at all nine gaps, from $-0.0332$ at one year to $-0.1618$ at nine years. 
This is the clearest release-conditional temporal-loss pattern in our full matched analysis: once projects and labels are fixed, later-year code receives lower mean F1 under R2024 at every gap. Because the full gap analysis also contains source years predating R2024, we interpret this result as temporal sensitivity; prospective aging is supported specifically by comparisons in which the code follows the frozen detector condition. 

The magnitude is not monotonic, however. Gap~6 ($-0.0445$) is smaller than gap~5 ($-0.1376$), so the data do not support a constant decay rate. R2021 and R2026 are negative at gaps one through seven, after which R2021 turns positive at gaps eight and nine and R2026 at gap nine. Those reversals rest on three year pairs for R2021 (two at gap~8 and one at gap~9) and on one pair for R2026, and gap~9 is the single 2016--2025 comparison. They do not cancel the shorter-gap losses, but they show that one monotone aging law does not fit all releases. R2024 is also the only historical condition that used buildless extraction; our data do not isolate whether that mode contributes to its consistent losses. Variation across gaps also mixes different year pairs and a changing case composition, so the gap means should not be read as a dose--response curve.

\subsection{Infer: Opposite Directions across Releases}\label{subsec:study-b-infer}

The matched Infer results invert the Study-A ordering. v1.1 and v1.3 are positive at all nine gaps, whereas v1.2, the best Study-A release, is negative at six. Table~\ref{tab:infer-gaps} adds the spread across year pairs. It is large relative to most means. At gap~5, v1.1 has mean $+0.0188$ but a pair standard deviation of 0.1032 and pair quantiles from $-0.1173$ to $+0.1335$, so year pairs of opposite sign cancel. The lower 2.5\% quantile is positive only at gaps 3, 4, 7, and 8 for v1.3 and at gaps 7 and 8 for v1.1; for v1.2 it is never positive.

\begin{table}[!tb]
\caption{Infer Study-B results by gap: mean $\dF$ over year pairs, the standard deviation (SD) of the pair means, and their 2.5\% and 97.5\% quantiles. Gap~9 has a single pair, so its spread is undefined. Blue cells mark a positive lower quantile.}
\label{tab:infer-gaps}
\centering\scriptsize
\setlength{\tabcolsep}{2.6pt}
\begin{tabular}{@{}r|rrrr|rrrr|rrrr@{}}
\toprule
& \multicolumn{4}{c|}{\textbf{v1.1}} & \multicolumn{4}{c|}{\textbf{v1.2}} & \multicolumn{4}{c}{\textbf{v1.3}}\\
\textbf{Gap} & \textbf{Mean} & \textbf{SD} & \textbf{2.5\%} & \textbf{97.5\%} & \textbf{Mean} & \textbf{SD} & \textbf{2.5\%} & \textbf{97.5\%} & \textbf{Mean} & \textbf{SD} & \textbf{2.5\%} & \textbf{97.5\%}\\
\midrule
1 & $+$.0025 & .0283 & $-$.0417 & $+$.0480 & $-$.0115 & .0252 & $-$.0625 & $+$.0129 & $+$.0049 & .0234 & $-$.0265 & $+$.0413\\
2 & $+$.0043 & .0589 & $-$.0978 & $+$.0774 & $-$.0142 & .0407 & $-$.0864 & $+$.0334 & $+$.0178 & .0347 & $-$.0155 & $+$.0835\\
3 & $+$.0270 & .0609 & $-$.0377 & $+$.1191 & $-$.0165 & .0410 & $-$.0772 & $+$.0280 & $+$.0394 & .0373 & \pc{$+$.0040} & $+$.0947\\
4 & $+$.0379 & .0774 & $-$.0844 & $+$.1038 & $-$.0005 & .0320 & $-$.0476 & $+$.0407 & $+$.0785 & .0551 & \pc{$+$.0026} & $+$.1532\\
5 & $+$.0188 & .1032 & $-$.1173 & $+$.1335 & $+$.0307 & .0522 & $-$.0394 & $+$.0855 & $+$.0674 & .0786 & $-$.0506 & $+$.1370\\
6 & $+$.0075 & .0459 & $-$.0343 & $+$.0666 & $+$.0153 & .0376 & $-$.0200 & $+$.0580 & $+$.0518 & .0840 & $-$.0167 & $+$.1598\\
7 & $+$.0805 & .0574 & \pc{$+$.0407} & $+$.1416 & $+$.0087 & .0264 & $-$.0118 & $+$.0364 & $+$.1099 & .0664 & \pc{$+$.0536} & $+$.1777\\
8 & $+$.1583 & .1240 & \pc{$+$.0749} & $+$.2416 & $-$.0009 & .0843 & $-$.0575 & $+$.0557 & $+$.1680 & .0943 & \pc{$+$.1047} & $+$.2314\\
9 & $+$.1912 & --- & --- & --- & $-$.0517 & --- & --- & --- & $+$.1370 & --- & --- & ---\\
\bottomrule
\end{tabular}
\end{table}

Variation between the 100 repetitions of a single pair is much smaller than variation between pairs. Table~\ref{tab:infer-reps} gives it for the two pairs we report at repetition level, the adjacent 2016$\rightarrow$2017 pair and the longest 2016$\rightarrow$2025 pair. Here the repetition standard deviation is at most 0.0462. The disagreement between Studies A and B therefore comes from which pairs and projects carry the comparison, not from sampling noise. The table also shows one notable decline that the gap means hide: under v1.2, the 2016$\rightarrow$2017 pair has mean $-0.0724$, and its entire 2.5--97.5\% repetition range is negative. The same pair is also negative throughout for v1.1 and v1.3, but only by $-0.0008$ and $-0.0108$ on average. The apparent growth of v1.1 and v1.3 at gaps seven to nine rests on three, two, and one pairs; we do not read it as compositional analysis improving on newer code.

\begin{table}[!tb]
\caption{Repetition-level spread of Infer $\dF$ for two year pairs (100 repetitions each).}
\label{tab:infer-reps}
\centering\footnotesize
\setlength{\tabcolsep}{6pt}
\zebra
\begin{tabular}{@{}llrrrr@{}}
\toprule
\textbf{Release} & \textbf{Year pair} & \textbf{Mean $\dF$} & \textbf{Repetition SD} & \textbf{2.5\%} & \textbf{97.5\%}\\
\midrule
v1.1 & 2016$\rightarrow$2017 & $-$.0008 & .0016 & $-$.0051 & $-$.0001\\
v1.1 & 2016$\rightarrow$2025 & $+$.1912 & .0197 & $+$.1756 & $+$.2185\\
v1.2 & 2016$\rightarrow$2017 & \nc{$-$.0724} & .0092 & $-$.0898 & $-$.0532\\
v1.2 & 2016$\rightarrow$2025 & $-$.0517 & .0462 & $-$.1407 & $+$.0269\\
v1.3 & 2016$\rightarrow$2017 & $-$.0108 & .0045 & $-$.0183 & $-$.0010\\
v1.3 & 2016$\rightarrow$2025 & $+$.1370 & .0198 & $+$.1204 & $+$.1648\\
\bottomrule
\end{tabular}
\end{table}

\subsection{Semgrep: Pair Heterogeneity Dominates the Gap Mean}\label{subsec:study-b-semgrep}

Semgrep is negative only at gap~4 ($-0.0032$). Table~\ref{tab:sg-gaps} splits each gap mean into its precision and recall components and gives the range of pair means and the support. Precision changes are small and positive at every gap (0.0034 to 0.0248), whereas recall changes are negative at gaps 4 to 7. The selected and evaluated columns show how many cases the label-conflict gate removes after sampling: at gap~4, the smallest side has 110 selected and 99 evaluated cases.

\begin{table}[!tb]
\caption{Semgrep Study-B results by gap: mean change in precision, recall, and F1, the range of the year-pair mean $\dF$, matched projects per pair, and cases per year side before (selected) and after (evaluated) the gate.}
\label{tab:sg-gaps}
\centering\footnotesize
\setlength{\tabcolsep}{3.6pt}
\zebra
\begin{tabular}{@{}rrrrrrrrrr@{}}
\toprule
\textbf{Gap} & \textbf{Pairs} & \textbf{$\Delta$Prec.} & \textbf{$\Delta$Recall} & \textbf{$\dF$} & \textbf{Min pair} & \textbf{Max pair} & \textbf{Projects} & \textbf{Selected/side} & \textbf{Evaluated/side}\\
\midrule
1 & 9 & $+$.0060 & $+$.0084 & $+$.0073 & $-$.0108 & $+$.0361 & 32--113 & 482--1,492 & 464--1,492\\
2 & 8 & $+$.0078 & $+$.0165 & $+$.0124 & $-$.0180 & $+$.0617 & 24--89 & 404--1,374 & 393--1,374\\
3 & 7 & $+$.0097 & $+$.0081 & $+$.0091 & $-$.0241 & $+$.0537 & 23--53 & 298--754 & 290--754\\
4 & 6 & $+$.0115 & \nc{$-$.0196} & \nc{$-$.0032} & $-$.0348 & $+$.0361 & 27--55 & 110--580 & 99--580\\
5 & 5 & $+$.0034 & \nc{$-$.0028} & $+$.0057 & $-$.0740 & $+$.0976 & 27--34 & 102--352 & 92--352\\
6 & 4 & $+$.0178 & \nc{$-$.0007} & $+$.0094 & $-$.0243 & $+$.0438 & 21--28 & 114--284 & 110--284\\
7 & 3 & $+$.0248 & \nc{$-$.0044} & $+$.0148 & $-$.0602 & $+$.0729 & 19--27 & 74--174 & 72--174\\
8 & 2 & $+$.0179 & $+$.0557 & $+$.0382 & $-$.0390 & $+$.1154 & 14--20 & 50--92 & 49--92\\
9 & 1 & $+$.0105 & $+$.2595 & $+$.1284 & $+$.1284 & $+$.1284 & 12 & 62 & 60--62\\
\bottomrule
\end{tabular}
\end{table}

\begin{figure}[!tb]
\centering
\includegraphics[width=0.62\linewidth]{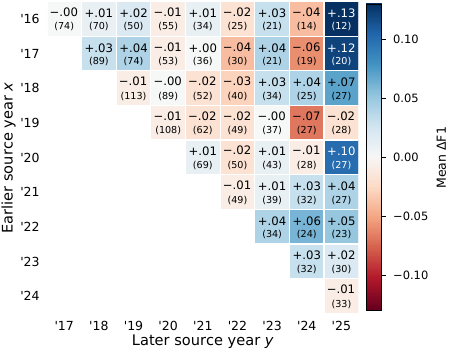}
\caption{Semgrep Study-B results for all 45 year pairs. Each cell shows the mean $\dF$ over 100 repetitions (two decimals) and, in parentheses, the number of matched projects. Red cells are negative (lower F1 in the later year), blue cells positive.}
\label{fig:heatmap}
\end{figure}

The gap means hide large pair-level differences, which Figure~\ref{fig:heatmap} shows for all 45 pairs together with the number of matched projects. Support is largest for adjacent years (up to 113 projects for 2018$\rightarrow$2019) and smallest for the longest gaps (12 projects for 2016$\rightarrow$2025). The three most negative pairs all end in 2024, and seven of the nine pairs that end in 2025 are positive. Table~\ref{tab:sg-extreme-pairs} lists the three most negative and three most positive pairs. At gap~5, pair means span $-0.0740$ to $+0.0976$ around a mean of $+0.0057$. For 2017$\rightarrow$2024 and 2016$\rightarrow$2024, the repetition range even crosses zero, whereas none of the three positive pairs does. The matched-sample F1 values also differ from the population values of Study~A: the 2016$\rightarrow$2025 pair starts at 0.4489, below the Study-A value for 2016 (0.4558), and ends at 0.5773, above every annual Study-A F1.

\begin{table}[!tb]
\caption{The three most negative and three most positive Semgrep year pairs. Early and late F1 are the matched-sample values; the last column is the 2.5--97.5\% range over the pair's 100 repetitions.}
\label{tab:sg-extreme-pairs}
\centering\footnotesize
\setlength{\tabcolsep}{5pt}
\zebra
\begin{tabular}{@{}lrrrrrr@{}}
\toprule
\textbf{Year pair} & \textbf{Projects} & \textbf{Selected/year} & \textbf{Early F1} & \textbf{Late F1} & \textbf{Mean $\dF$} & \textbf{Repetition range}\\
\midrule
2019$\rightarrow$2024 & 27 & 106 & .5978 & .5238 & \nc{$-$.0740} & $-$.1126 to $-$.0251\\
2017$\rightarrow$2024 & 19 & 74 & .5656 & .5054 & \nc{$-$.0602} & $-$.1291 to $+$.0124\\
2016$\rightarrow$2024 & 14 & 50 & .5264 & .4874 & \nc{$-$.0390} & $-$.1330 to $+$.0622\\
2016$\rightarrow$2025 & 12 & 62 & .4489 & .5773 & \pc{$+$.1284} & $+$.0068 to $+$.2395\\
2017$\rightarrow$2025 & 20 & 92 & .4388 & .5542 & \pc{$+$.1154} & $+$.0457 to $+$.1882\\
2020$\rightarrow$2025 & 27 & 102 & .4318 & .5294 & \pc{$+$.0976} & $+$.0111 to $+$.1755\\
\bottomrule
\end{tabular}
\end{table}

Support also shrinks quickly with the gap. Gap~1 pairs have 32--113 projects and 464--1,492 evaluated cases per side; gap~9 has 12 projects and 60--62 cases per side. Figure~\ref{fig:study-b}(c) and Table~\ref{tab:sg-gaps} show that the gap-9 value is driven by recall ($+0.2595$) while precision barely moves ($+0.0105$): the 2025 side receives more positives, but they are not concentrated on vulnerable cases. Since Semgrep's rules are identical on both sides, we treat the $+0.1284$ as a property of one 12-project comparison, not as evidence that distance in time improves detection. Matching fixes projects and label counts, but vulnerability type, module, scan scope, and analysis quality still vary within a pair.

\finding{B}{Exact project-by-label matching reveals a consistent later-year loss for CodeQL R2024, mostly negative short-gap changes for R2021 and R2026, positive changes at every gap for Infer v1.1 and v1.3, small mixed changes for Infer v1.2, and mostly small changes for Semgrep outside the sparse longest gaps. Sparse long gaps and large pair-to-pair differences make a tool-wide ``F1 lost per year of code age'' unsupportable.}

\section{RQ3: Within-Project Change (Study C)}\label{sec:study-c}

\begin{figure}[!tb]
\centering
\includegraphics[width=\linewidth]{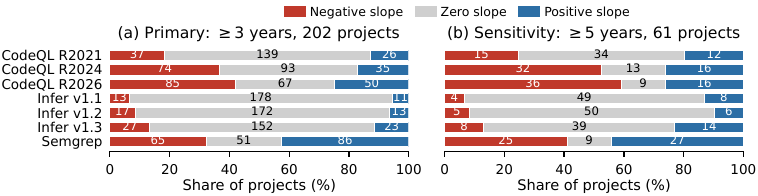}
\caption{Direction of project-level F1 slopes in Study~C. Red, gray, and blue mark negative, zero, and positive slopes; numbers inside bars are project counts.}
\label{fig:slopes}
\end{figure}

Study~C asks a narrower question than Study~B: as the same project evolves, does its F1 change? Table~\ref{tab:study-c} reports the distribution of project-level F1 slopes and the pooled cohort metrics, and Figure~\ref{fig:slopes} shows the slope directions as shares. To read the first row: under CodeQL R2021, 37 of the 202 primary projects have a negative slope, 139 a zero slope, and 26 a positive slope, and the mean slope is $-0.00170$ F1 per year. Table~\ref{tab:cohort-metrics} adds the cohort precision, recall, and FPR. Slope signs are not significance tests, and the 61 five-year projects also belong to the primary cohort.

\begin{table}[!tb]
\caption{Study-C project slopes and pooled cohort metrics. N/Z/P count projects with negative, zero, and positive F1 slopes; ``Mean/yr'' is the equal-project mean slope. Cohort F1 and AUROC are equal-year means of pooled annual metrics (Semgrep AUROC not computed for cohorts). The cohorts are nested.}
\label{tab:study-c}
\centering\footnotesize
\setlength{\tabcolsep}{3pt}
\zebra
\begin{tabular}{@{}l|rrrrr|rrrrr@{}}
\toprule
& \multicolumn{5}{c|}{\textbf{Primary cohort ($\geq$3 years, 202 projects)}} & \multicolumn{5}{c}{\textbf{Sensitivity cohort ($\geq$5 years, 61 projects)}}\\
\textbf{Condition} & \textbf{N/Z/P} & \textbf{N share} & \textbf{Mean/yr} & \textbf{F1} & \textbf{AUROC} & \textbf{N/Z/P} & \textbf{N share} & \textbf{Mean/yr} & \textbf{F1} & \textbf{AUROC}\\
\midrule
CodeQL R2021 & 37/139/26 & 18.3\% & \nc{$-$.00170} & .1141 & .4996 & 15/34/12 & 24.6\% & \pc{+.00039} & .1135 & .4999\\
CodeQL R2024 & 74/93/35 & 36.6\% & \nc{$-$.01926} & .1918 & .5003 & 32/13/16 & 52.5\% & \nc{$-$.01247} & .1929 & .5012\\
CodeQL R2026 & 85/67/50 & 42.1\% & \nc{$-$.01657} & .2760 & .4981 & 36/9/16 & 59.0\% & \nc{$-$.01095} & .2801 & .4985\\
Infer v1.1 & 13/178/11 & 6.4\% & \nc{$-$.00267} & .0511 & .4988 & 4/49/8 & 6.6\% & \pc{+.00538} & .0490 & .4990\\
Infer v1.2 & 17/172/13 & 8.4\% & \nc{$-$.00673} & .1722 & .5273 & 5/50/6 & 8.2\% & \nc{$-$.00353} & .1838 & .5286\\
Infer v1.3 & 27/152/23 & 13.4\% & \nc{$-$.00514} & .1231 & .4983 & 8/39/14 & 13.1\% & \pc{+.00939} & .1106 & .4986\\
Semgrep & 65/51/86 & 32.2\% & \pc{+.01426} & .4820 & --- & 25/9/27 & 41.0\% & \pc{+.00286} & .4919 & ---\\
\bottomrule
\end{tabular}
\end{table}

\begin{table}[!tb]
\caption{Study-C cohort precision, recall, and FPR (equal-year means of pooled annual values). For CodeQL and for Infer v1.1 and v1.3, recall and FPR are close, as in Study~A. Semgrep has no cohort FPR.}
\label{tab:cohort-metrics}
\centering\footnotesize
\setlength{\tabcolsep}{6pt}
\zebra
\begin{tabular}{@{}l|rrr|rrr@{}}
\toprule
& \multicolumn{3}{c|}{\textbf{Primary cohort}} & \multicolumn{3}{c}{\textbf{Sensitivity cohort}}\\
\textbf{Condition} & \textbf{Precision} & \textbf{Recall} & \textbf{FPR} & \textbf{Precision} & \textbf{Recall} & \textbf{FPR}\\
\midrule
CodeQL R2021 & .5030 & .0677 & .0686 & .4962 & .0687 & .0689\\
CodeQL R2024 & .5073 & .1251 & .1255 & .5121 & .1257 & .1240\\
CodeQL R2026 & .4977 & .1942 & .1982 & .4996 & .1992 & .2021\\
Infer v1.1 & .4813 & .0279 & .0302 & .4750 & .0269 & .0289\\
Infer v1.2 & .7196 & .1015 & .0475 & .7238 & .1105 & .0541\\
Infer v1.3 & .4890 & .0730 & .0765 & .4872 & .0646 & .0675\\
Semgrep & .4928 & .4753 & --- & .4919 & .4961 & ---\\
\bottomrule
\end{tabular}
\end{table}

\subsection{CodeQL: Higher Level, More Declining Projects}\label{subsec:study-c-codeql}

The number of projects with negative descriptive F1 slopes (i.e., \textit{declining projects}) 
doubles from R2021 to R2024 and grows again in R2026 (37, 74, 85), and the mean slope is far more negative under the two later releases ($-0.00170$, $-0.01926$, $-0.01657$). At the same time, pooled cohort F1 rises from 0.1141 to 0.2760 while AUROC stays within 0.0019 of 0.5. Table~\ref{tab:cohort-metrics} shows the reason for the rising level: as in Study~A, recall and FPR grow together and stay almost equal in every release (for example 0.1942 and 0.1982 for R2026). Level and trajectory are different properties. A release can flag more cases overall and still lose its signal faster as individual projects evolve. The nested five-year cohort keeps the ordering (15, 32, and 36 of 61 projects decline), so the difference between R2021 and the later releases does not come from short trajectories alone. The cohort metrics and the mean project slope use different weights and should not be read as two estimates of the same quantity.

\subsection{Infer: Flat because Empty}\label{subsec:study-c-infer}

Most Infer slopes are zero: 178, 172, and 152 of 202 primary projects. Table~\ref{tab:zero-slopes} decomposes these flat trajectories and shows that they rarely represent stable detection. For v1.1, 176 of the 178 flat projects have F1 equal to zero in every usable year, and only two have a constant nonzero F1; the numbers are 167 of 172 for v1.2 and 144 of 152 for v1.3. A flat slope thus usually means that Infer never yields a label-level signal for the project. A zero F1 does not mean that Infer reported nothing: the project may still receive findings on fixed cases or outside the scored files.

\begin{table}[!tb]
\caption{Decomposition of zero Infer F1 slopes. ``All-zero'': F1 is zero in every usable year; ``Const.\,$>$0'': the same nonzero F1 in every year; ``Nonconst.\,$\approx$0'': F1 varies, but the slope stays within the zero tolerance.}
\label{tab:zero-slopes}
\centering\footnotesize
\setlength{\tabcolsep}{6pt}
\zebra
\begin{tabular}{@{}llrrrr@{}}
\toprule
\textbf{Release} & \textbf{Cohort} & \textbf{Zero slopes} & \textbf{All-zero} & \textbf{Const.\,$>$0} & \textbf{Nonconst.\,$\approx$0}\\
\midrule
v1.1 & Primary & 178 & 176 & 2 & 0\\
v1.1 & Sensitivity & 49 & 48 & 1 & 0\\
v1.2 & Primary & 172 & 167 & 3 & 2\\
v1.2 & Sensitivity & 50 & 49 & 0 & 1\\
v1.3 & Primary & 152 & 144 & 8 & 0\\
v1.3 & Sensitivity & 39 & 38 & 1 & 0\\
\bottomrule
\end{tabular}
\end{table}

Every primary and sensitivity project has a fitted F1 slope under all three releases. AUROC slopes are less complete, because AUROC is undefined when one label is missing from a project-year after exclusions: three primary projects and one sensitivity project have no AUROC trend under v1.1 and v1.3, while v1.2 has all of them. Projects with a nonzero F1 slope use between three and nine usable years (five to nine in the sensitivity cohort), and their endpoints cover all of 2016--2025.

\begin{figure}[!tb]
\centering
\includegraphics[width=\linewidth]{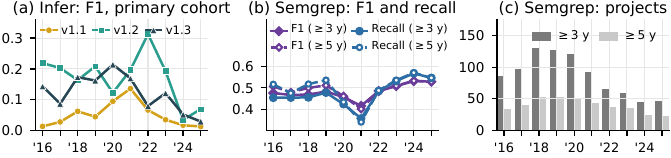}
\caption{Study-C annual cohort profiles by source-code year (x-axis), pooled over projects and averaged over 100 repetitions. (b) Solid lines: primary cohort; dashed lines with open markers: nested sensitivity cohort. (c) The number of contributing Semgrep projects shrinks sharply after 2020.}
\label{fig:study-c}
\end{figure}

Among nonzero trajectories, both directions occur under every release. The primary mean slopes are mildly negative, while the sensitivity cohort turns v1.1 and v1.3 positive ($+0.00538$ and $+0.00939$). That sign change reflects which projects supply long histories rather than an improvement, because the restricted cohort is selected on the number of usable years. Figure~\ref{fig:study-c}(a) shows the pooled primary-cohort F1 by year: the releases cross repeatedly, v1.2 peaks at about 0.31 in 2022, and all three fall to between about 0.01 and 0.07 by 2025. The cohort precision of v1.2 (0.7196) is far above that of the other releases, again because v1.2 flags few fixed cases (Table~\ref{tab:cohort-metrics}).

\subsection{Semgrep: A Rising Curve That Projects Do Not Share}\label{subsec:study-c-semgrep}

\begin{figure}[!tb]
\centering
\includegraphics[width=\linewidth]{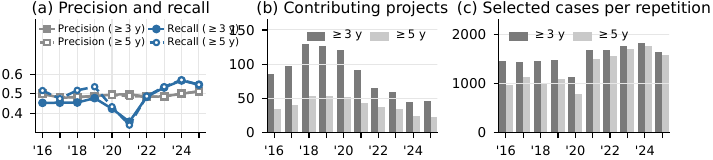}
\caption{Semgrep Study-C cohorts by source-code year (x-axis). (a) Pooled precision and recall; solid lines: primary cohort, dashed lines with open markers: sensitivity cohort. (b) Projects that contribute to each year. (c) Selected cases per repetition. Fewer projects carry more cases in later years.}
\label{fig:sg-studyc}
\end{figure}

Table~\ref{tab:sg-studyc} gives the Semgrep cohort values for every year, and Figure~\ref{fig:sg-studyc} plots their precision, recall, and support. The primary cohort falls from F1=0.4826 in 2019 to 0.4155 in 2021 and then rises to 0.5321 in 2024; the five-year cohort follows the same shape (0.5118, 0.4007, 0.5335). Two facts temper this rise. First, it is carried by recall, which dips to 0.3565 in 2021 and peaks at 0.5683 in 2024, while precision stays within 0.479--0.513. Second, the number of contributing projects falls from 130 in 2018 to 45 in 2024, while the number of selected cases per repetition rises from 1,460 to 1,830. In 2024, 1,830 cases come from 45 projects (about 41 per project), against 1,454 cases from 86 projects in 2016 (about 17): later years reflect fewer, larger projects. The annual values pool cases across projects; they are not equal-project means and do not imply that most projects followed the same path over the decade.

\begin{table}[!tb]
\caption{Semgrep Study-C annual values for both cohorts: contributing projects, selected cases per repetition (balanced before the label-conflict gate), the range of evaluated cases over repetitions, and pooled precision, recall, and F1 averaged over 100 repetitions.}
\label{tab:sg-studyc}
\centering\footnotesize
\setlength{\tabcolsep}{3pt}
\zebra
\begin{tabular}{@{}l|rrrrrr|rrrrrr@{}}
\toprule
& \multicolumn{6}{c|}{\textbf{Primary cohort ($\geq$3 years)}} & \multicolumn{6}{c}{\textbf{Sensitivity cohort ($\geq$5 years)}}\\
\textbf{Year} & \textbf{Proj.} & \textbf{Sel.} & \textbf{Eval.} & \textbf{Prec.} & \textbf{Recall} & \textbf{F1} & \textbf{Proj.} & \textbf{Sel.} & \textbf{Eval.} & \textbf{Prec.} & \textbf{Recall} & \textbf{F1}\\
\midrule
2016 & 86 & 1,454 & 1,415--1,425 & .5006 & .4536 & .4759 & 34 & 980 & 972--974 & .4976 & .5165 & .5068\\
2017 & 97 & 1,442 & 1,432--1,433 & .4796 & .4543 & .4666 & 41 & 1,132 & 1,124 & .4833 & .4753 & .4793\\
2018 & 130 & 1,460 & 1,430--1,433 & .4828 & .4553 & .4687 & 53 & 1,004 & 980--982 & .4828 & .5179 & .4998\\
2019 & 127 & 1,478 & 1,447--1,452 & .4876 & .4778 & .4826 & 53 & 1,100 & 1,075--1,078 & .4896 & .5361 & .5118\\
2020 & 121 & 1,138 & 1,094--1,096 & .4954 & .4235 & .4566 & 52 & 790 & 752--753 & .4930 & .4336 & .4614\\
2021 & 92 & 1,682 & 1,672--1,673 & .4979 & .3565 & .4155 & 43 & 1,508 & 1,498--1,499 & .4886 & .3396 & .4007\\
2022 & 66 & 1,688 & 1,687--1,688 & .4849 & .4848 & .4848 & 37 & 1,552 & 1,551--1,552 & .4848 & .4879 & .4863\\
2023 & 59 & 1,774 & 1,774 & .4866 & .5323 & .5084 & 35 & 1,706 & 1,706 & .4859 & .5359 & .5097\\
2024 & 45 & 1,830 & 1,828 & .5002 & .5683 & .5321 & 25 & 1,772 & 1,770 & .5007 & .5710 & .5335\\
2025 & 46 & 1,652 & 1,652 & .5124 & .5466 & .5289 & 23 & 1,582 & 1,582 & .5127 & .5468 & .5292\\
\midrule
Total & 869 & 15,598 & & & & & 396 & 13,126 & & & & \\
\bottomrule
\end{tabular}
\end{table}

\begin{figure}[!tb]
\centering
\includegraphics[width=0.74\linewidth,trim=20 0 0 0,clip]{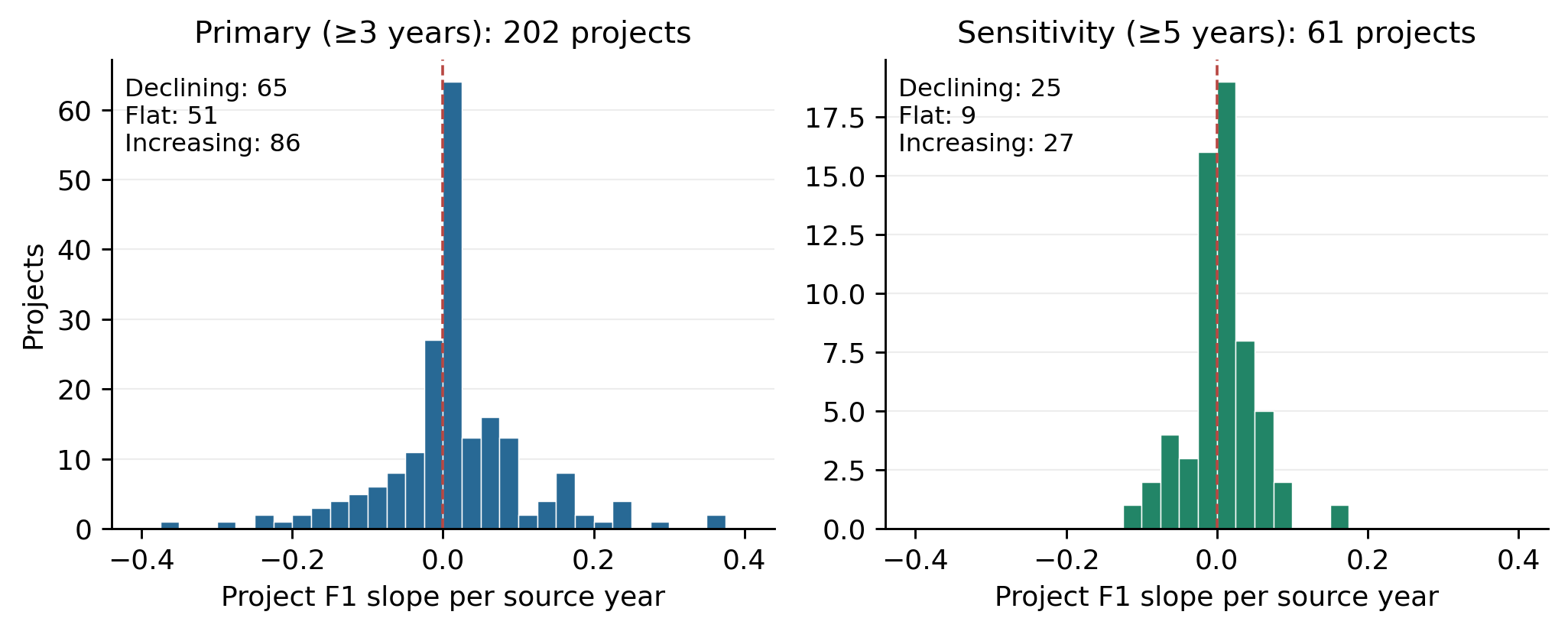}
\caption{Distribution of Semgrep project F1 slopes (F1 per source year) in the primary and sensitivity cohorts; the dashed line marks zero. Slopes within $\pm10^{-10}$ count as flat.}
\label{fig:semgrep-slopes}
\end{figure}

At the project level, 86 primary projects (42.6\%) have positive slopes, 65 (32.2\%) negative ones, and 51 (25.2\%) flat ones, with a mean of $+0.0143$ F1 per year. A higher pooled F1 in later years therefore does not mean that every project improved. In the five-year cohort the split is nearly even (27 versus 25) and the mean drops to $+0.0029$. Figure~\ref{fig:semgrep-slopes} shows the full slope distributions. Most slopes lie within $\pm0.1$ F1 per year. The primary cohort has tails beyond $\pm0.3$, whereas no project in the five-year cohort exceeds $\pm0.2$; extreme slopes thus come from short histories.

Nine project-year pairs tie for the largest Semgrep decline, from F1=1.0 to 0.0 (Table~\ref{tab:semgrep-extremes}). In all of them, every early cell has exactly one true positive and one true negative per repetition, so a single changed prediction erases the whole F1, and in these selected cells the counts do not change across the 100 repetitions. The late cells differ. MuJS 2022 has two false negatives and one false positive, libxml2 2020 has four false negatives and one false positive, Varnish 2019 has two false negatives, and the others have one false negative and no false positive. The same F1 drop therefore aggregates different error patterns. The representative cases in the table are vulnerable cases from the early and late cells; they are not the same vulnerability seen twice, and the source year of a snapshot can differ from the year in its CVE identifier (the 2020 Varnish case carries CVE-2021-36740). Because MilkyTracker's and Varnish's pairs share their start year, the nine rows are not independent. After removing shared cells, these pairs give the 42-case queue of \S\ref{sec:failure-analysis}, with 7 TP, 2 FP, 14 FN, and 19 TN.

\begin{table}[!tb]
\caption{The nine tied largest Semgrep Study-C declines (F1 1.0$\rightarrow$0.0) with a representative vulnerable case from each endpoint. The late cell gives TP/FP/FN/TN per repetition (every early cell is 1/0/0/1). Diagnostics: P = partial parsing, T = rule timeout, S = syntax error, --- = none.}
\label{tab:semgrep-extremes}
\centering\scriptsize
\setlength{\tabcolsep}{3pt}
\zebra
\begin{tabular}{@{}llc>{\raggedright\arraybackslash}p{0.34\linewidth}>{\raggedright\arraybackslash}p{0.34\linewidth}@{}}
\toprule
\textbf{Project} & \textbf{Years} & \textbf{Late cell} & \textbf{Early representative (TP): CVE, findings, diag., file} & \textbf{Late representative (FN): CVE, diag., file}\\
\midrule
MuJS & 2018$\rightarrow$2022 & 0/1/2/1 & CVE-2018-6191; 1; P, T; \texttt{jsparse.c} & CVE-2022-44789; ---; \texttt{jsobject.c}\\
FreeRDP & 2021$\rightarrow$2025 & 0/0/1/1 & CVE-2021-37594; 2; P; \texttt{timezone.c} & CVE-2025-68118; P; \texttt{certificate\_data.c}\\
libreswan & 2019$\rightarrow$2020 & 0/0/1/1 & CVE-2019-12312; 9; P, T; \texttt{kernel.c} & CVE-2020-1763; P; \texttt{ikev1.c}\\
MilkyTracker & 2019$\rightarrow$2020 & 0/0/1/1 & CVE-2019-14464; 2; ---; \texttt{LoaderXM.cpp} & CVE-2020-15569; ---; \texttt{ChannelMixer.cpp}\\
MilkyTracker & 2019$\rightarrow$2022 & 0/0/1/1 & CVE-2019-14464; 2; ---; \texttt{LoaderXM.cpp} & CVE-2022-34927; T; \texttt{SectionSamples.cpp}\\
OpenBSD & 2019$\rightarrow$2024 & 0/0/1/1 & CVE-2019-8460; 21; P, S, T; \texttt{drm\_linux.c} & CVE-2025-30334; S; \texttt{if\_wg.c}\\
Varnish & 2017$\rightarrow$2019 & 0/0/2/2 & CVE-2017-8807; 6; P; \texttt{cache\_acceptor.c} & CVE-2019-20637; ---; \texttt{vdef.h}\\
Varnish & 2017$\rightarrow$2020 & 0/0/1/1 & CVE-2017-8807; 6; P; \texttt{cache\_acceptor.c} & CVE-2021-36740; P; \texttt{cache\_http2.h}\\
libxml2 & 2019$\rightarrow$2020 & 0/1/4/3 & CVE-2019-19956; 25; P, T; \texttt{relaxng.c} & PRIMEVUL-2020-0e1a\ldots; S; \texttt{parser.c}\\
\bottomrule
\end{tabular}
\end{table}

\finding{C}{Within-project evidence separates detection level, trajectory, and the presence of any signal. Later CodeQL releases have higher F1 but more declining projects; most flat Infer projects never produce a positive F1; and Semgrep's positive mean slope nearly disappears when longer histories are required. The largest drops start from early cells with only two cases.}

\section{RQ4: Failure Mechanisms behind the Largest Drops}\label{sec:failure-analysis}

Studies A--C show where scores change; RQ4 asks what the recorded evidence says about why. Table~\ref{tab:mechanisms} summarizes the mechanisms we found, each with concrete cases and the limit of what the evidence shows. The selected cases are extremes, so the table describes which mechanisms exist, not how often they occur. The 42 Semgrep cases and the small, shared cells of the other queues cannot estimate the prevalence of any cause in the full population.

\begin{table}[!tb]
\caption{Mechanisms observed in the selected drops (RQ4) and execution records (RQ5). ``Observed'' does not imply population prevalence.}
\label{tab:mechanisms}
\centering\footnotesize
\setlength{\tabcolsep}{3.5pt}
\renewcommand{\arraystretch}{1.08}
\zebra
\begin{tabular}{@{}>{\raggedright\arraybackslash}p{0.13\linewidth}>{\raggedright\arraybackslash}p{0.6\linewidth}>{\raggedright\arraybackslash}p{0.23\linewidth}@{}}
\toprule
\textbf{Mechanism} & \textbf{Concrete evidence (tool)} & \textbf{What the evidence does not show}\\
\midrule
Scope mismatch & HAProxy 2019 alerts in \texttt{proto\_htx.c}, fix in \texttt{mux\_h2.c} (CodeQL); libiec61850 alerts in \texttt{control.c}, fix in \texttt{cotp.c} (Infer); 11 of 20 vulnerable/fixed groups scored on disjoint files (Semgrep) & That off-repair alerts are unrelated to the CVE; data flow can cross files\\
Incidental positive & 17 \texttt{memset} findings in libxml2 \texttt{relaxng.c}; two \texttt{memset} findings in MilkyTracker \texttt{XModule.h}; \texttt{atoi} and \texttt{sizeof}-pointer findings (Semgrep); \texttt{DEAD\_STORE} warnings (Infer) & That the warnings are false; only that they do not reach the repair\\
Fixed-side warning & HAProxy 2023 overflow check present before and after the fix (CodeQL R2024); 10 file-level hits on Slurm's fixed 2020 snapshot (Infer v1.2); six \texttt{sizeof}-pointer findings on the MuJS fixed side (Semgrep) & That the warning is wrong for every defect\\
Rule/query mismatch & Nagios NULL entry guard vs.\ nullness predicates; BlueZ bonding check vs.\ authentication queries (CodeQL); Varnish CRLF-length invariant vs.\ API patterns (Semgrep) & That no configured rule could ever match\\
Trace mismatch & libxml2 use-after-free traces start at the C assignment \texttt{delete = cur;} (Semgrep) & The parser's internal cause\\
Parse/extract limits & Partial parsing and file-wide syntax errors (Semgrep); missing headers, incomplete types (CodeQL); empty or partial capture (Infer) & Which construct was lost\\
Rule-specific timeout & Two string-mismatch rules time out on MilkyTracker \texttt{SectionSamples.cpp}, while the repair is in \texttt{LoaderXM.cpp} (Semgrep) & The cause of a miss in another file or on the other label\\
Build failure & No compiler seen by a traced build; buildless discovery that ignores GCC (CodeQL); unmaterialized pinned submodules (Infer) & The cause of every historical failure row\\
Query cost & Two range-analysis queries consume a 12-hour budget on BlueZ (CodeQL) & Universal nontermination or OOM\\
Repeated snapshot & Two libxml2 identifiers select the same commit pair and \texttt{parser.c} change (Semgrep) & Two independent confirmations\\
Small cells & Every selected early cell has one TP and one TN (all tools) & Population-wide deterioration\\
\bottomrule
\end{tabular}
\end{table}

\subsection{The Changed-File Proxy Can Manufacture a Drop}\label{subsec:proxy}

\head{CodeQL.} Table~\ref{tab:drops} lists the selected CodeQL and Infer drops. Each release column gives F1 at the earlier and later endpoint, and the last two columns repeat the selecting release's F1 under the two location ablations. In the first row, for example, CodeQL R2021 falls from 1 to 0 on HAProxy 2019$\rightarrow$2022, but both endpoints are already 0 when only alerts in the repaired file count. Among the 39 scored CodeQL findings in the selected cases, 22 lie outside the verified repair file, 11 lie in the repair file but outside the changed hunk, and only six touch a changed line. These counts come from 27 extreme cases and do not estimate a population false-positive rate. The three HAProxy ties (\S\ref{subsec:motivating}) and BlueZ 2016$\rightarrow$2020 all disappear once alerts must lie in the repaired file: in each, the earlier positive alert lies outside the paired repair file. For the HAProxy pairs, the later releases did not improve the signal; the earlier signal simply disappeared. Nagios is the exception. Its 2016 signal survives even the changed-line restriction under all three releases, and its 2018 NULL-dereference repair is genuinely missed (\S\ref{subsec:no-finding}). For Unbound, a 2024 case lacks a paired repair, so we claim no ablated value for its later endpoint.

\begin{table}[!tb]
\caption{Selected maximum drops for CodeQL and Infer: F1 at the earlier$\rightarrow$later endpoint for each release, and the selecting release's F1 when only scored alerts in a verified repair file or on a changed repair line are kept. Red marks a drop; n/r: no ablation was computed for this supplementary pair; n/a: no paired repair at the later endpoint; ``no drop'': both endpoints are equal under the ablated score.}
\label{tab:drops}
\centering\scriptsize
\setlength{\tabcolsep}{3.2pt}
\zebra
\begin{tabular}{@{}lllcccll@{}}
\toprule
\textbf{Tool} & \textbf{Project, years} & \textbf{Selected by} & \textbf{2021 release} & \textbf{2024 release} & \textbf{2026 release} & \textbf{Repair file} & \textbf{Repair line}\\
\midrule
CodeQL & HAProxy 2019$\rightarrow$2022 & R2021 & \nc{1$\rightarrow$0} & 0$\rightarrow$0 & 0$\rightarrow$0 & 0$\rightarrow$0 & 0$\rightarrow$0\\
CodeQL & HAProxy 2019$\rightarrow$2023 & R2021 & \nc{1$\rightarrow$0} & \pc{0$\rightarrow$.67} & 0$\rightarrow$0 & 0$\rightarrow$0 & 0$\rightarrow$0\\
CodeQL & HAProxy 2019$\rightarrow$2024 & R2021 & \nc{1$\rightarrow$0} & 0$\rightarrow$0 & 0$\rightarrow$0 & 0$\rightarrow$0 & 0$\rightarrow$0\\
CodeQL & Nagios 2016$\rightarrow$2018 & R2021 & \nc{1$\rightarrow$0} & \nc{1$\rightarrow$0} & \nc{1$\rightarrow$0} & \nc{1$\rightarrow$0} & \nc{1$\rightarrow$0}\\
CodeQL & Nagios 2016$\rightarrow$2017 & R2024 & \nc{1$\rightarrow$.67} & \nc{1$\rightarrow$0} & \nc{1$\rightarrow$0} & n/r & n/r\\
CodeQL & BlueZ 2016$\rightarrow$2020 & R2024 & 0$\rightarrow$0 & \nc{1$\rightarrow$0} & \nc{1$\rightarrow$0} & 0$\rightarrow$0 & 0$\rightarrow$0\\
CodeQL & BlueZ 2016$\rightarrow$2023 & R2024 & 0$\rightarrow$0 & \nc{1$\rightarrow$0} & \nc{1$\rightarrow$0} & n/r & n/r\\
CodeQL & Unbound 2020$\rightarrow$2024 & R2024 & 0$\rightarrow$0 & \nc{1$\rightarrow$0} & \nc{1$\rightarrow$0} & 0$\rightarrow$n/a & n/a\\
\midrule
Infer & libiec61850 2020$\rightarrow$2023 & all & \nc{1$\rightarrow$0} & \nc{1$\rightarrow$0} & \nc{1$\rightarrow$0} & 0$\rightarrow$0 & 0$\rightarrow$0\\
Infer & firejail 2019$\rightarrow$2022 & v1.2 & 0$\rightarrow$0 & \nc{1$\rightarrow$0} & 0$\rightarrow$0 & \nc{1$\rightarrow$0} & 0$\rightarrow$0\\
Infer & acrn-hypervisor 2020$\rightarrow$2021 & v1.2 & 0$\rightarrow$0 & \nc{1$\rightarrow$0} & 0$\rightarrow$0 & \nc{1$\rightarrow$0} & 0$\rightarrow$0\\
Infer & Slurm 2016$\rightarrow$2020 & v1.2 & 0$\rightarrow$0 & \nc{1$\rightarrow$0} & 0$\rightarrow$0 & \nc{1$\rightarrow$0} & 0$\rightarrow$0\\
Infer & Slurm 2017$\rightarrow$2020 & v1.2 & 0$\rightarrow$0 & \nc{1$\rightarrow$0} & 0$\rightarrow$0 & \nc{1$\rightarrow$0} & 0$\rightarrow$0\\
Infer & mosquitto 2018$\rightarrow$2021 & v1.3 & 0$\rightarrow$0 & 0$\rightarrow$0 & \nc{1$\rightarrow$0} & no drop & no drop\\
Infer & mosquitto 2018$\rightarrow$2024 & v1.3 & 0$\rightarrow$0 & 0$\rightarrow$0 & \nc{1$\rightarrow$0} & no drop & no drop\\
Infer & folly 2018$\rightarrow$2019 & v1.3 & 0$\rightarrow$0 & 0$\rightarrow$0 & \nc{1$\rightarrow$0} & no drop & no drop\\
Infer & folly 2018$\rightarrow$2021 & v1.3 & 0$\rightarrow$0 & 0$\rightarrow$0 & \nc{1$\rightarrow$0} & no drop & no drop\\
\bottomrule
\end{tabular}
\end{table}

Table~\ref{tab:codeql-cells} shows the confusion counts behind the CodeQL endpoints. They are the same in all 100 repetitions, so each count holds per repetition and is not multiplied by 100. Similar F1 changes hide distinct patterns. Under R2024, HAProxy 2023 has one true and one false positive (F1=2/3); under R2026 it loses the true positive and keeps the false positive. Nagios goes from one TP and one TN in 2016 to one FN and one TN in 2018 under every release. BlueZ 2020 is the only selected cell with four cases per repetition. We also checked whether the final security filter could have removed an existing alert. Of the 81 case-by-release records, 68 have a zero score and no changed-file finding of any kind before that filter, so the filter did not discard an alert in a scored file.

\begin{table}[!tb]
\caption{Confusion counts (TP, FP, FN, TN) per repetition for the CodeQL cells behind Table~\ref{tab:drops}. FP means disagreement with the fixed label, not a manual judgment.}
\label{tab:codeql-cells}
\centering\footnotesize
\setlength{\tabcolsep}{6pt}
\zebra
\begin{tabular}{@{}lcccc@{}}
\toprule
\textbf{Project and year} & \textbf{Cases} & \textbf{R2021} & \textbf{R2024} & \textbf{R2026}\\
\midrule
HAProxy 2019 & 2 & (1, 0, 0, 1) & (0, 0, 1, 1) & (0, 0, 1, 1)\\
HAProxy 2023 & 2 & (0, 0, 1, 1) & (1, 1, 0, 0) & (0, 1, 1, 0)\\
Nagios 2016 & 2 & (1, 0, 0, 1) & (1, 0, 0, 1) & (1, 0, 0, 1)\\
Nagios 2018 & 2 & (0, 0, 1, 1) & (0, 0, 1, 1) & (0, 0, 1, 1)\\
BlueZ 2016 & 2 & (0, 0, 1, 1) & (1, 0, 0, 1) & (1, 0, 0, 1)\\
BlueZ 2020 & 4 & (0, 0, 2, 2) & (0, 0, 2, 2) & (0, 0, 2, 2)\\
Unbound 2020 & 2 & (0, 0, 1, 1) & (1, 0, 0, 1) & (1, 0, 0, 1)\\
\bottomrule
\end{tabular}
\end{table}

\head{Infer.} Infer exposes a complementary failure. Of the 620 scored findings in its selected cases, 610 lie in a verified repair file, so file-level filtering barely changes the score; yet only two lie on a changed repair line. Of the 27 selected pair--release combinations, 11 show a drop under the original score, four still show it under the repair-file restriction (firejail, acrn-hypervisor, and both Slurm pairs, all under v1.2), and none survives the changed-line restriction. File-level agreement is therefore necessary but not sufficient: an alert elsewhere in a large repaired file can create the early true positive or the later false positive that drives a drop. The scored findings are also not one kind of security finding. They include dead stores, buffer overruns, integer overflows, null dereferences, and large allocations. Table~\ref{tab:infer-endpoints} gives the raw scores at every endpoint, and Table~\ref{tab:infer-locations} shows where the scoring findings lie.

\begin{table}[!tb]
\caption{Infer changed-file scores at the selected endpoints, as vulnerable/fixed case scores per release. ``(e)'' marks an \texttt{empty\_capture} observation: the run produced a legal report, but capture recorded no source files. }
\label{tab:infer-endpoints}
\centering\footnotesize
\setlength{\tabcolsep}{6pt}
\zebra
\begin{tabular}{@{}llcccl@{}}
\toprule
\textbf{Project} & \textbf{Year} & \textbf{v1.1} & \textbf{v1.2} & \textbf{v1.3} & \textbf{What the endpoint shows}\\
\midrule
libiec61850 & 2020 & 2/0 & 2/0 & 2/0 & nonempty positive vulnerable endpoint in all releases\\
libiec61850 & 2023 & 0/0 & 0/0 & 0/0 & zero score in all releases; captures are nonempty\\
firejail & 2019 & 0/0 (e) & 3/0 & 0/0 (e) & only v1.2 captured and scored the snapshot\\
firejail & 2022 & 0/0 (e) & 0/0 & 0/0 (e) & zero score; v1.1 and v1.3 captured nothing\\
Slurm & 2016 & 0/0 (e) & 298/0 & 0/0 (e) & only v1.2 captured and scored the snapshot\\
Slurm & 2017 & 0/0 (e) & 298/0 & 0/0 (e) & same vulnerable code as 2016 (CVE-2016-10030)\\
Slurm & 2020 & 0/0 (e) & 0/10 & 0/0 (e) & v1.2 keeps ten file-level hits on the fixed case\\
\bottomrule
\end{tabular}
\end{table}

\begin{table}[!tb]
\caption{Where the scoring Infer findings lie, and what the verified repair changed.}
\label{tab:infer-locations}
\centering\scriptsize
\setlength{\tabcolsep}{3pt}
\zebra
\begin{tabular}{@{}>{\raggedright\arraybackslash}p{0.13\linewidth}>{\raggedright\arraybackslash}p{0.29\linewidth}>{\raggedright\arraybackslash}p{0.2\linewidth}>{\raggedright\arraybackslash}p{0.31\linewidth}@{}}
\toprule
\textbf{Endpoint} & \textbf{Primary finding locations} & \textbf{Score and issue types} & \textbf{Verified repair change}\\
\midrule
libiec61850 2020, vulnerable & \texttt{control.c}:1223 and 1382, in \texttt{Control\_readAccessControlObject} and \texttt{Control\_writeAccessControlObject} & 2; \texttt{DEAD\_STORE} & \texttt{cotp.c}:723 adds a packet-size guard before the error exit\\
firejail 2019, vulnerable & \texttt{fs\_lib.c}:265, \texttt{preproc.c}:156, \texttt{sandbox.c}:947 & 3; two \texttt{INTEGER\_\allowbreak OVERFLOW\_L2}, one \texttt{INFERBO\_\allowbreak ALLOC\_\allowbreak MAY\_\allowbreak BE\_BIG} & reorganizes seccomp paths, cleanup, and read-only protection across four files\\
Slurm 2016, vulnerable & \texttt{req.c}, e.g., lines 294, 315, 346, 466 & 298 of 11,875 findings; buffer overrun, integer overflow & \texttt{req.c} adds \texttt{\_open\_as\_other} and replaces a direct open\\
Slurm 2020, fixed & \texttt{x11\_util.c}, ten lines between 180 and 281 & 10 of 9,236 findings; null dereference, buffer overrun, dead store & \texttt{x11\_util.c} creates a protected temporary xauth file\\
\bottomrule
\end{tabular}
\end{table}

libiec61850 shows the mechanism in detail. All three Infer releases report the same two \texttt{DEAD\_STORE} findings in the 2020 vulnerable snapshot, about an unused write to \texttt{nextVarName} in \texttt{src/iec61850/server/mms\_mapping/control.c}. The paired CVE-2020-15158 repair adds the guard \texttt{if (self->packetSize <= bufPos) goto exit\_error} in \texttt{src/mms/iso\_cotp/cotp.c}, and the recorded vulnerable commit is the parent of the fixed commit. The two scores nevertheless refer to different commit comparisons. The vulnerable case's changed-file list describes that snapshot's own parent-to-commit change in \texttt{control.c}, whereas the fixed case's list describes the repair in \texttt{cotp.c}. The case score is two in every release, although the total number of findings in the three reports differs a lot. Under the repair-file restriction, the 2020 endpoint becomes zero for every release, and the ``drop'' vanishes. The case shows that the original contrast depends on the proxy. It does not show a checker regression, and we did not judge whether the two dead stores are real defects in the code.

Slurm shows the opposite situation. The 2016 and 2017 vulnerable endpoints hold the same vulnerable code for CVE-2016-10030, whose repair adds \texttt{\_open\_as\_other} in \texttt{src/slurmd/slurmd/req.c}, opens the file after changing credentials in a child process, and replaces the direct open in \texttt{\_prolog\_error}. Under v1.2, the vulnerable report holds 11,875 findings, 298 of them buffer-overrun and integer-overflow warnings in \texttt{req.c}. The fixed 2020 endpoint (CVE-2020-27746) holds 9,236 findings, ten of them in \texttt{x11\_util.c}, where the repair creates a mode-protected temporary xauth file, unlinks it, and frees the argument array. That 2020 cell has one false positive and one false negative per repetition. At file level the drop persists. At line level the vulnerable score becomes zero and the fixed score two, so both endpoints have F1=0 and the comparison carries no information. The firejail 2019 case (CVE-2019-12589) is similar: its report holds 35 findings, three of which hit the scored files, and none of the three lies on a changed line. The repair removes standalone seccomp paths, introduces a seccomp directory, creates it, and makes it read-only. For acrn-hypervisor, a single vulnerable-side hit carries the drop. In the later acrn-hypervisor (2021) and mosquitto (2024) cells, each label has two cases per repetition.

\head{Semgrep.} Semgrep's queue shows the same problem at the level of benchmark construction. Its 42 cases form 20 vulnerable/fixed groups that share project, year, and identifier, plus two unmatched FreeRDP records. In 19 of the 20 groups, the vulnerable target is the parent of the fixing commit, but only nine groups score the same changed files on both sides; eleven use disjoint files, ten of them with adjacent commits. A vulnerable case can be scored on the files changed \emph{into} that snapshot, while the fixed case is scored on the files changed \emph{by} the fix. In MuJS CVE-2022-30975, for example, the vulnerable commit is the parent of the fixing commit, but the scored vulnerable file \texttt{pp.c} already contains the \texttt{if (f) fclose(f)} guard, because the recorded diff added it. The fix itself adds \texttt{if (prog)} checks before AST-pointer dereferences in \texttt{jsP\_dumpsyntax} and \texttt{jsP\_dumplist} in \texttt{jsdump.c}. The false negative therefore cannot test whether Semgrep would detect the missing guard, and the fixed-side false positive comes from six \texttt{sizeof-pointer-type} findings unrelated to the repair. A different file set does not by itself mean broken commit lineage. It means that a valid label can be paired with a scored file that does not contain the repaired operation.

\subsection{A Finding Can Be Real yet Irrelevant to the Label}\label{subsec:incidental}

\head{Semgrep.} All nine positive Semgrep cases in the selected queue (seven TP and two FP) are label-level: none of them reaches the labelled repair mechanism. Table~\ref{tab:semgrep-queue} summarizes the queue by project. Most early positives come from generic \texttt{memset}, string-copy, or \texttt{strcat} rules, and others from \texttt{atoi}, stat-before-action ordering, or \texttt{sizeof}-pointer rules, in code that the fix does not touch.
\begin{itemize}
\item \emph{Varnish CVE-2017-8807.} Six generic \texttt{memset} findings appear in a transport-identification change. The fix adds \texttt{if (l > ll) l = ll} before a \texttt{memcpy} in \texttt{cache\_fetch.c}, which lies outside the scope of the vulnerable case.
\item \emph{MilkyTracker CVE-2019-14464.} The case scans \texttt{LoaderXM.cpp} and \texttt{XModule.h} after a change that adds instrument and sample-count checks; the two findings are \texttt{memset} calls in the header. The fixing commit adds S3M instrument and pattern-count checks in \texttt{LoaderS3M.cpp}. The commits are adjacent; the file sets do not overlap.
\item \emph{OpenBSD CVE-2019-8460.} Twenty-one \texttt{memset} and \texttt{sizeof}-pointer findings come from a broad file-offset locking change. The fixing record, whose parent is the vulnerable target, adds a limit of 128 SACK holes in two other files. The locking diff must not be read as the CVE repair.
\item \emph{MuJS CVE-2018-6191.} One \texttt{strcat} finding at line 42 of \texttt{jsparse.c} comes from a change to recursion-depth bookkeeping. The fixing record clamps an exponent in \texttt{jsdtoa.c}, a file absent from the vulnerable scope.
\item \emph{FreeRDP CVE-2021-37594.} Two generic copy and memory findings come from a \texttt{timezone.c} change that replaces \texttt{getenv}. The fixed record selected for the same year belongs to a different CVE (CVE-2021-37595, bounds checks in \texttt{wf\_cliprdr.c}), so the two are independent cases rather than a before and after view of one repair.
\item \emph{libreswan CVE-2019-12312.} Nine findings (string and memory functions, \texttt{atoi}, and stat-before-action ordering) come from a hardware-offload change across three files. The fixing record changes \texttt{ikev2\_send.c}, and the two commits are not in a direct parent relation; this is the one group of 20 that fails that check.
\item \emph{libxml2 CVE-2016-3709.} The second false positive is a \texttt{memset} finding on the fixed side, unrelated to the URI and HTML escaping change in \texttt{HTMLtree.c}.
\end{itemize}

\begin{table}[!tb]
\caption{Semgrep qualitative queue (42 cases) by project. V/F: vulnerable/fixed cases; outcomes: TP/FP/FN/TN.}
\label{tab:semgrep-queue}
\centering\footnotesize
\setlength{\tabcolsep}{3.2pt}
\zebra
\begin{tabular}{@{}lllc>{\raggedright\arraybackslash}p{0.6\linewidth}@{}}
\toprule
\textbf{Project} & \textbf{Years} & \textbf{V/F} & \textbf{Outcomes} & \textbf{Principal evidence}\\
\midrule
MuJS & '18, '22 & 3/3 & 1/1/2/2 & Incidental \texttt{strcat}; already-guarded scanned \texttt{pp.c}; separate \texttt{jsdump.c} repair; property-semantics FN\\
FreeRDP & '21, '25 & 2/2 & 1/0/1/2 & Different 2021 identifiers; certificate-length fix with partial parsing\\
libreswan & '19, '20 & 2/2 & 1/0/1/2 & Generic findings; no 2019 parent link; state access near a macro diagnostic\\
MilkyTracker & '19, '20, '22 & 3/3 & 1/0/2/3 & Disjoint scope; lifetime defect outside scope; string-rule timeouts\\
OpenBSD & '19, '24 & 2/2 & 1/0/1/2 & Positives away from the SACK repair; \texttt{mbuf} layout fix with syntax errors\\
Varnish & '17, '19, '20 & 4/4 & 1/0/3/4 & Disjoint repairs; macro API observation; aligned HTTP length invariants\\
libxml2 & '19, '20 & 5/5 & 1/1/4/4 & UAF traces from C assignments; syntax errors; escaping FP in another file\\
\bottomrule
\end{tabular}
\end{table}

The libxml2 CVE-2019-19956 case is the most telling. The vulnerable record scores \texttt{relaxng.c} after a change that adds NULL checks; the selected fixing commit is a memory-leak repair in \texttt{parser.c}. Its 25 findings comprise 17 generic \texttt{memset} warnings and eight \texttt{local-variable-new-delete} findings, the latter reporting a use after a C++ \texttt{delete}. Figure~\ref{fig:code}(c) shows what the traces actually start from: \texttt{xmlRelaxNGCleanupTree} declares \texttt{xmlNodePtr cur, delete;}, and every trace names an assignment \texttt{delete = cur;} (line 7418 for five traces, 7062 for three) as the release, whereas the rule's sources are \texttt{delete \$LOCAL\_VAR} and \texttt{delete[] \$LOCAL\_VAR}. The trace sinks are later accesses to \texttt{cur}. The real cleanup happens elsewhere, through \texttt{xmlUnlinkNode(delete)} and \texttt{xmlFreeNode(delete)}. The use-after-free reading is thus not supported at the nominated locations, which is consistent with C code being matched by a C++ pattern. The case still counts as a true positive, because any qualifying finding suffices. The same file also records two rule timeouts and partial parsing at line 667.

\begin{figure}[!tb]
\centering
\begin{minipage}[t]{0.345\linewidth}\vspace{0pt}
\begin{tcolorbox}[equal height group=pre,colback=vulnback,boxrule=0pt,arc=1pt,left=1mm,right=0.5mm,top=0.5mm,bottom=0.5mm,title={\scriptsize\textbf{(a) MilkyTracker, before fix}},coltitle=black,colbacktitle=vulnback!70!red!20,toptitle=0.2mm,bottomtitle=0.2mm]
\begin{lstlisting}[style=snip]
205 if (mixer)
206   delete mixer;
...
208 if (player)
210   if (mixer->isActive() &&
        !mixer->isDeviceRemoved(player))
211     mixer->removeDevice(player);
212   delete player;
\end{lstlisting}
\end{tcolorbox}\vspace{-1.5mm}
\begin{tcolorbox}[equal height group=post,colback=fixback,boxrule=0pt,arc=1pt,left=1mm,right=0.5mm,top=0.5mm,bottom=0.5mm,title={\scriptsize\textbf{after fix}},coltitle=black,colbacktitle=fixback!70!green!20,toptitle=0.2mm,bottomtitle=0.2mm]
\begin{lstlisting}[style=snip]
206 if (player)
208   if (mixer && mixer->isActive() &&
        !mixer->isDeviceRemoved(player))
209     mixer->removeDevice(player);
210   delete player;
...
213 if (mixer)
214   delete mixer;
\end{lstlisting}
\end{tcolorbox}
\end{minipage}\hfill
\begin{minipage}[t]{0.375\linewidth}\vspace{0pt}
\begin{tcolorbox}[equal height group=pre,colback=vulnback,boxrule=0pt,arc=1pt,left=1mm,right=0.5mm,top=0.5mm,bottom=0.5mm,title={\scriptsize\textbf{(b) Varnish, before fix}},coltitle=black,colbacktitle=vulnback!70!red!20,toptitle=0.2mm,bottomtitle=0.2mm]
\begin{lstlisting}[style=snip]
131 if (!vct_iscrlf(r, htc->rxbuf_e)) {
...
136 q = r;
137 assert(r < htc->rxbuf_e);
138 r = vct_skipcrlf(r, htc->rxbuf_e);
139 if (r >= htc->rxbuf_e)
140   break;
\end{lstlisting}
\end{tcolorbox}\vspace{-1.5mm}
\begin{tcolorbox}[equal height group=post,colback=fixback,boxrule=0pt,arc=1pt,left=1mm,right=0.5mm,top=0.5mm,bottom=0.5mm,title={\scriptsize\textbf{after fix}},coltitle=black,colbacktitle=fixback!70!green!20,toptitle=0.2mm,bottomtitle=0.2mm]
\begin{lstlisting}[style=snip]
131 i = vct_iscrlf(r, htc->rxbuf_e);
132 if (i == 0) {
...
137 q = r;
138 r += i;
139 assert(r <= htc->rxbuf_e);
140 if (r == htc->rxbuf_e)
141   break;
\end{lstlisting}
\end{tcolorbox}
\end{minipage}\hfill
\begin{minipage}[t]{0.255\linewidth}\vspace{0pt}
\begin{tcolorbox}[equal height group=pre,colback=ruleback,boxrule=0pt,arc=1pt,left=1mm,right=0.5mm,top=0.5mm,bottom=0.5mm,title={\scriptsize\textbf{(c) libxml2: trace sources}},coltitle=black,colbacktitle=ruleback!80!blue!15,toptitle=0.2mm,bottomtitle=0.2mm]
\begin{lstlisting}[style=snip]
7037 xmlNodePtr cur, delete;
...
7062 delete = cur; (3 traces)
...
7418 delete = cur; (5 traces)
\end{lstlisting}
\end{tcolorbox}\vspace{-1.5mm}
\begin{tcolorbox}[equal height group=post,colback=ruleback,boxrule=0pt,arc=1pt,left=1mm,right=0.5mm,top=0.5mm,bottom=0.5mm,title={\scriptsize\textbf{rule source patterns}},coltitle=black,colbacktitle=ruleback!80!blue!15,toptitle=0.2mm,bottomtitle=0.2mm]
\begin{lstlisting}[style=snip]
delete $LOCAL_VAR
delete[] $LOCAL_VAR
\end{lstlisting}
\end{tcolorbox}
\end{minipage}
\caption{Source-level evidence from the Semgrep queue (red: pre-fix; green: post-fix; gray-blue: trace or rule). (a) A real use-after-free in \texttt{PlayerGeneric.cpp} that the vulnerable case never scans. (b) An aligned no-finding case: the fix keeps the CRLF length and accepts the exact buffer end. (c) Traces that start from a C assignment, matched by C++ \texttt{delete} patterns. Indentation simplified; ``...'' marks omitted lines.}
\label{fig:code}
\end{figure}

Table~\ref{tab:rule-source} sets the inspected rules against what they actually test. It explains why the positives above say little about the repairs. Most rules check for an API call or a local pattern. The reviewed patches instead change how pointer position, length, allocation, and state relate.

\begin{table}[!tb]
\caption{What the inspected Semgrep rules test, and what that means for the selected cases.}
\label{tab:rule-source}
\centering\footnotesize
\setlength{\tabcolsep}{3.5pt}
\zebra
\begin{tabular}{@{}>{\raggedright\arraybackslash}p{0.25\linewidth}>{\raggedright\arraybackslash}p{0.33\linewidth}>{\raggedright\arraybackslash}p{0.37\linewidth}@{}}
\toprule
\textbf{Rule or rule class} & \textbf{What the rule tests} & \textbf{Consequence for the selected cases}\\
\midrule
\texttt{insecure-use-memset} & Presence of a \texttt{memset} call; no sensitive data required & Explains many incidental TP and FP signals; in \texttt{xmllint.c} the repair itself adds a \texttt{memset}\\
\texttt{insecure-use-strcat-fn}, \texttt{insecure-use-\allowbreak string-copy-fn} & Calls to \texttt{strcat}/\texttt{strncat} or \texttt{strcpy}/\texttt{strncpy} & Do not test exponent bounds, protocol framing, or property-definition semantics\\
\texttt{sizeof-pointer-type} & Selected local \texttt{sizeof}/pointer expressions & MuJS fixed-side findings do not show the missing AST-pointer guards\\
\texttt{null-library-function} & Enumerated library return values used in restricted dereference or call patterns & Not a general checker for every nullable project object, but the configuration does contain null-use checks\\
\texttt{local-variable-\allowbreak new-delete} & A deletion source followed by use of a local pointer & The eight libxml2 traces nominate assignments to a C identifier, not a release\\
String-mismatch rules & Taint between narrow and wide string APIs & Their MilkyTracker timeouts say nothing about the loader bounds repair\\
Protocol and layout invariants in the patches & Relations among pointer, length, allocation, and state & Absent from the inspected generic API patterns; other rules remain untested\\
\bottomrule
\end{tabular}
\end{table}

\head{CodeQL and Infer.} The same gap between finding and label occurs for the other tools. Infer's libiec61850 and firejail positives were discussed above. For CodeQL, changes in the query pack alter which proxy alerts exist. The vulnerable BlueZ CVE-2016-7837 case scores 0, 4, and 4 across releases. Under R2024, three of the four alerts come from \texttt{cpp/missing-check-scanf} at lines 3052, 3058, and 3070 of \texttt{src/device.c}, and one from \texttt{cpp/overrunning-write} at line 403. The scanf query follows an uninitialized \texttt{scanf}-like output to an unguarded use within the same function, and it is absent from the saved R2021 query list. The fix, however, bounds a loop in \texttt{tools/csr.c}. Rule selection thus contributes to the score change, but extraction differences and incomplete invocation records prevent assigning the whole change to it. Unbound CVE-2020-28935 scores 0, 8, and 4. The eight R2024 alerts are four \texttt{cpp/use-after-free} results in the generated \texttt{util/configparser.c} and four \texttt{cpp/path-injection} results in \texttt{util/configlexer.c}, while the fix hardens pidfile handling in \texttt{daemon/unbound.c}. The 2024 Unbound cell also contains different vulnerabilities from the 2020 cell, so the contrast mixes alert availability with defect composition; it is not one CVE becoming harder to detect as time passes.

\finding{Q1}{Across all three tools, the selected label-level positives often come from alerts whose location or semantics do not reach the labelled repair. File-level agreement is especially misleading for Infer, where 610 of 620 selected findings lie in a repair file but only 2 touch a changed line. Temporal F1 can therefore change because incidental alerts come and go, even when coverage of repair mechanisms does not.}

\subsection{No Finding Has Several Possible Explanations}\label{subsec:no-finding}

A zero score can mean that the relevant file was not scanned, that the parser or extractor did not represent a needed construct, that no rule or query encodes the defect, or that the analysis encodes it but reported no path. The selected evidence separates these layers where possible.

\head{Semgrep.} The Semgrep queue has 33 no-finding cases: 14 false negatives and 19 true negatives. Of the 14 false negatives, five have changed-file scopes disjoint from their fixing record (two of these also carry diagnostics), seven have aligned scopes with diagnostics, and two have aligned scopes without diagnostics; nine of the 14 thus carry at least one diagnostic. A true negative here only means that no finding appeared on a fixed case; it does not show full analysis of the file.

\emph{Scope.} MilkyTracker CVE-2020-15569 belongs to the disjoint group. Figure~\ref{fig:code}(a) shows a genuine lifetime defect: the destructor deletes \texttt{mixer} at line 206 and then calls \texttt{mixer->isActive()} and \texttt{mixer->isDeviceRemoved(player)} at line 210; the fix adds a null check and moves the deletion after the last use. The vulnerable case, however, scores \texttt{ChannelMixer.cpp} after its parent's change to \texttt{mixerNumActiveChannels}, with no finding and no diagnostic. The false negative cannot tell whether Semgrep's lifetime rules would catch the destructor. MilkyTracker CVE-2022-34927 is also disjoint: it scans \texttt{SectionSamples.cpp}, where two narrow/wide string-mismatch rules time out, while the repair checks \texttt{instr[y].size} before computing \texttt{instr[y].size - 4} in \texttt{LoaderXM.cpp}. The timeouts show that those rules did not finish on that file, but they cannot explain a miss in another file. Varnish CVE-2019-20637 shows a third variant. The case scores \texttt{include/vdef.h}, whose \texttt{bstrcpy} macro (lines 60--64) contains an assertion followed by \texttt{strcpy(dst, src)}; no finding and no diagnostic appear, although the string-copy rule targets exactly that call. The adjacent fixing commit resets \texttt{req->err\_code} and \texttt{req->err\_reason} in \texttt{cache\_req\_fsm.c}, so the macro is not the repair. Why the copy rule stays silent inside the macro is an open question; macro handling is one hypothesis.

\emph{Aligned, no diagnostics.} The two aligned, diagnostic-free cases are the cleanest candidates for a rule--target mismatch. In Varnish CVE-2019-15892 (Figure~\ref{fig:code}(b)), both snapshots scan \texttt{cache\_http1\_proto.c}. The fix keeps the length returned by \texttt{vct\_iscrlf}, advances the pointer by exactly that length, changes the assertion to allow equality with the buffer end, and stops there. The inspected API-presence patterns do not express this pointer--length invariant. MuJS CVE-2022-44789 replaces \texttt{js\_setproperty} calls with \texttt{js\_defproperty} calls in \texttt{jsobject.c}, for example \texttt{js\_setproperty(J, -2, "value")} with \texttt{js\_defproperty(J, -2, "value", 0)}, a JavaScript property-semantics distinction that no inspected rule models. Both sides of both cases have zero findings. They remain candidate explanations; they do not prove that no rule in the configuration could ever detect these defects.

\emph{Aligned, with diagnostics.} Here the diagnostics weaken the negative result without naming the missed construct.
\begin{itemize}
\item OpenBSD CVE-2025-30334 changes \texttt{mbuf} allocation in \texttt{if\_wg.c} (allocate \texttt{len + max\_hdr}, align the payload, set lengths explicitly); both sides report syntax errors from line 1.
\item libreswan CVE-2020-1763 changes an error-reporting path around line 2132 of \texttt{ikev1.c}. The vulnerable side reports partial parsing at a token-pasting macro near line 276. Only the fixed side records string-mismatch rule timeouts, so the timeouts cannot explain the vulnerable-side miss.
\item FreeRDP CVE-2025-68118 tightens a hostname-length check in \texttt{certificate\_data.c}: the patch passes the buffer capacity minus one, rejects a length that reaches the capacity, and bounds a \texttt{strdup}. Both sides report partial parsing near a formatted string at line 87. The inspected copy patterns do not test such producer--capacity relations, and whether rule coverage or parsing caused the miss remains unresolved.
\item Varnish CVE-2021-36740 parses Content-Length in the HTTP/2 code, counts body bytes, and rejects mismatches; both sides report partial parsing around generated headers and enumerations.
\item libxml2 CVE-2020-24977 zeroes four bytes at the end of the input buffer in \texttt{xmllint.c} before encoding entities. The repair itself adds a \texttt{memset} call, which the generic \texttt{memset} rule should match, yet no finding is reported; the file carries file-spanning syntax errors. The miss can therefore not be blamed on the absence of an unsafe-API pattern.
\item libxml2 PRIMEVUL-2020-0e1a49c89076 and CVE-2020-7595 select the same target and fixing commit and the same end-of-input change in \texttt{parser.c}, under two identifiers. Their four records are one observation, not two confirmations.
\end{itemize}

\head{CodeQL.} Nagios 2018 contains a conditional NULL path. The query handler \texttt{qh\_input} initializes \texttt{query} to NULL and assigns it only if the request contains a space, so requests such as \texttt{\#core}, \texttt{\#help}, or \texttt{\#echo} pass NULL to handlers whose old versions dereference it. The fix adds NULL guards at the three handler entries. This is a static path in the source; we did not reproduce an exploit. Neither saved suite selects \texttt{MissingNullTest} or \texttt{InconsistentNullnessTesting}, and the selected \texttt{cpp/redundant-null-check-simple} query requires a dereference \emph{followed} by a NULL test, a different pattern. To test whether adding a nullness query would help, we copied the exact old and new bodies of \texttt{qh\_echo}, \texttt{qh\_help}, and \texttt{qh\_core} into an isolated R2024 fixture with a synthetic dispatcher that passes NULL and with positive controls. All four input files matched the archived source bytes, and extraction succeeded for all three translation units. \texttt{MissingNullTest} reported its local-null control and \texttt{cpp/redundant-null-check-simple} its dereference-before-check control, but neither reported any of the six original bodies. The positive controls of \texttt{InconsistentNullnessTesting} did not report, so we draw no conclusion about that query. A diagnostic query returned 31 nullness/use tuples in the six bodies, including 18 dereferences of \texttt{buf} across all of them, and none satisfied the \texttt{maybeNull} predicate. The library assumes pointers valid by default and follows local control flow, so the caller's NULL argument does not reach the callback parameters. Adding the query alone would not catch this pattern.

Extraction may also play a part. For the 2018 handler file \texttt{base/query-handler.c}, the R2024 results record 38 extraction warnings, including a missing \texttt{include/config.h}. One falls inside \texttt{qh\_echo} (lines 25--42), five inside \texttt{qh\_help} (370--402), none inside \texttt{qh\_core} (404--470), and 32 elsewhere. The diagnostics also report an incomplete \texttt{struct query\_handler} at the dispatch line (227) and at the callback storage (328), so extraction problems touch the NULL path itself, although we cannot show that IR edges were lost. The recovered R2026 run reports 101 partial and 36 successful extraction commands; its score stays zero, and its 79 finding signatures match the frozen record exactly.

For BlueZ CVE-2020-0556, the fix adds a bonding check in \texttt{hog\_accept}: the fixed code calls \texttt{device\_is\_bonded} and returns \texttt{-ECONNREFUSED} for an unbonded device before \texttt{bt\_hog\_attach}. Both roles have zero changed-file hits in every release, despite unrelated findings elsewhere. We inspected two authentication queries. The first, \nolinkurl{cpp/user-controlled-bypass}, requires tainted data to reach a condition that contains a hard-coded address; the second, \nolinkurl{cpp/tainted-permissions-check}, requires it to reach a modeled privilege operation. A missing bonding policy provides neither. Under R2024 the scored file \texttt{hog.c} also carries eleven extraction problem notifications per role.

BlueZ CVE-2020-27153 shows a scoring-file mapping limit. The vulnerable case is scored on \texttt{mesh/cfgmod-server.c}, while the verified repair changes \texttt{src/shared/att.c}. No release reports an alert on \texttt{att.c} for either role, so the mapping did not hide an existing alert. The upstream fix addresses an invalid read in \texttt{enable\_ccc\_callback} after the client has been torn down during an earlier disconnect callback: \texttt{bt\_gatt\_client\_free} frees a characteristic through \texttt{notify\_chrc\_free}, and the callback later dereferences it. The saved R2021 query list excludes the use-after-free and double-free queries, and R2024 includes both, yet R2024 reports nothing on \texttt{att.c}. Whether the free reaches the later use through the queue callbacks and client-owned fields is exactly what the query would have to model, and we did not isolate the failing predicate.

Across the paired CodeQL repairs, extraction warnings on the repair file occur in 8 of 26 R2021 and 20 of 26 R2024 records, and all 27 selected R2026 result files report partial extraction. Either the presence or the absence of a warning on a file is weak evidence, because a file can be partly extracted without a warning on it. Table~\ref{tab:codeql-open} summarizes what the evidence supports for each CodeQL no-signal case and what would settle the remaining question.

\begin{table}[!tb]
\caption{CodeQL cases without a scoring signal: what the evidence supports and what remains open.}
\label{tab:codeql-open}
\centering\footnotesize
\setlength{\tabcolsep}{3.5pt}
\zebra
\begin{tabular}{@{}>{\raggedright\arraybackslash}p{0.18\linewidth}>{\raggedright\arraybackslash}p{0.4\linewidth}>{\raggedright\arraybackslash}p{0.38\linewidth}@{}}
\toprule
\textbf{Case} & \textbf{Supported explanation} & \textbf{Open question; deciding check}\\
\midrule
Nagios 2018 NULL repair & Positive controls pass; the repaired parameter accesses fail \texttt{maybeNull}; the target file has missing-header diagnostics & Whether any nullness analysis models the caller-to-handler NULL path; query inclusion alone is insufficient in the fixture\\
BlueZ 2020 callback repair & Client-data lifetime path in the source; R2021 excludes the memory queries & Why R2024's use-after-free query stays silent; trace \texttt{free(chrc)} to the later use\\
BlueZ 2020 bonding repair & The repair adds an authorization policy that the two inspected queries do not encode & Coverage by the full suite; compare an explicit bonding query on the same database\\
HAProxy 2023 under R2026 & Recovered results are identity-validated & Source repair versus query or IR contribution is not isolated\\
BlueZ 2016, Unbound 2020 & Score and query-list differences are observed & Query selection versus extraction or semantic change; needs compatible databases\\
\bottomrule
\end{tabular}
\end{table}

\head{Infer.} Infer's no-finding endpoints include valid empty reports. For firejail and Slurm, the v1.1 and v1.3 endpoints are \texttt{empty\_capture} observations with score zero (Table~\ref{tab:infer-endpoints}), so their ``0$\rightarrow$0'' rows in Table~\ref{tab:drops} reflect missing capture rather than evaluated code; only v1.2 captured those snapshots. Keeping empty and partial captures in the evaluation avoids a stricter completion rule for newer releases, but it lets capture coverage enter measured effectiveness.

\section{RQ5: Operational Sustainability}\label{sec:operations}

RQ5 looks at the runs themselves: whether they finish, what a ``success'' contains, and what they cost. The three tools record different things, so we report each one in its own terms before comparing.

\subsection{CodeQL: Status Moves in Both Directions}\label{subsec:reliability}

Table~\ref{tab:codeql-rel}(a) reports final outcomes on deduplicated targets, and Table~\ref{tab:codeql-rel}(b) follows the same target identities across releases. The two historical releases end with 676 final failures (394 timeouts and 282 database failures), which cover 567 distinct targets, because 109 targets fail under both. These are target counts. Several cases can share a target, so the 290 R2024 timeout targets correspond to the 332 timeout cases of Table~\ref{tab:populations}. Of 19,780 exact historical targets, 281 succeed under R2021 but fail under R2024, while 177 move the other way. A newer release is therefore not uniformly more executable. The 281 regressions consist of 277 timeouts and four database failures. The four database failures are the vulnerable and fixed targets of KeepKey CVE-2022-30330 and pdns CVE-2018-1046; their R2024 database step ran for only 1.8--2.4 seconds, and the same targets succeed under R2021 and R2026. No raw attempt or error output survives for them, so their cause stays unknown. R2026 reaches final success on all 19,820 targets, but its recovered July checkpoint contains 295 failures, so ``zero final failures'' describes delivered coverage, not first-pass robustness.

\begin{table}[!tb]
\caption{CodeQL reliability. (a) Final outcomes on deduplicated targets. (b) Status transitions of identical targets (target, commit, role, project, year) between releases. S: success; F: failure.}
\label{tab:codeql-rel}
\centering\footnotesize
\begin{minipage}[t]{0.44\linewidth}\centering
\setlength{\tabcolsep}{2.2pt}
\zebra
\begin{tabular}{@{}lrrrrr@{}}
\multicolumn{6}{l}{\textbf{(a) Final outcomes}}\\
\toprule
& \textbf{Targets} & \textbf{Success} & \textbf{Timeout} & \textbf{DB fail} & \textbf{Rate}\\
\midrule
R2021 & 19,780 & 19,494 & 104 & 182 & 1.446\%\\
R2024 & 19,780 & 19,390 & 290 & 100 & 1.972\%\\
R2026 & 19,820 & 19,820 & 0 & 0 & 0.000\%\\
\bottomrule
\end{tabular}
\end{minipage}\hfill
\begin{minipage}[t]{0.54\linewidth}\centering
\setlength{\tabcolsep}{3pt}
\zebra
\begin{tabular}{@{}lrrrr@{}}
\multicolumn{5}{l}{\textbf{(b) Same-target transitions}}\\
\toprule
\textbf{Earlier $\rightarrow$ later} & \textbf{S$\rightarrow$F} & \textbf{F$\rightarrow$S} & \textbf{F$\rightarrow$F} & \textbf{S$\rightarrow$S}\\
\midrule
R2021 $\rightarrow$ R2024 final & \nc{281} & 177 & 109 & 19,213\\
R2021 $\rightarrow$ R2026 checkpoint & \nc{278} & 269 & 17 & 19,216\\
R2024 $\rightarrow$ R2026 checkpoint & \nc{277} & 372 & 18 & 19,113\\
R2021 $\rightarrow$ R2026 final & 0 & 286 & 0 & 19,494\\
R2024 $\rightarrow$ R2026 final & 0 & 390 & 0 & 19,390\\
\bottomrule
\end{tabular}
\end{minipage}
\end{table}

A second view compares different snapshots of the same project under one release. We found 161 pairs in which an older snapshot succeeds and a newer one fails, spread over 41 release-by-project combinations. In GPAC under R2021, for example, a fixed 2018 target succeeds and a distinct fixed 2019 target ends in a database failure; in QEMU, a fixed 2016 target succeeds and a fixed 2020 target fails during database creation after 4.6 seconds, before analysis starts. Neither record keeps a command, an error message, or resource data, so these pairs show where the failure happened but cannot say why. They concern different targets and must not be added to the exact-target counts.

\head{Build and discovery mechanisms.} The historical archive keeps mostly coarse final statuses. It lacks the selected attempt, the full command, error output, exit code, disk, and machine load for each final record. A recovered raw ledger of 6,138 attempts contains 56 rows with direct diagnostics, and they point to two mechanisms.

The first concerns traced builds. Six R2021 attempts end with return code 32 and the message that no source code was seen and extracted. Their build-tracer logs list only shell utilities (\texttt{sh}, \texttt{mkdir}, \texttt{find}, \texttt{grep}) and no compiler call, and their object directories are empty. The saved build wrapper runs its compiler fallback only when a recursive search finds no object file or compile-command file. Both affected projects track prebuilt objects in their repositories: the two Sandboxie commits contain 20 tracked \texttt{.obj} files and the two js-compute-runtime commits contain three tracked \texttt{.o} files, and neither has a build file that the wrapper's main branches recognize. In a local control with one compilable C file, adding one tracked object file from each project and then removing it changed the number of compiler invocations from 1 to 0 to 1, while the wrapper exited successfully every time. The wrapper thus mistakes the presence of a build artifact for evidence of fresh compilation.

The second concerns buildless extraction. Fifty R2024 attempts report that no suitable C++ compiler was found, although the targets contain C++ files (322 in each of four audited targets). CLI 2.20.0 looks only for \texttt{cl}, \texttt{clang}, and \texttt{clang++}, and never for \texttt{gcc} or \texttt{g++}. In a controlled experiment on one synthetic C file, a GCC-only environment failed (exit 2), adding Clang succeeded and extracted the file, and removing Clang failed again. The R2026 extractor, by contrast, finds \texttt{gcc} and \texttt{g++} after looking for Clang. These local experiments establish the mechanisms; they do not recover the historical environment of every failure. In particular, the 50 attempts were not final failures: the same targets end in 46 final successes and 4 timeouts, so they may be earlier attempts that were later replaced.

\head{Linking raw attempts to final records.} All 56 raw diagnostics can be tied to a target, but not to the attempt that produced the final record. The six no-source rows agree with final database failures; of the 50 compiler-not-found rows, 4 agree with final timeouts and 46 with final successes. Across three attempt sources, 34 attempts overlap only 30 of the 676 final-failure keys, and 24 of those attempts succeeded although the final index records a failure. When successful and failed attempts coexist on the same inputs, an identity match cannot tell which execution produced the final row. We therefore treat the mechanisms in Table~\ref{tab:codeql-evidence} as supported only at the level stated there.

\begin{table}[!tb]
\caption{Candidate causes of CodeQL execution failures and the strength of the evidence for each.}
\label{tab:codeql-evidence}
\centering\footnotesize
\setlength{\tabcolsep}{3pt}
\zebra
\begin{tabular}{@{}>{\raggedright\arraybackslash}p{0.16\linewidth}>{\raggedright\arraybackslash}p{0.36\linewidth}>{\raggedright\arraybackslash}p{0.22\linewidth}>{\raggedright\arraybackslash}p{0.2\linewidth}@{}}
\toprule
\textbf{Candidate cause} & \textbf{Observed fact} & \textbf{Missing evidence} & \textbf{Status}\\
\midrule
Build system and extraction & No-source errors; tracer logs show no compiler call; empty object directories & Archived workspace, build output, clean-build control & Raw mechanism supported; final attribution open\\
Compiler discovery & 50 compiler-not-found rows in buildless mode on targets with C++ files & Selected attempt; historical PATH and compiler versions & Raw mechanism supported; final cause unconfirmed\\
Dependencies and mirrors & No direct dependency or mirror failure in the audited records & Dependency, submodule, and checkout fields & Hypothesis only\\
Extractor or query support & No final diagnostic names this class & Error output; same-target control & Not supported as a final cause\\
Timeout, memory, disk, load & 394 timeouts clustered at two phase limits & Final RSS, disk, load, process tree & Timeout observed; cause open\\
Runner path, cache, temp & Raw commands show autobuild and a PATH error & Final environment and cache state & Raw level only\\
\bottomrule
\end{tabular}
\end{table}

\head{Timeouts and query cost.} The 394 historical timeouts split into 342 analysis-stage cutoffs near 1,800 seconds and 52 phase cutoffs near 7,200 seconds; 47 successful analyses ran for at least 1,700 seconds, so the cap also censors slow successes. The cutoffs tell us at which stage a run stopped but not why it was slow, and without memory, disk, and load records we cannot upgrade a timeout to an out-of-memory or contention claim. BlueZ CVE-2020-0556 shows what can lie behind a timeout. Analysis took 83.4 and 88.3 seconds (vulnerable and fixed) under R2021 and 176.2 and 185.2 seconds under R2024. Under R2026 both targets timed out in the checkpoint run, and a later retry on the existing databases timed out again after 43,200 seconds, with the watchdog reporting no stall. The evaluator logs show that 96 of the 98 queries in the suite already had cached results. Only \texttt{cpp/overrun-write} and \texttt{cpp/invalid-pointer-deref} were still evaluating the recursive modulus analysis of the range-analysis library, which tracks facts of the form ``expression equals a base plus an offset modulo a modulus''. Near the end, the fixed run was at iteration 18,850 and the vulnerable run at 18,972, still producing new facts. Removing one of the two queries left the other in the same recursion; removing both produced valid results from the same databases. The selected recovery used fresh buildless databases and a reduced query set. It finished in 61.97 and 61.27 seconds, with 152 successful and 312 partial extraction commands out of 464 in each role. Faster completion here came with a smaller query set and partial extraction, so the recovered result is not equivalent coverage.

\head{Recovery hides fragility.} Table~\ref{tab:codeql-recovery} breaks down the 295 R2026 checkpoint failures and the recovery runs that later turned them into successes on the same commit and role. All 219 database exit-2 failures used changed-file-scoped buildless extraction; a local preflight routed 214 of them to a header-driver recovery and five to a standalone-source recovery. The 15 unavailable inputs were commits missing from the checkpoint mirror, all of which were present in the later mirror. Twelve recovery runs failed on the way (11 analysis timeouts and one database failure) before a later run succeeded. Success can also coexist with incomplete extraction. In 45 local validation attempts, partial-extraction diagnostics appeared in 8 of 23 R2021 and 22 of 22 R2024 runs, and one R2024 htmldoc log shows 81 successful and 27 failed compilation commands out of 108.

\begin{table}[!tb]
\caption{The 295 R2026 checkpoint failures: what failed, and which recovery run delivered the final success.}
\label{tab:codeql-recovery}
\centering\footnotesize
\setlength{\tabcolsep}{4pt}
\begin{minipage}[t]{0.48\linewidth}\centering
\zebra
\begin{tabular}{@{}lr@{}}
\multicolumn{2}{l}{\textbf{(a) Checkpoint failure}}\\
\toprule
\textbf{Failure type} & \textbf{Count}\\
\midrule
Database exit 2 (buildless, changed-file scope) & 219\\
Analysis timeout & 48\\
Database input unavailable & 15\\
Analysis exit 2 & 11\\
Analysis killed (memory not confirmed) & 1\\
Database timeout & 1\\
\bottomrule
\end{tabular}
\end{minipage}\hfill
\begin{minipage}[t]{0.48\linewidth}\centering
\zebra
\begin{tabular}{@{}lr@{}}
\multicolumn{2}{l}{\textbf{(b) Recovery that delivered success}}\\
\toprule
\textbf{Recovery run} & \textbf{Count}\\
\midrule
Header-driver recovery & 217\\
First standalone-source tranche & 43\\
Continuation of that tranche & 23\\
BlueZ reduced-suite retry & 11\\
Scoped retry (one target) & 1\\
\midrule
Fresh database and analysis / cached reuse & 285 / 10\\
\bottomrule
\end{tabular}
\end{minipage}
\end{table}

\head{Membership history.} The target set changed once during the study: 40 target identities were removed and 40 added. Sixteen results supplied for removed identities were quarantined (11 had succeeded and 5 had failed), and 89 Chromium targets missing from one delivery remained in the dataset; together with the 40 new identities they accounted for 129 results that had to be produced later. No commit or identity mismatch remains among the current records. Because 11 of the retired results had succeeded, the removal cannot be read as CodeQL failing on those targets.

\subsection{Infer: Build Availability and Empty Success}\label{subsec:infer-reliability}

Infer v1.1 ends with 232 build failures and 8 analysis timeouts; v1.3 ends with 108 build failures and 11 capture timeouts. Table~\ref{tab:infer-fail} gives the units. The 359 release-by-case failure records cover 251 unique cases, 108 of which fail under both releases, and they are documented by 1,318 attempt records. The dominant diagnosis is unmaterialized pinned submodules (216 v1.1 and 92 v1.3 rows). It is a property of the project, its dependencies, and the environment rather than of the checkers. The remaining build failures are 15 rows per release whose capture database was still absent after continuation and one row per release with a missing dependency or build strategy. These labels name the selected diagnosis; they do not prove a unique root cause. Figure~\ref{fig:exclusions}(a) shows how the failures spread over the years.

\begin{table}[!tb]
\caption{Infer final failures and their evidence. Each failure has one direct failure packet; attempt-history records include earlier attempts on the same cases.}
\label{tab:infer-fail}
\centering\footnotesize
\setlength{\tabcolsep}{5pt}
\zebra
\begin{tabular}{@{}lrrrrr@{}}
\toprule
\textbf{Release} & \textbf{Failure records} & \textbf{Build failed} & \textbf{Timeout (stage)} & \textbf{Attempt records} & \textbf{Unique cases}\\
\midrule
v1.1 & 240 & 232 & 8 (analyze) & 962 & 240\\
v1.3 & 119 & 108 & 11 (capture) & 356 & 119\\
Union & 359 & 340 & 19 & 1,318 & 251\\
\bottomrule
\end{tabular}
\end{table}

\begin{table}[!tb]
\caption{Infer original runner status vs.\ final selected status (case-by-release rows). v1.2 is omitted because it uses an older ledger.}
\label{tab:infer-status}
\centering\footnotesize
\setlength{\tabcolsep}{8pt}
\zebra
\begin{tabular}{@{}llrr@{}}
\toprule
\textbf{Original status} & \textbf{Final} & \textbf{v1.1} & \textbf{v1.3}\\
\midrule
build\_failed & build\_failed & 217 & 93\\
build\_failed & success & 1,389 & 1,513\\
empty\_capture & build\_failed & 15 & 15\\
empty\_capture & success & 5,347 & 4,120\\
success & success & 125 & 786\\
partial capture & success & 12,887 & 13,444\\
timeout & success & 12 & 18\\
timeout & timeout & 8 & 11\\
\bottomrule
\end{tabular}
\end{table}

Table~\ref{tab:infer-status} shows that a final ``success'' summarizes very different histories. Only 125 v1.1 and 786 v1.3 cases succeeded in their original run without a partial-capture flag. Most final successes come from partial capture (12,887 and 13,444) or from recovery after a build failure (1,389 and 1,513) or an empty capture (5,347 and 4,120). In 5,716 v1.1 and 4,460 v1.3 cases, the selected result comes from an analysis that was resumed from preserved capture data. A recovered empty capture yields a legal report, possibly empty, and therefore a score, but it does not show that the source tree was analyzed. Infer thus has three different notions of coverage: a report that can be scored, a capture that recorded some source files, and a capture that recorded the case's changed files. None of them shows that the vulnerable code was analyzed.

Two same-project contrasts illustrate the point. They are drawn from 119 older-success/later-failure contrasts across 18 projects, which are diagnostic observations rather than controlled pairs. Under v1.1, a QEMU 2021 vulnerable case ends as a success with score zero, but its original attempt recorded zero source files and zero procedures, and its log says there was nothing to compile or analyze. The 2022 case fails at setup after about 1.3 seconds: its record lists 23 pinned submodule links, and the recovery history notes missing local objects and disabled remote URLs. Under v1.3, a Linux 2019 fixed case completes with partial capture (16 source files and 253 procedures), but capture of the changed \texttt{floppy.c} was not established; its setup logs also report missing libelf and OpenSSL headers. The 2020 case hits the 7,200-second capture deadline after 7,200.2 seconds and is killed (exit code $-15$). That is a deadline; it does not show an infinite loop or a checker defect. Both pairs concern whether Infer can observe the code, not whether its checkers aged. The v1.2 archive reports no final failures, but its 49 legacy records (38 timeouts, 10 workspace failures, one build failure) follow a different ledger schema, and only 12 map to current case identities, so we do not rank v1.2 reliability against the other releases.

\subsection{Semgrep: Success with Degraded Coverage}\label{subsec:semgrep-reliability}

\begin{figure}[!tb]
\centering
\includegraphics[width=0.52\linewidth]{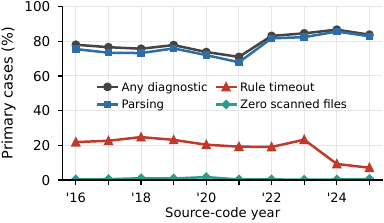}
\caption{Share of primary Semgrep cases with each diagnostic type per source-code year (x-axis; types overlap). Diagnostics are present in most cases in every year, under a successful run status.}
\label{fig:semgrep-diag}
\end{figure}

All 20,000 Semgrep runs report success, including the 113 catalog cases (109 primary) that scanned zero files. A status-based measure would show perfect executability, and there is no old-success/new-failure transition to analyze. A successful status, however, only means that the command returned a result record. It does not show complete parsing, complete rule evaluation, or complete coverage of the target. Table~\ref{tab:semgrep-diag} gives the diagnostics. Of the 19,744 primary cases, 15,590 (78.96\%) carry at least one diagnostic: 13,150 record partial parsing, 2,603 syntax errors, 647 other syntax errors, and 3,738 rule timeouts. A further 109 cases scan zero files, and 4,045 have neither a diagnostic nor an empty scope. Figure~\ref{fig:semgrep-diag} shows that diagnostics are common in every year: the share of cases with any diagnostic ranges from about 71\% (2021) to more than 85\% (2024), almost entirely driven by parsing problems, while the share with rule timeouts stays near 20\% until 2023 and drops below 10\% in 2024 and 2025.

\begin{table}[!tb]
\caption{Semgrep diagnostics. Left: prevalence in the catalog and the primary population (categories overlap). Right: mutually exclusive strata of the primary population; zero-target cases are assigned first, and ``parsing'' covers partial parsing and both kinds of syntax error. The strata sum to the Study-A confusion counts.}
\label{tab:semgrep-diag}
\centering\footnotesize
\begin{minipage}[t]{0.32\linewidth}\centering\scriptsize
\setlength{\tabcolsep}{2.4pt}
\zebra
\begin{tabular}{@{}lrr@{}}
\toprule
\textbf{Diagnostic} & \textbf{Catalog} & \textbf{Primary}\\
\midrule
Any diagnostic & 15,765 & 15,590\\
Partial parsing & 13,274 & 13,150\\
Syntax error & 2,641 & 2,603\\
Other syntax error & 655 & 647\\
Rule timeout & 3,792 & 3,738\\
Zero targets & 113 & 109\\
None, $>$0 targets & 4,122 & 4,045\\
\bottomrule
\end{tabular}
\end{minipage}\hfill
\begin{minipage}[t]{0.66\linewidth}\centering\scriptsize
\setlength{\tabcolsep}{2.2pt}
\zebra
\begin{tabular}{@{}lrrrrrrrrrr@{}}
\toprule
\textbf{Stratum} & \textbf{n} & \textbf{Share} & \textbf{TP} & \textbf{FP} & \textbf{FN} & \textbf{TN} & \textbf{Prec.} & \textbf{F1} & \textbf{FPR} & \textbf{AUC}\\
\midrule
No diagnostics & 4,045 & 20.5\% & 640 & 658 & 1,371 & 1,376 & .493 & .387 & .324 & .498\\
Parsing only & 11,852 & 60.0\% & 2,527 & 2,644 & 3,376 & 3,305 & .489 & .456 & .444 & .493\\
Timeout only & 366 & 1.9\% & 102 & 96 & 86 & 82 & .515 & .529 & .539 & .506\\
Parsing + timeout & 3,372 & 17.1\% & 1,320 & 1,335 & 371 & 346 & .497 & .608 & .794 & .504\\
Zero targets & 109 & 0.6\% & 0 & 0 & 70 & 39 & 0 & 0 & 0 & .500\\
\bottomrule
\end{tabular}
\end{minipage}
\end{table}

The strata resist a simple correction. Cases with both parsing and timeout diagnostics have the highest F1 (0.6075) and the highest FPR (0.7942), while cases without diagnostics have F1 0.3868; AUROC is close to 0.5 in all strata. Precision tells the same story from another angle: it lies between 0.489 and 0.515 in every non-empty stratum. Diagnostics change how much a case is flagged, while the share of flags on vulnerable cases stays the same. Diagnosed cases differ in files, projects, and finding density, so the higher F1 of some strata is an association with a different case mix, not evidence that diagnostics improve detection. Excluding them would keep only one fifth of the corpus, a different population, and would not reveal a clean estimate. The AUROC of 0.5 for the zero-target stratum only reflects identical zero scores on both labels.

The recorded diagnostics name specific constructs: token-pasting macros in libreswan, formatted strings in FreeRDP, generated headers and enumerations in Varnish, file-spanning syntax errors in OpenBSD and libxml2, and rule timeouts in MilkyTracker. Within a project, scan quality can change between years. MilkyTracker's two 2019 and two 2020 cases have no diagnostic, while one of its two 2022 cases reports two string-rule timeouts; Varnish's four 2019 cases have no diagnostic, while both 2020 cases report partial parsing around generated HTTP/2 constructs. Early FreeRDP, libreswan, OpenBSD, and libxml2 cases already carry diagnostics, though, so their later no-finding results are not a uniform move from complete analysis to failure. Naming the exact construct or rule that was lost would need diagnostic detail that the recorded scans do not keep.

\subsection{Cost on Successful Runs}\label{subsec:efficiency}

Table~\ref{tab:cost} collects the available cost measurements. The tools record different fields under different policies, so the table is an envelope per condition, not a ranking.

\begin{table}[!tb]
\caption{Recorded cost of successful runs (seconds unless noted). ``Tail'' is the 95th percentile, except for Infer (90th). Cached CodeQL phases and missing fields are excluded, not counted as zero. Infer rows use all selected attempts after the identity gate, including resumed analyses and label-conflict cases, since cost does not depend on the label; Semgrep rows use its primary population.}
\label{tab:cost}
\centering\scriptsize
\setlength{\tabcolsep}{2.6pt}
\zebra
\begin{tabular}{@{}llrrrrl@{}}
\toprule
\textbf{Condition} & \textbf{Phase / measure} & \textbf{n} & \textbf{Median} & \textbf{Tail} & \textbf{Max} & \textbf{Note}\\
\midrule
CodeQL R2021 & database / analysis wall & 19,213 & 4.6 / 17.3 & 172.64 / 540.58 & --- & matched targets\\
CodeQL R2024 & database / analysis wall & 19,213 & 5.3 / 22.3 & 178.68 / 277.00 & --- & ratio 1.227 / 1.173\\
CodeQL R2026 & database / analysis wall & 11,789 / 11,826 & 7.80 / 73.26 & 114.16 / 3,359.36 & --- & invoked phases only\\
CodeQL R2026 & database / analysis CPU & 7,858 / 7,866 & 16.07 / 150.69 & 580.29 / 1,310.78 & --- & 66.7\% / 66.5\% of invoked\\
CodeQL R2026 & database / analysis RSS (MiB) & 7,858 / 7,866 & 644 / 3,869 & 3,966 / 11,835 & --- / 38,580 & per phase\\
Infer v1.1 & capture / analyze wall & 19,713 & 8.85 / 0.36 & 65.20 / 1.97 & 11,334 / 21,362 & selected attempts\\
Infer v1.3 & capture / analyze wall & 19,833 & 19.67 / 0.52 & 109.47 / 3.88 & 7,359 / 33,466 & selected attempts\\
Semgrep & scan wall & 19,314 & 62.46 & 394.83 & 3,504.56 & 430 missing (cached)\\
Semgrep & scan CPU (user + system) & 19,314 & 5.10 & 168.53 & 9,298.75 & per scan\\
Semgrep & peak RSS (MiB) & 19,314 & 183.30 & 230.30 & 820.90 & per scan\\
\bottomrule
\end{tabular}
\end{table}

\head{CodeQL.} On the 19,213 targets that succeeded under both historical releases, R2024 analysis is slower than R2021 on 12,582 targets (65.5\%), faster on 6,531, and equal on 100; the median per-target ratio is 1.173. Yet the R2024 95th percentile is lower (277.00 vs.\ 540.58 seconds). Typical and tail behavior thus move in opposite directions. Each of these targets ran on the same machine in both releases (13,637 on one host and 5,576 on another), and the ratio is computed per target before taking the median. The matched set excludes the 458 targets that changed status, so it is biased toward targets that both releases could finish. R2026 has a median analysis time of 73.26 seconds but a 95th percentile of 3,359.36 seconds and a single-phase RSS maximum of 38,580 MiB, which is the peak of one phase rather than of the whole process tree. Of its phase slots, 8,031 database and 7,994 analysis slots were served from cache (about 40\%), and their current cost is unknown. CPU and memory were recorded for only two thirds of the invoked phases. We do not rank R2026 against the historical timings, because its recovery policies, query coverage, and cache use differ.

\head{Infer.} Infer's medians are small, but recovery budgets produce long maxima. Table~\ref{tab:infer-timing} separates all selected attempts from the original attempts that needed no resumed analysis; the two populations are not paired, so their difference is not a speedup estimate. The two longest analyses ran under the 43,200-second recovery budget: an ACPICA 2017 fixed case took 21,362 seconds under v1.1 and a jsish 2020 fixed case 33,466 seconds under v1.3, both with partial capture and a clean exit. The longest captures belong to one PyTorch 2024 fixed case under a 21,600-second recovery budget. Under v1.1, 587 seconds of CMake build and 10,747 seconds of Infer capture produced an empty capture that a resumed analysis later turned into a success with score 5; under v1.3, 586 seconds of CMake and 6,773 seconds of capture covered the changed path directly. Rare long analyses are the largest maxima, so the data do not support a blanket claim that capture dominates the tail. v1.2 lacks comparable timing, and no Infer release records CPU time or peak memory in its final ledger.

\begin{table}[!tb]
\caption{Infer wall time of selected attempts (seconds). ``All selected'' uses every selected attempt, with resumed analyses timed from the continuation; ``original only'' excludes resumed rows.}
\label{tab:infer-timing}
\centering\footnotesize
\setlength{\tabcolsep}{6pt}
\zebra
\begin{tabular}{@{}llrrrr@{}}
\toprule
\textbf{Release, population} & \textbf{Phase} & \textbf{Attempts} & \textbf{Median} & \textbf{90th pct.} & \textbf{Maximum}\\
\midrule
v1.1, all selected & capture & 19,713 & 8.85 & 65.20 & 11,334.17\\
v1.1, original only & capture & 14,009 & 22.32 & 66.24 & 2,918.44\\
v1.1, all selected & analyze & 19,713 & 0.36 & 1.97 & 21,361.92\\
v1.1, original only & analyze & 14,009 & 0.27 & 0.88 & 21,361.92\\
v1.3, all selected & capture & 19,833 & 19.67 & 109.47 & 7,359.40\\
v1.3, original only & capture & 15,383 & 41.44 & 121.48 & 7,359.40\\
v1.3, all selected & analyze & 19,833 & 0.52 & 3.88 & 33,466.22\\
v1.3, original only & analyze & 15,383 & 0.35 & 2.98 & 33,466.22\\
\bottomrule
\end{tabular}
\end{table}

\begin{figure}[!tb]
\centering
\includegraphics[width=0.66\linewidth]{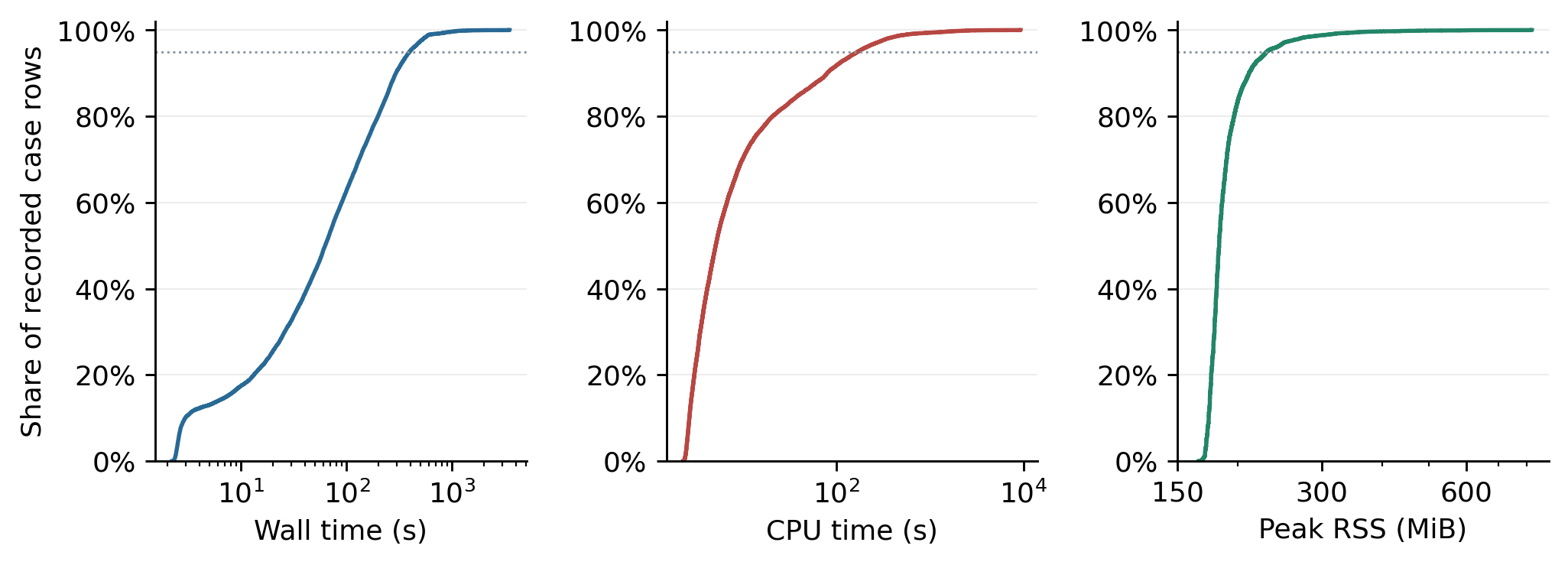}
\caption{Empirical cumulative distributions of Semgrep wall time, CPU time, and peak RSS (19,314 primary cases; logarithmic x-axes for time). The dotted line marks the 95th percentile.}
\label{fig:semgrep-ecdf}
\end{figure}

\begin{figure}[!tb]
\centering
\includegraphics[width=\linewidth]{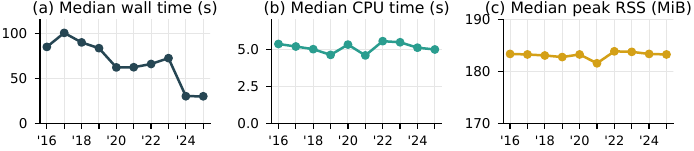}
\caption{Yearly medians of Semgrep wall time, CPU time, and peak RSS by source-code year (x-axis). Cases per year with resource fields range from 1,794 (2020) to 2,000.}
\label{fig:sg-resources}
\end{figure}

\head{Semgrep.} Semgrep's resource fields cover 19,314 primary cases (97.82\%). The 430 missing cases all fall in 2016--2020, and every one of them carries a recorded cache-return message. Wall time is measured around the scan command; CPU time (user plus system) and peak RSS come from GNU \texttt{time} and describe a single scan without summing parallel scans. Figure~\ref{fig:semgrep-ecdf} shows heavy right tails for wall and CPU time and a narrow memory band. The 99th percentiles are 625.99 seconds of wall time, 572.04 CPU seconds, and 310.80 MiB. The maximum CPU time (9,298.75 seconds) exceeds the maximum wall time (3,504.56 seconds). Figure~\ref{fig:sg-resources} shows the yearly medians. Median wall time peaks at 100.3 seconds for the 2017 code and falls to about 30 seconds for 2024 and 2025, while median CPU time (4.59--5.55 seconds) and memory (181.6--183.9 MiB) hardly move. The drop in wall time is therefore more plausibly tied to runner conditions and caching, which we cannot separate, than to less analysis work, and we do not read it as a speedup on newer code. The resource records also do not exist per Study-B or Study-C draw, so we make no matched efficiency claim.

\finding{Q2}{Final success is necessary but not sufficient evidence of reliability. CodeQL targets change status in both directions, and recovery can change the query set and extraction scope; Infer recovers thousands of empty or failed captures into scored successes; and Semgrep reports success for every run although 78.96\% of cases carry parsing or timeout diagnostics. Costs are heavy-tailed, and the recorded fields differ too much across tools to support a cross-tool ranking.}

\section{Discussion}\label{sec:discussion}

\subsection{What SAST Sustainability Means in Practice}\label{subsec:synthesis}

Temporal sustainability is a relation among a detector condition, a code population, a measurement scope, and an execution environment; it is not a number attached to a tool name. Three distinctions organize our evidence. First, \emph{level and trajectory differ}. Later CodeQL releases score higher but lose ground within projects faster than R2021, and Infer v1.2 has the highest level while v1.1 and v1.3 have the more positive matched-gap profiles. Second, \emph{the three designs answer different questions}. Calendar-year populations, matched pairs, and within-project slopes disagree in several places (for example, Infer's ordering flips between Studies A and B), and the disagreements we examined trace back to which projects and cases carry each comparison. No single curve substitutes for the three. Third, \emph{label agreement, mechanism evidence, and operational completion are separate outcomes}. A changed-file warning may be unrelated to the repair, and a final success may follow retries, query exclusion, or an empty capture. Treating every degraded run as a negative prediction would mix reliability into effectiveness; ignoring degraded runs would overstate coverage.

One pattern runs through almost every condition. 
Across many of our release and year comparisons, higher case-label F1 accompanies increases in both recall and FPR, while count-based AUROC remains near 0.5. 
Infer v1.2 is the exception, because it flags few fixed cases. A reader who looks at F1 alone, or at recall alone, would conclude that newer tools or newer code are handled better. Read together with FPR and AUROC, the same numbers say that the tools flag more of everything.

\subsection{Implications}\label{subsec:implications}

\head{For tool builders.} Release regression suites should replay project histories, not only fixed benchmarks, and report repair-aligned detections, front-end coverage (parse, extraction, and capture rates), query-level runtime tails, and same-target status transitions. Recall should always be paired with a fixed-snapshot control, so that broader alerting is not mistaken for better discrimination; our CodeQL and Infer results show how easily the two diverge. Dashboards should distinguish projects with stable nonzero performance from projects that never produce a signal. Recovery runs should report which queries or files they dropped; for BlueZ, per-query time budgets with explicit reporting of skipped queries would have made a silent timeout visible. Build wrappers should not take the presence of an object file as proof of compilation, and buildless extractors should report which compilers they searched for.

\head{For practitioners.} Before upgrading the analyzer in a CI gate, run the candidate release on fixed commits from the organization's own history and compare repair-relevant warnings, analyzed scope, diagnostics, status, and runtime with the current release. Default configurations are common in practice~\cite{bennett2024developers}, but our results show that the configuration, extraction mode, and recovery policy can change outcomes as much as the engine version. A run that ends in ``success'' should be checked for parse and capture diagnostics before its silence is trusted. SAST should remain one control among tests, fuzzing, review, and runtime defenses; historical scores do not show zero-day coverage.

\head{For benchmark designers and researchers.} A temporal benchmark should store the vulnerable parent, the fix, verified repair hunks, the source year, a canonical project identity, a build recipe, the detector condition, and attempt-level provenance. Scoring should use comparable semantic scopes on both labels; our Semgrep queue shows that adjacent commits can still be scored on disjoint files, and that two identifiers can point at one snapshot. Broad proxies should be calibrated with blinded mechanism review, and the inference unit should be explicit: 100 resamples are not 100 detections, nested cohorts are not additive, and a single year pair is one comparison.

\subsection{Why This Matters for Zero-Day Discovery}\label{subsec:zeroday}

SAST is most valuable before a defect has a name: a general taint query, abstract domain, or bounds rule can expose an instance that later becomes a CVE. Our results identify three ways in which this transfer to new code can fail. The defect idiom can fall outside the rule or query predicate (Nagios, Varnish); the relevant code can never reach the analysis because of build, capture, parsing, or scoping problems (MilkyTracker, QEMU, OpenBSD); or unrelated warnings can mask the absence of mechanism coverage (libxml2, libiec61850). We do not equate later-year CVEs with zero-days. What we measure is a necessary condition for prospective discovery: rule semantics that remain applicable, together with observable analysis coverage, as code evolves.

\subsection{What the Design Leaves Open}\label{subsec:open}

Several analyses would sharpen these results, and we did not run them. Study~B pools cases within each pair, so large projects weigh more; a project-level variant that computes $\dF$ per project and then averages projects would show whether a few projects drive a pair. Study~C reports mean slopes and their signs; median slopes with uncertainty per project, and a pooled model with a random effect per project (for example, a logistic model of per-case correctness on source year), would turn the descriptive slopes into an estimate. Our year-pair ranges describe resampling variation; they are not confidence intervals. A hierarchical model with project, CWE, scope, and condition effects is the natural way to obtain intervals for the population of projects. Finally, reliability and cost were measured on the stored Study-A runs only. Following failure rates and run time along the Study-C trajectories would need resource records per resampled case, which the archives do not keep. Each extension can reuse the frozen memberships and results in our package.

\subsection{Threats to Validity}\label{subsec:threats}

\head{Construct validity.} Our score is a proxy: an in-scope alert counts as a detection without CWE agreement or repair overlap, and F1 on a balanced corpus is not the precision a developer would see. We therefore report AUROC and FPR beside F1 and calibrate the selected extremes with repair-file and repair-line ablations; the ablations cover the selected cases, not all 20,000. Count-based AUROC is a weak notion of ranking, but it treats all tools alike. Labels, source years, and CVE-to-repair links come from the corpus. We exclude the 256 case identities that carry both labels from label-based evaluation for all tools and checked parent-to-fix relations for the inspected cases, but label noise known from such datasets~\cite{croft2023data} may remain, and commit adjacency does not validate every CVE-to-repair binding. The CodeQL score depends on a security-relevance filter whose effect we checked only for the selected cases, where it removed no alert in a scored file.

\head{Internal validity.} A detector condition bundles engine, rules, extraction mode, recovery, and cache state, so release differences are not pure version effects; CodeQL R2024, for example, is also the only buildless historical condition. Apart from the shared label-conflict gate, each tool has its own exclusion gate. The common-intersection results preserve every ordering we report, but they keep only cases that every release completed and may favor easy targets. Infer keeps the highest-scoring completed candidate within each reuse group. This rule can raise both true and false positives, its effect can differ between releases with different numbers of candidates, and the common intersection does not control it; we have not quantified it. Matching fixes projects and label counts but not CWE, module, code size, or analysis quality, and in some pairs one project supplies most of the matched cases. Cost figures describe operating ranges, not tool speed, because machine load and caching were not controlled.

\head{Conclusion validity.} Year pairs share projects, repetitions resample a fixed pool, and the Study-C cohorts are nested, so significance tests that assume independence would overstate certainty. We report spread and support instead, rely on sign patterns that recur across many pairs or survive the sensitivity cohort, and draw no conclusion from gap~9, which rests on one pair. Repetition-level spread is available only for selected pairs (two Infer pairs and six Semgrep pairs).

\head{External validity.} We study open-source C/C++ code, three analyzers, and default or locally curated rules on changed files. Other languages, commercial rules, proprietary code, and whole-repository scans may behave differently. 
The source years cover 2016–2025. The complete year-pair analysis intentionally includes code that predates some detector conditions and therefore characterizes temporal sensitivity across code vintages, rather than necessarily prospective aging. Only comparisons between code populations following the frozen detector's release can provide direct evidence of prospective aging.

\head{Qualitative and operational evidence.} The RQ4 queues are selected extremes; they show that a mechanism exists, not how often. Missing historical intermediate representations, environments, and links to the selected attempt prevent a unique causal attribution for many historical failures, so we grade each claim by its evidence. CPU, memory, retries, cache state, queueing, and machine load are recorded incompletely. Language-model agents assisted source and log inspection, and the cross-check was not blind. We therefore tied every claim to source, diff, rule, query, or log evidence that others can inspect.

\section{Related Work}\label{sec:related-work}

\head{SAST effectiveness on real vulnerabilities.} Cross-sectional studies agree that SAST misses many real vulnerabilities. Li et al.\ report that seven Java SAST tools together leave 70.9\% of real-world vulnerabilities undetected~\cite{li2023comparison}; Lipp et al.\ find that C analyzers miss 47--80\% of 192 real vulnerabilities~\cite{lipp2022empirical}; Habib and Pradel find that three Java bug detectors catch only a small fraction of 594 real bugs~\cite{habib2018many}; and Bennett et al.\ measure individual detection rates of 11.2--26.5\% before extending Semgrep's rules~\cite{bennett2024semgrep}. A synthetic Java benchmark shows high precision but low recall~\cite{esposito2024extensive}, rule-based and learning-based approaches have been compared directly~\cite{croft2021empirical}, and NIST's SATE exercises evaluate tools on injected and collected bugs~\cite{delaitre2023sate}. Charoenwet et al.\ study SAST warnings in secure code review~\cite{charoenwet2024secure}. All of these studies evaluate one point in time. We show that a reasonable score can coexist with a finding count that barely separates vulnerable from fixed code, and that a fixed detector's score moves as the code ages.

\head{Static analysis in practice.} Industrial reports describe what keeps analyzers useful at scale~\cite{bessey2010few,sadowski2018lessons,distefano2019scaling}, and developer studies document distrust of false positives, workflow friction, and context-dependent use~\cite{johnson2013developers,christakis2016developers,vassallo2020developers,imtiaz2019challenges,beller2016analyzing}. Practitioners fear false negatives most and rarely benchmark tools~\cite{ami2024false}, many keep default configurations~\cite{bennett2024developers}, and work on actionable-alert identification and false-positive mitigation~\cite{heckman2011actionable,kang2022detecting,guo2023mitigating} focuses on warning volume and triage. We add a temporal concern: settings that teams rarely revisit, such as extraction mode and recovery policy, can decide whether new code is analyzed at all, and large warning volumes can produce high label-level scores with chance-level discrimination.

\head{Longitudinal studies of static analysis.} Earlier longitudinal work follows the warning population: how statically detected vulnerabilities decay~\cite{dipenta2008evolution}, how security warnings persist in C++ projects~\cite{aloraini2019warnings}, how PMD warnings evolve in Apache projects~\cite{trautsch2020warning}, and how warning history can guide prioritization~\cite{kim2007which}. The closest work, by Noirot Ferrand et al., runs 114 CodeQL versions on 3,993 CVEs and finds that detections are not monotonic across versions~\cite{ferrand2026longitudinal}. They vary the detector; we fix it and vary the age of the code, for three analyzers, with exact project--label matching, within-project trajectories, operational ledgers, and repair-level checks. Because detectors and code evolve together, the two designs complement each other.

\head{Temporal evaluation in security and defect prediction.} Building on concept-drift research~\cite{gama2014survey}, malware studies showed that classifiers age and that mixing time periods inflates results~\cite{jordaney2017transcend,pendlebury2019tesseract,yang2021cade,arp2022dos}. Learning-based vulnerability and bug detectors degrade on realistic or shifted data~\cite{chakraborty2022deep,he2022distribution,risse2024uncovering,ding2025vulnerability}, vulnerability prediction depends on realistic labelling over time~\cite{jimenez2019importance}, and defect-prediction conclusions change with the evaluation period~\cite{bangash2020time}. These works motivate chronological, composition-aware evaluation. We bring it to SAST, where parsers, builds, rule selection, and recovery can drift without any learned model. 
The concept of sustainability in security defense was earlier proposed~\cite{cai2018towards} 
and later studied~\cite{cai2020assessing} for malware detection, including characterization concerning its underlying driving mechanisms due to software evolution~\cite{cai2020embracing}, but all for mobile apps, as opposed to the sustainability of 
static security analyzers as focused in this paper.  

\head{Vulnerability data and buildability.} Vulnerability-fix datasets such as Big-Vul, Devign, CVEfixes, DiverseVul, and PrimeVul~\cite{fan2020bigvul,zhou2019devign,bhandari2021cvefixes,chen2023diversevul,ding2025vulnerability} supply the vulnerable/fixed pairs that designs like ours depend on, and studies of security patches and vulnerability lifetimes~\cite{li2017large,alexopoulos2022vulnerabilities} characterize the repairs. Data-quality studies report label noise and duplication in such datasets~\cite{croft2023data}; our label-conflict gate, scope-alignment analysis, and the duplicate libxml2 snapshot show the same concern for SAST scoring. Work on rebuilding past snapshots~\cite{tufano2017there,maesbermejo2022revisiting} explains why build-dependent analyzers such as Infer and traced CodeQL lose historical targets. Graph-based query frameworks such as code property graphs~\cite{yamaguchi2014modeling} and analyzers that add LLM-inferred specifications to CodeQL~\cite{li2025iris} inherit the front-end and query dependencies that we show to be time-sensitive.

\section{Conclusion}\label{sec:conclusion}

This paper studies whether SAST detectors remain effective, executable, and affordable as the code they analyze evolves. Using yearly populations, project-matched year pairs, and within-project trajectories over a decade of C/C++ vulnerabilities, 
we found that temporal score changes can arise from broader alerting, shifting evaluation scope, and alerts unrelated to the labeled repair, while selected genuine misses arise from rule/query mismatch and incomplete analysis coverage. 
These results show that label-level score changes alone cannot establish corresponding changes in vulnerability-mechanism coverage. 
SAST sustainability should therefore be assessed by replaying fixed detector configurations on project histories, with repair-level evidence and full reporting of analysis coverage. Future work will extend the study to other languages, commercial and learning-based detectors, and labels finer than the changed-file proxy.

\section*{Data Availability}\label{sec:data-availability}

Our replication package (\url{https://anonymous.4open.science/r/sast-sustainability}) contains the frozen case ledger, the canonical project map, Study-B/C membership files, normalized findings, the reliability registries, analysis and figure scripts, a data dictionary, and a claim-to-evidence manifest that links every number in this paper to its source artifact. Third-party source snapshots and tool distributions are referenced by commit and version and redistributed only where licenses permit.

\bibliographystyle{ACM-Reference-Format}
\bibliography{paper-fse27-arXiv}

@article{ferrand2026longitudinal,
  author  = {Jean-Charles {Noirot Ferrand} and Kyle Domico and Yohan Beugin and Patrick McDaniel},
  title   = {Longitudinal Analyses of {SAST} Tools: A {CodeQL} Case Study},
  journal = {arXiv preprint arXiv:2605.07900},
  year    = {2026},
  doi     = {10.48550/arXiv.2605.07900}
}

@inproceedings{li2023comparison,
  author    = {Kaixuan Li and Sen Chen and Lingling Fan and Ruitao Feng and Han Liu and Chengwei Liu and Yang Liu and Yixiang Chen},
  title     = {Comparison and Evaluation on Static Application Security Testing ({SAST}) Tools for {Java}},
  booktitle = {Proceedings of the 31st ACM Joint European Software Engineering Conference and Symposium on the Foundations of Software Engineering (ESEC/FSE)},
  pages     = {921--933},
  year      = {2023},
  publisher = {ACM},
  doi       = {10.1145/3611643.3616262}
}

@inproceedings{lipp2022empirical,
  author    = {Stephan Lipp and Sebastian Banescu and Alexander Pretschner},
  title     = {An Empirical Study on the Effectiveness of Static {C} Code Analyzers for Vulnerability Detection},
  booktitle = {Proceedings of the 31st ACM SIGSOFT International Symposium on Software Testing and Analysis (ISSTA)},
  pages     = {544--555},
  year      = {2022},
  publisher = {ACM},
  doi       = {10.1145/3533767.3534380}
}

@inproceedings{habib2018many,
  author    = {Andrew Habib and Michael Pradel},
  title     = {How Many of All Bugs Do We Find? {A} Study of Static Bug Detectors},
  booktitle = {Proceedings of the 33rd IEEE/ACM International Conference on Automated Software Engineering (ASE)},
  pages     = {317--328},
  year      = {2018},
  publisher = {ACM},
  doi       = {10.1145/3238147.3238213}
}

@inproceedings{bennett2024semgrep,
  author    = {G. Bennett and T. Hall and E. Winter and S. Counsell},
  title     = {Semgrep*: Improving the Limited Performance of Static Application Security Testing ({SAST}) Tools},
  booktitle = {Proceedings of the 28th International Conference on Evaluation and Assessment in Software Engineering (EASE)},
  pages     = {614--623},
  year      = {2024},
  publisher = {ACM},
  doi       = {10.1145/3661167.3661262}
}

@inproceedings{bennett2024developers,
  author    = {G. Bennett and T. Hall and S. Counsell and T. Shippey},
  title     = {Do Developers Use Static Application Security Testing ({SAST}) Tools Straight Out of the Box? {A} Large-Scale Empirical Study},
  booktitle = {Proceedings of the 18th ACM/IEEE International Symposium on Empirical Software Engineering and Measurement (ESEM)},
  pages     = {454--460},
  year      = {2024},
  publisher = {ACM},
  doi       = {10.1145/3674805.3690750}
}

@inproceedings{esposito2024extensive,
  author    = {Matteo Esposito and Valentina Falaschi and Davide Falessi},
  title     = {An Extensive Comparison of Static Application Security Testing Tools},
  booktitle = {Proceedings of the 28th International Conference on Evaluation and Assessment in Software Engineering (EASE)},
  pages     = {69--78},
  year      = {2024},
  publisher = {ACM},
  doi       = {10.1145/3661167.3661199}
}

@inproceedings{croft2021empirical,
  author    = {Roland Croft and Dominic Newlands and Ziyu Chen and M. Ali Babar},
  title     = {An Empirical Study of Rule-Based and Learning-Based Approaches for Static Application Security Testing},
  booktitle = {Proceedings of the 15th ACM/IEEE International Symposium on Empirical Software Engineering and Measurement (ESEM)},
  pages     = {1--12},
  year      = {2021},
  publisher = {ACM},
  doi       = {10.1145/3475716.3475781}
}

@techreport{delaitre2023sate,
  author      = {Aurelien Delaitre and Paul E. Black and Damien Cupif and Guillaume Haben and {Loembe Alex-Kevin} and Vadim Okun and Yann Prono},
  title       = {{SATE VI} Report: Bug Injection and Collection},
  institution = {National Institute of Standards and Technology},
  type        = {NIST Special Publication},
  number      = {500-341},
  year        = {2023},
  url         = {https://nvlpubs.nist.gov/nistpubs/SpecialPublications/NIST.SP.500-341.pdf}
}

@inproceedings{charoenwet2024secure,
  author    = {Wachiraphan Charoenwet and Patanamon Thongtanunam and Van-Thuan Pham and Christoph Treude},
  title     = {An Empirical Study of Static Analysis Tools for Secure Code Review},
  booktitle = {Proceedings of the 33rd ACM SIGSOFT International Symposium on Software Testing and Analysis (ISSTA)},
  pages     = {691--703},
  year      = {2024},
  publisher = {ACM},
  doi       = {10.1145/3650212.3680313}
}

@inproceedings{johnson2013developers,
  author    = {Brittany Johnson and Yoonki Song and Emerson Murphy-Hill and Robert Bowdidge},
  title     = {Why Don't Software Developers Use Static Analysis Tools to Find Bugs?},
  booktitle = {Proceedings of the 35th International Conference on Software Engineering (ICSE)},
  pages     = {672--681},
  year      = {2013},
  publisher = {IEEE},
  doi       = {10.1109/ICSE.2013.6606613}
}

@inproceedings{christakis2016developers,
  author    = {Maria Christakis and Christian Bird},
  title     = {What Developers Want and Need from Program Analysis: An Empirical Study},
  booktitle = {Proceedings of the 31st IEEE/ACM International Conference on Automated Software Engineering (ASE)},
  pages     = {332--343},
  year      = {2016},
  publisher = {ACM},
  doi       = {10.1145/2970276.2970347}
}

@article{vassallo2020developers,
  author  = {Carmine Vassallo and Sebastiano Panichella and Fabio Palomba and Sebastian Proksch and Harald C. Gall and Andy Zaidman},
  title   = {How Developers Engage with Static Analysis Tools in Different Contexts},
  journal = {Empirical Software Engineering},
  volume  = {25},
  number  = {2},
  pages   = {1419--1457},
  year    = {2020},
  doi     = {10.1007/s10664-019-09750-5}
}

@inproceedings{imtiaz2019challenges,
  author    = {Nasif Imtiaz and Akond Rahman and Effat Farhana and Laurie Williams},
  title     = {Challenges with Responding to Static Analysis Tool Alerts},
  booktitle = {Proceedings of the 16th International Conference on Mining Software Repositories (MSR)},
  pages     = {245--249},
  year      = {2019},
  publisher = {IEEE},
  doi       = {10.1109/MSR.2019.00049}
}

@inproceedings{beller2016analyzing,
  author    = {Moritz Beller and Radjino Bholanath and Shane McIntosh and Andy Zaidman},
  title     = {Analyzing the State of Static Analysis: A Large-Scale Evaluation in Open Source Software},
  booktitle = {Proceedings of the 23rd IEEE International Conference on Software Analysis, Evolution, and Reengineering (SANER)},
  pages     = {470--481},
  year      = {2016},
  publisher = {IEEE},
  doi       = {10.1109/SANER.2016.105}
}

@inproceedings{ami2024false,
  author    = {Amit Seal Ami and Kevin Moran and Denys Poshyvanyk and Adwait Nadkarni},
  title     = {``{False} Negative -- That One Is Going to Kill You'': Understanding Industry Perspectives of Static Analysis Based Security Testing},
  booktitle = {Proceedings of the 2024 IEEE Symposium on Security and Privacy (S\&P)},
  year      = {2024},
  publisher = {IEEE},
}

@article{heckman2011actionable,
  author  = {Sarah Heckman and Laurie Williams},
  title   = {A Systematic Literature Review of Actionable Alert Identification Techniques for Automated Static Code Analysis},
  journal = {Information and Software Technology},
  volume  = {53},
  number  = {4},
  pages   = {363--387},
  year    = {2011},
  doi     = {10.1016/j.infsof.2010.12.007}
}

@inproceedings{kang2022detecting,
  author    = {Hong Jin Kang and Khai Loong Aw and David Lo},
  title     = {Detecting False Alarms from Automatic Static Analysis Tools: How Far Are We?},
  booktitle = {Proceedings of the 44th International Conference on Software Engineering (ICSE)},
  year      = {2022},
  publisher = {ACM},
  doi       = {10.1145/3510003.3510214}
}

@article{guo2023mitigating,
  author  = {Zhaoqiang Guo and Tingting Tan and Shiran Liu and Xutong Liu and Wei Lai and Yibiao Yang and Yanhui Li and Lin Chen and Wei Dong and Yuming Zhou},
  title   = {Mitigating False Positive Static Analysis Warnings: Progress, Challenges, and Opportunities},
  journal = {IEEE Transactions on Software Engineering},
  volume  = {49},
  number  = {12},
  pages     = {5154--5188},
  year    = {2023},
  doi     = {10.1109/TSE.2023.3329667}
}

@inproceedings{kim2007which,
  author    = {Sunghun Kim and Michael D. Ernst},
  title     = {Which Warnings Should {I} Fix First?},
  booktitle = {Proceedings of the 6th Joint Meeting of the European Software Engineering Conference and the ACM SIGSOFT Symposium on the Foundations of Software Engineering (ESEC/FSE)},
  pages     = {45--54},
  year      = {2007},
  publisher = {ACM},
  doi       = {10.1145/1287624.1287633}
}

@inproceedings{dipenta2008evolution,
  author    = {Massimiliano {Di Penta} and Luigi Cerulo and Lerina Aversano},
  title     = {The Evolution and Decay of Statically Detected Source Code Vulnerabilities},
  booktitle = {Proceedings of the 8th IEEE International Working Conference on Source Code Analysis and Manipulation (SCAM)},
  pages     = {101--110},
  year      = {2008},
  publisher = {IEEE},
  doi       = {10.1109/SCAM.2008.20}
}

@article{aloraini2019warnings,
  author  = {Bushra Aloraini and Meiyappan Nagappan and Daniel M. German and Shinpei Hayashi and Yoshiki Higo},
  title   = {An Empirical Study of Security Warnings from Static Application Security Testing Tools},
  journal = {Journal of Systems and Software},
  volume  = {158},
  pages   = {110427},
  year    = {2019},
  doi     = {10.1016/j.jss.2019.110427}
}

@article{trautsch2020warning,
  author  = {Alexander Trautsch and Steffen Herbold and Jens Grabowski},
  title   = {A Longitudinal Study of Static Analysis Warning Evolution and the Effects of {PMD} on Software Quality in {Apache} Open Source Projects},
  journal = {Empirical Software Engineering},
  volume  = {25},
  number  = {6},
  pages   = {5137--5192},
  year    = {2020},
  doi     = {10.1007/s10664-020-09880-1}
}

@article{tufano2017there,
  author  = {Michele Tufano and Fabio Palomba and Gabriele Bavota and Massimiliano {Di Penta} and Rocco Oliveto and Andrea {De Lucia} and Denys Poshyvanyk},
  title   = {There and Back Again: Can You Compile That Snapshot?},
  journal = {Journal of Software: Evolution and Process},
  volume  = {29},
  number  = {4},
  pages   = {e1838},
  year    = {2017},
  doi     = {10.1002/smr.1838}
}

@article{maesbermejo2022revisiting,
  author  = {Michel Maes-Bermejo and Micael Gallego and Francisco Gort{\'a}zar and Gregorio Robles and Jes{\'u}s M. Gonz{\'a}lez-Barahona},
  title   = {Revisiting the Building of Past Snapshots -- A Replication and Reproduction Study},
  journal = {Empirical Software Engineering},
  volume  = {27},
  number  = {3},
  pages   = {65},
  year    = {2022},
  doi     = {10.1007/s10664-022-10117-6}
}

@inproceedings{jordaney2017transcend,
  author    = {Roberto Jordaney and Kumar Sharad and Santanu K. Dash and Zhi Wang and Davide Papini and Ilia Nouretdinov and Lorenzo Cavallaro},
  title     = {Transcend: Detecting Concept Drift in Malware Classification Models},
  booktitle = {Proceedings of the 26th USENIX Security Symposium},
  pages     = {625--642},
  year      = {2017},
  publisher = {USENIX Association}
}

@inproceedings{pendlebury2019tesseract,
  author    = {Feargus Pendlebury and Fabio Pierazzi and Roberto Jordaney and Johannes Kinder and Lorenzo Cavallaro},
  title     = {{TESSERACT}: Eliminating Experimental Bias in Malware Classification across Space and Time},
  booktitle = {Proceedings of the 28th USENIX Security Symposium},
  pages     = {729--746},
  year      = {2019},
  publisher = {USENIX Association}
}

@inproceedings{yang2021cade,
  author    = {Limin Yang and Wenbo Guo and Qingying Hao and Arridhana Ciptadi and Ali Ahmadzadeh and Xinyu Xing and Gang Wang},
  title     = {{CADE}: Detecting and Explaining Concept Drift Samples for Security Applications},
  booktitle = {Proceedings of the 30th USENIX Security Symposium},
  pages     = {2327--2344},
  year      = {2021},
  publisher = {USENIX Association}
}

@inproceedings{arp2022dos,
  author    = {Daniel Arp and Erwin Quiring and Feargus Pendlebury and Alexander Warnecke and Fabio Pierazzi and Christian Wressnegger and Lorenzo Cavallaro and Konrad Rieck},
  title     = {Dos and Don'ts of Machine Learning in Computer Security},
  booktitle = {Proceedings of the 31st USENIX Security Symposium},
  pages     = {3971--3988},
  year      = {2022},
  publisher = {USENIX Association}
}

@article{gama2014survey,
  author  = {Jo{\~a}o Gama and Indr{\.e} {\v{Z}}liobait{\.e} and Albert Bifet and Mykola Pechenizkiy and Abdelhamid Bouchachia},
  title   = {A Survey on Concept Drift Adaptation},
  journal = {ACM Computing Surveys},
  volume  = {46},
  number  = {4},
  pages   = {44:1--44:37},
  year    = {2014},
  doi     = {10.1145/2523813}
}

@inproceedings{he2022distribution,
  author    = {Jingxuan He and Luca Beurer-Kellner and Martin Vechev},
  title     = {On Distribution Shift in Learning-Based Bug Detectors},
  booktitle = {Proceedings of the 39th International Conference on Machine Learning (ICML)},
  series    = {Proceedings of Machine Learning Research},
  volume    = {162},
  pages     = {8559--8580},
  year      = {2022},
  publisher = {PMLR}
}

@inproceedings{risse2024uncovering,
  author    = {Niklas Risse and Marcel B{\"o}hme},
  title     = {Uncovering the Limits of Machine Learning for Automatic Vulnerability Detection},
  booktitle = {Proceedings of the 33rd USENIX Security Symposium},
  pages     = {4247--4264},
  year      = {2024},
  publisher = {USENIX Association}
}

@article{chakraborty2022deep,
  author  = {Saikat Chakraborty and Rahul Krishna and Yangruibo Ding and Baishakhi Ray},
  title   = {Deep Learning Based Vulnerability Detection: Are We There Yet?},
  journal = {IEEE Transactions on Software Engineering},
  volume  = {48},
  number  = {9},
  pages   = {3280--3296},
  year    = {2022},
  doi     = {10.1109/TSE.2021.3087402}
}

@inproceedings{ding2025vulnerability,
  author    = {Yangruibo Ding and Yanjun Fu and Omniyyah Ibrahim and Chawin Sitawarin and Xinyun Chen and Basel Alomair and David Wagner and Baishakhi Ray and Yizheng Chen},
  title     = {Vulnerability Detection with Code Language Models: How Far Are We?},
  booktitle = {Proceedings of the 47th IEEE/ACM International Conference on Software Engineering (ICSE)},
  year      = {2025},
  publisher = {IEEE},
  doi       = {10.1109/ICSE55347.2025.00038}
}

@inproceedings{jimenez2019importance,
  author    = {Matthieu Jimenez and Renaud Rwemalika and Mike Papadakis and Federica Sarro and Yves {Le Traon} and Mark Harman},
  title     = {The Importance of Accounting for Real-World Labelling When Predicting Software Vulnerabilities},
  booktitle = {Proceedings of the 27th ACM Joint Meeting on European Software Engineering Conference and Symposium on the Foundations of Software Engineering (ESEC/FSE)},
  year      = {2019},
  publisher = {ACM},
  doi       = {10.1145/3338906.3338941}
}

@article{bangash2020time,
  author  = {Abdul Ali Bangash and Hareem Sahar and Abram Hindle and Karim Ali},
  title   = {On the Time-Based Conclusion Stability of Cross-Project Defect Prediction Models},
  journal = {Empirical Software Engineering},
  volume  = {25},
  number  = {6},
  pages   = {5047--5083},
  year    = {2020},
  doi     = {10.1007/s10664-020-09878-9}
}

@inproceedings{fan2020bigvul,
  author    = {Jiahao Fan and Yi Li and Shaohua Wang and Tien N. Nguyen},
  title     = {A {C/C++} Code Vulnerability Dataset with Code Changes and {CVE} Summaries},
  booktitle = {Proceedings of the 17th International Conference on Mining Software Repositories (MSR)},
  pages     = {508--512},
  year      = {2020},
  publisher = {ACM},
  doi       = {10.1145/3379597.3387501}
}

@inproceedings{zhou2019devign,
  author    = {Yaqin Zhou and Shangqing Liu and Jingkai Siow and Xiaoning Du and Yang Liu},
  title     = {Devign: Effective Vulnerability Identification by Learning Comprehensive Program Semantics via Graph Neural Networks},
  booktitle = {Advances in Neural Information Processing Systems 32 (NeurIPS)},
  year      = {2019}
}

@inproceedings{bhandari2021cvefixes,
  author    = {Guru Bhandari and Amara Naseer and Leon Moonen},
  title     = {{CVEfixes}: Automated Collection of Vulnerabilities and Their Fixes from Open-Source Software},
  booktitle = {Proceedings of the 17th International Conference on Predictive Models and Data Analytics in Software Engineering (PROMISE)},
  year      = {2021},
  publisher = {ACM},
  doi       = {10.1145/3475960.3475985}
}

@inproceedings{chen2023diversevul,
  author    = {Yizheng Chen and Zhoujie Ding and Lamya Alowain and Xinyun Chen and David Wagner},
  title     = {{DiverseVul}: A New Vulnerable Source Code Dataset for Deep Learning Based Vulnerability Detection},
  booktitle = {Proceedings of the 26th International Symposium on Research in Attacks, Intrusions and Defenses (RAID)},
  year      = {2023},
  publisher = {ACM},
  doi       = {10.1145/3607199.3607242}
}

@inproceedings{croft2023data,
  author    = {Roland Croft and M. Ali Babar and M. Mehdi Kholoosi},
  title     = {Data Quality for Software Vulnerability Datasets},
  booktitle = {Proceedings of the 45th IEEE/ACM International Conference on Software Engineering (ICSE)},
  pages     = {121--133},
  year      = {2023},
  publisher = {IEEE},
  doi       = {10.1109/ICSE48619.2023.00022}
}

@inproceedings{li2017large,
  author    = {Frank Li and Vern Paxson},
  title     = {A Large-Scale Empirical Study of Security Patches},
  booktitle = {Proceedings of the 2017 ACM SIGSAC Conference on Computer and Communications Security (CCS)},
  pages     = {2201--2215},
  year      = {2017},
  publisher = {ACM},
  doi       = {10.1145/3133956.3134072}
}

@inproceedings{alexopoulos2022vulnerabilities,
  author    = {Nikolaos Alexopoulos and Manuel Brack and Jan Philipp Wagner and Tim Grube and Max M{\"u}hlh{\"a}user},
  title     = {How Long Do Vulnerabilities Live in the Code? {A} Large-Scale Empirical Measurement Study on {FOSS} Vulnerability Lifetimes},
  booktitle = {Proceedings of the 31st USENIX Security Symposium},
  pages     = {359--376},
  year      = {2022},
  publisher = {USENIX Association}
}

@inproceedings{bilge2012before,
  author    = {Leyla Bilge and Tudor Dumitra{\c{s}}},
  title     = {Before We Knew It: An Empirical Study of Zero-Day Attacks in the Real World},
  booktitle = {Proceedings of the 2012 ACM Conference on Computer and Communications Security (CCS)},
  pages     = {833--844},
  year      = {2012},
  publisher = {ACM},
  doi       = {10.1145/2382196.2382284}
}

@article{lehman1980programs,
  author  = {Meir M. Lehman},
  title   = {Programs, Life Cycles, and Laws of Software Evolution},
  journal = {Proceedings of the IEEE},
  volume  = {68},
  number  = {9},
  pages   = {1060--1076},
  year    = {1980},
  doi     = {10.1109/PROC.1980.11805}
}

@article{fawcett2006introduction,
  author  = {Tom Fawcett},
  title   = {An Introduction to {ROC} Analysis},
  journal = {Pattern Recognition Letters},
  volume  = {27},
  number  = {8},
  pages   = {861--874},
  year    = {2006},
  doi     = {10.1016/j.patrec.2005.10.010}
}

@article{stuart2010matching,
  author  = {Elizabeth A. Stuart},
  title   = {Matching Methods for Causal Inference: A Review and a Look Forward},
  journal = {Statistical Science},
  volume  = {25},
  number  = {1},
  pages   = {1--21},
  year    = {2010},
  doi     = {10.1214/09-STS313}
}

@article{bessey2010few,
  author  = {Al Bessey and Ken Block and Ben Chelf and Andy Chou and Bryan Fulton and Seth Hallem and Charles Henri-Gros and Asya Kamsky and Scott McPeak and Dawson Engler},
  title   = {A Few Billion Lines of Code Later: Using Static Analysis to Find Bugs in the Real World},
  journal = {Communications of the ACM},
  volume  = {53},
  number  = {2},
  pages   = {66--75},
  year    = {2010},
  doi     = {10.1145/1646353.1646374}
}

@article{sadowski2018lessons,
  author  = {Caitlin Sadowski and Edward Aftandilian and Alex Eagle and Liam Miller-Cushon and Ciera Jaspan},
  title   = {Lessons from Building Static Analysis Tools at {Google}},
  journal = {Communications of the ACM},
  volume  = {61},
  number  = {4},
  pages   = {58--66},
  year    = {2018},
  doi     = {10.1145/3188720}
}

@article{distefano2019scaling,
  author  = {Dino Distefano and Manuel F{\"a}hndrich and Francesco Logozzo and Peter W. O'Hearn},
  title   = {Scaling Static Analyses at {Facebook}},
  journal = {Communications of the ACM},
  volume  = {62},
  number  = {8},
  pages   = {62--70},
  year    = {2019},
  doi     = {10.1145/3338112}
}

@inproceedings{demoor2007ql,
  author    = {Oege de Moor and Mathieu Verbaere and Elnar Hajiyev and Pavel Avgustinov and Torbj{\"o}rn Ekman and Neil Ongkingco and Damien Sereni and Julian Tibble},
  title     = {Keynote Address: {.QL} for Source Code Analysis},
  booktitle = {Proceedings of the 7th IEEE International Working Conference on Source Code Analysis and Manipulation (SCAM)},
  year      = {2007},
  publisher = {IEEE},
}

@inproceedings{avgustinov2016ql,
  author    = {Pavel Avgustinov and Oege de Moor and Michael Peyton Jones and Max Sch{\"a}fer},
  title     = {{QL}: Object-Oriented Queries on Relational Data},
  booktitle = {Proceedings of the 30th European Conference on Object-Oriented Programming (ECOOP)},
  series    = {LIPIcs},
  volume    = {56},
  pages     = {2:1--2:25},
  year      = {2016},
  publisher = {Schloss Dagstuhl--Leibniz-Zentrum f{\"u}r Informatik},
  doi       = {10.4230/LIPIcs.ECOOP.2016.2}
}

@article{calcagno2011compositional,
  author  = {Cristiano Calcagno and Dino Distefano and Peter W. O'Hearn and Hongseok Yang},
  title   = {Compositional Shape Analysis by Means of Bi-Abduction},
  journal = {Journal of the ACM},
  volume  = {58},
  number  = {6},
  pages   = {1--66},
  year    = {2011},
  doi     = {10.1145/2049697.2049700}
}

@inproceedings{calcagno2015moving,
  author    = {Cristiano Calcagno and Dino Distefano and J{\'e}r{\'e}my Dubreil and Dominik Gabi and Pieter Hooimeijer and Martino Luca and Peter W. O'Hearn and Irene Papakonstantinou and Jim Purbrick and Dulma Rodriguez},
  title     = {Moving Fast with Software Verification},
  booktitle = {NASA Formal Methods (NFM)},
  series    = {LNCS},
  volume    = {9058},
  pages     = {3--11},
  year      = {2015},
  publisher = {Springer},
  doi       = {10.1007/978-3-319-17524-9_1}
}

@inproceedings{ohearn2018continuous,
  author    = {Peter W. O'Hearn},
  title     = {Continuous Reasoning: Scaling the Impact of Formal Methods},
  booktitle = {Proceedings of the 33rd Annual ACM/IEEE Symposium on Logic in Computer Science (LICS)},
  year      = {2018},
  publisher = {ACM},
  doi       = {10.1145/3209108.3209109}
}

@inproceedings{yamaguchi2014modeling,
  author    = {Fabian Yamaguchi and Nico Golde and Daniel Arp and Konrad Rieck},
  title     = {Modeling and Discovering Vulnerabilities with Code Property Graphs},
  booktitle = {Proceedings of the 2014 IEEE Symposium on Security and Privacy (S\&P)},
  pages     = {590--604},
  year      = {2014},
  publisher = {IEEE},
  doi       = {10.1109/SP.2014.44}
}

@inproceedings{li2025iris,
  author    = {Ziyang Li and Saikat Dutta and Mayur Naik},
  title     = {{IRIS}: {LLM}-Assisted Static Analysis for Detecting Security Vulnerabilities},
  booktitle = {Proceedings of the 13th International Conference on Learning Representations (ICLR)},
  year      = {2025}
}

@misc{semgrep2026glossary,
  author       = {{Semgrep}},
  title        = {Static Analysis and Rule-Writing Glossary},
  year         = {2026},
  howpublished = {Semgrep documentation, \url{https://semgrep.dev/docs/writing-rules/glossary}},
  note         = {Accessed 2026-10-01}
}

@inproceedings{cai2020embracing,
  title={Embracing mobile app evolution via continuous ecosystem mining and characterization},
  author={Cai, Haipeng},
  booktitle={Proceedings of the IEEE/ACM 7th International Conference on Mobile Software Engineering and Systems},
  pages={31--35},
  year={2020}
}

@article{cai2020assessing,
  title={Assessing and improving malware detection sustainability through app evolution studies},
  author={Cai, Haipeng},
  journal={ACM Transactions on Software Engineering and Methodology (TOSEM)},
  volume={29},
  number={2},
  pages={1--28},
  year={2020},
  publisher={ACM, New York, NY, USA}
}

@inproceedings{cai2018towards,
  title={Towards sustainable android malware detection},
  author={Cai, Haipeng and Jenkins, John},
  booktitle={Proceedings of the 40th International Conference on Software Engineering: Companion Proceeedings},
  pages={350--351},
  year={2018}
}

\end{document}